\documentclass[pdflatex, sn-aps]{sn-jnl}% American Physical Society (APS) Reference Style
\jyear{2022}%

\usepackage{graphicx}
\usepackage[american]{babel}
\usepackage{url}
\usepackage{amsmath}
\usepackage{rotating}
\usepackage{multirow}
\usepackage{booktabs}
\usepackage{parskip}

\usepackage{longtable}

\newcommand{\osm}{online supplement }
\newcommand{\osmnosp}{online supplement}

\begin{document}
\pagenumbering{arabic}
\title[Conceptions of honesty]{From Alternative conceptions of honesty to
alternative facts in communications by U.S. politicians}


\author[1,2]{\fnm{Jana} \sur{Lasser}}

\author[1,3]{\fnm{Segun T.} \sur{Aroyehun}}

\author[4]{\fnm{Fabio} \sur{Carrella}}

\author[4]{\fnm{Almog} \sur{Simchon}}

\author[1,2,3]{\fnm{David} \sur{Garcia}}

\author*[4,5,6]{\fnm{Stephan} \sur{Lewandowsky}}\email{stephan.lewandowsky@bristol.ac.uk}

\affil[1]{\orgname{Graz University of Technology}, \orgaddress{\street{Inffeldgasse 16C}, \city{Graz}, \postcode{8010}, \country{Austria}}}

\affil[2]{\orgname{Complexity Science Hub Vienna}, \orgaddress{\street{Josefst\"adterstr. 39}, \city{Vienna}, \postcode{1080},  \country{Austria}}}

\affil[3]{ \orgname{University of Konstanz}, \orgaddress{\street{}, \city{Kosntanz},  \country{Germany}}}

\affil[4]{ \orgname{University of Bristol}, \orgaddress{\street{}, \city{Bristol},  \country{UK}}}

\affil[5]{\orgname{University of Western Australia}, \orgaddress{\city{Crawley}, \postcode{6009}, \state{W.A.}, \country{Australia}}}

\affil[6]{\orgname{University of Potsdam}, \orgaddress{\city{Potsdam},  \country{Germany}}}

\abstract{The spread of online misinformation on social media is increasingly perceived as a problem for societal cohesion and democracy. The role of political leaders in this process has attracted less research attention, even though politicians who ``speak their mind'' are perceived by segments of the public as authentic and honest even if their statements are unsupported by evidence. Analyzing communications by members of the U.S. Congress on Twitter between 2011 and 2022, we show that politicians’ conception of honesty has undergone a distinct shift, with authentic belief-speaking that may be decoupled from evidence becoming more prominent and more differentiated from explicitly evidence-based truth seeking. We show that for Republicans---but not Democrats---an increase of belief-speaking of 10\% is associated with a decrease of 12.8 points of quality (NewsGuard scoring system) in the sources shared in a tweet. Conversely, an increase in truth-seeking language is associated with an increase in quality of sources for both parties. The results support the hypothesis that the current dissemination of misinformation in political discourse is in part driven by an alternative understanding of truth and honesty that emphasizes invocation of subjective belief at the expense of reliance on evidence.}

%\keywords{misinformation, Keyword2, Keyword3, Keyword4}



\maketitle

\section{Introduction}\label{sec1}

\pagenumbering{arabic}
\setcounter{page}{2}

Numerous indicators suggest that democracy is in retreat worldwide \citep[e.g.,][]{FreedomHouse20,FreedomHouse21}. Although symptoms and causes of this democratic backsliding are difficult to tease apart, the widespread dissemination of misinformation---on social media, in hyperpartisan news sites, and in political discourse---is undoubtedly a challenge to democracies \citep{lewandowsky2020wilful}. There is increasing evidence that exposure to misinformation can cause people to  change their behavior \citep[e.g.,][]{Loomba21}.  Exposure to misinformation has been identified as a contributing cause of voting for populist parties in Italy \citep{Cantarella20} and has been causally linked to ethnic hate crimes in Germany (\citep{muller2021fanning}; for a review of causal effects, see \citep{LorenzSpreen22}). Note that we use ``misinformation'' as an umbrella term to refer to any information that people consume and which later on turns out to be false. Misinformation can be spread unintentionally, when communicators mistakenly believe some item of information to be true, or it can be spread intentionally, for example in pursuit of a political agenda. Intentionally disseminated misinformation is often referred to as ``disinformation''. The psychological and cognitive consequences of disinformation are indistinguishable from those of unintentional misinformation, and we therefore use the latter term throughout.

Misinformation has several troubling psychological attributes. First, misinformation lingers in memory even if people acknowledge, believe, and try to adhere to a correction \citep{Lewandowsky17b}. Even though people may adjust their factual beliefs in response to corrections \citep{Wood18}, their political behaviors and attitudes may be largely unaffected \citep{Swire17,Swire19}. Second, perhaps most concerningly, in some circumstances people may even come to value overt dishonesty as a signal of ``authenticity'' \citep{Hahl18}. A politician who routinely and blatantly misinforms the public is overtly violating the established societal norm of being accurate and truthful. Within a populist logic, this norm violation identifies the politician as an enemy of the ``establishment'' and, by implication, an authentic champion of ``the people''---dishonesty and misinformation thus become a sign of distinction \cite{Hahl18}. For example, polls have shown that around 75\% of Republicans considered President Trump to be ``honest'' at various points throughout his presidency (e.g., NBC poll, April 2018). This perception of honesty is at odds with the records of fact checkers and the media, which have identified more than 30,000 false or misleading statements by Trump during his presidency (Washington Post fact checker). 

%The disconnect between accuracy and politicians' attractiveness to voters has also been established in behavioral experiments involving the American public~\citep{Swire17,Swire19}.

This discrepancy between factual accuracy and perceived honesty is, however, understandable if ``speaking one's mind'' on behalf of a constituency is considered a better marker of honesty than veracity. The idea that untrue statements can be ``honest'', provided they arise from authentic belief speaking, points to a distinct ontology of honesty that does not rely on the notion of evidence, but on a radically constructivist appeal to an intuitive shared experience as ``truth'' \citep{lewandowsky2020wilful}. There have been several attempts to characterize this ontology of truth and honesty and the stream of misinformation it gives rise to \citep[e.g.,][]{Lewandowsky17,lewandowsky2020wilful,McCright17}. A recent analysis of ontologies of political truth~(\cite{lewandowsky2020wilful}; see also \cite{Cooper21}) proposed two distinct conceptions of truth: ``belief-speaking'' and ``truth-seeking''. Belief-speaking relates only to the speaker's beliefs, thoughts, and feelings, without regard to factual accuracy. Truth-seeking, by contrast, relates to the search for accurate information and an updating of one's beliefs based on that information.

The first of these two ontologies echoes the radical constructivist ``truth'', based on intuition and feelings, that characterized 1930s fascism \citep[e.g.,][]{Varshizky12}. This conception of truth sometimes rejects the role of evidence outright. For example, Nazi ideology postulated the existence of an ``organic truth'' based on personal experience and intuition that can only be revealed through inner reflection but not external evidence \citep[e.g.,][]{Varshizky12,Voegelin00}. Contemporary variants of this conception of truth can be found in critical postmodern theory \cite{vanZoonen12} and right-wing populism \cite{Edis20,Waisbord18}. The second ontology, based on truth-seeking, aims to establish a shared evidence-based reality that is essential for the well-being of democracy \cite{Farrell18a}. This conception of truth aims to be dispassionate and does not admit appeals to emotion as a valid tool to adjudicate evidence, although it also does not preclude truth-finding from being highly contested and messy (\cite[][]{Uscinski13} vs. \cite{Amazeen15}). 

For democratic societies, a conception of truth that is based on ``belief-speaking'' alone can have painful consequences as democracy requires a body of common political knowledge in order to enable societal coordination \citep{Farrell18a}. For example, people in a democracy must share the knowledge that the electoral system is fair and that a defeat in one election does not prevent future wins. Without that common knowledge, democracy is at risk. The attempts by Donald Trump and his supporters to overturn the 2020 election results with baseless claims of electoral fraud have brought that risk into sharp focus \citep{Jacobson21}. To achieve a common body of knowledge, democratic discourse must go beyond belief-speaking. In particular, democratic politics requires truth-seeking by leaders---otherwise, they may choose to remain wilfully ignorant of embarrassing information, for example, by refusing briefings from experts that are critical of their favoured public-health policy. A corollary of this requirement is that the public considers truth-seeking by politicians as an indicator of honesty rather than (only) belief speaking.

Although truth and honesty are closely linked concepts, with honesty and truthfulness being nearly synonymous \cite{Williams+2002}, in the present context they need to be disentangled for clarity. We focus here primarily on conceptions of honesty, which refers to a virtuous human quality and a socially recognized norm, rather than truth, which refers to the quality of information about the world. Thus, the two ontologies of truth just introduced describe how the world can be known---namely either through applying intuition or seeking evidence, irrespective of the virtuous qualities (or lack thereof) of the beholder. Nonetheless, this ontological dichotomy maps nearly seamlessly into the different conceptions of honesty that we characterize as belief-speaking and truth-seeking, respectively. 

To date, there has been much concern but limited evidence about the increasing prevalence of belief-speaking at the expense of truth-seeking in American public and political life. We aim to explore this presumed shift in conceptions of truth and honesty by focusing on Twitter activity by members of both houses of the U.S. Congress. The U.S. is not only the world's leading democracy but it is also a crucible of the contemporary conflict between populism and liberal democracy and the intense partisan polarization it has entailed \cite{Graham20}. The choice of Twitter is driven by the fact that public outreach on Twitter has become one of the most important avenues of public-facing discourse by U.S. politicians in the last decade~\cite{barbera2019leads} and is frequently used by politicians for agenda-setting purposes~\cite{lewandowsky2020using}.

Our analysis addressed several research questions: Can we identify aspects of belief-speaking and truth-seeking in public-facing statements by members of Congress? And if so, how do these conceptions evolve over time? What partisan differences, if any, are there? Is the quality of shared information linked to the different conceptions of honesty? To answer these questions, we performed a computational analysis of an exhaustive dataset of tweets posted by U.S. politicians, detecting links to misinformation sources and analyzing text of tweets and news sources.

\section{Results}
\subsection*{Identifying different conceptions of honesty in political speech}\label{sec:identifying_honesty_components}
We first sought to identify the two components of truth and honesty --- belief-speaking and truth-seeking --- in public-facing political speech by elected U.S. officials. For our analyses, we collected a corpus of tweets from members of the U.S. Congress between January 1, 2011 and December 31, 2022. After removing retweets and duplicates, our corpus contained a total of 4,527,814 tweets (see 
%Section ``U.S. Congress member tweet corpus'' in the 
Methods for details). Twitter accounts were categorized by party affiliation.

To measure the conceptions of honesty in text, we created two dictionaries of words associated with each of the concepts. We followed a computational grounded theory approach~\cite{nelson2020computational} to incorporate both expert knowledge and computational pattern recognition. We started with a list of seed words for each conception, followed by computational expansion and iterative pruning and refinement through human input (see 
%also Section ``Honesty component keywords and validation'' in 
Methods for details). 

We validated the dictionaries in three steps. First, to validate the candidate keywords (selected by the authors), we created a survey on Prolific and asked participants ($N=51$) to rate each keyword's representativeness of the two honesty components on two separate Likert scales. We then ran paired $t$-tests between each word's representativeness ratings for belief-speaking and truth-seeking, respectively. Keywords that were rated as significantly more representative for belief-speaking (truth-seeking) were included in the belief-speaking (truth-seeking) dictionaries. The final dictionaries include a total of 37 keywords for each component and are provided in Table~\ref{tab:ext_tab1} (see 
%also Section ``Honesty component keywords and validation'' in the 
Methods and \osm Sections~S1 and S2 for details).
Following the distributed dictionary representation (DDR) approach~\cite{garten2018dictionaries}, we converted the keywords into vector embeddings using a pretrained algorithm (GloVe). Those representations capture nuanced contextual information and are amenable to a vector-similarity approach to establish overlap between each dictionary and the text or document of interest (see Methods for details).

In the second validation step, we applied the dictionaries to our tweet corpus and calculated the  semantic similarity $D_\mathrm{b}$ and $D_\mathrm{t}$ between the article and the belief-speaking and truth-seeking dictionaries, respectively (see 
%Section ``Identification of honesty components in text" in the 
Methods for details). A positive semantic similarity  means that a piece of text is more similar to the words contained in a dictionary, whereas a negative similarity means that it is more dissimilar. We then sampled tweets that had a high belief-speaking or truth-seeking similarity or were dissimilar to both honesty components. We again created a survey on Prolific with the same setup as described for the keyword validation. Using tweets for which a majority of human raters agreed that they were representative of ``belief-speaking'' or ``truth-seeking'' as ground-truth, we find satisfactory agreement between the computed belief-speaking and truth-seeking similarity scores and human ratings with $\mathrm{AUC}=0.824$ for belief-speaking and $\mathrm{AUC}=0.772$ for truth-seeking (see Methods and \osm Section~S3 for details).   

In the third validation step we applied the dictionaries to historic articles from the New York Times for three text categories: ``opinion'', ``politics'' and ``science'' (see 
%Section ``New York Times corpus'' in the 
Methods for details). We found that articles in the ``science'' category are more similar to truth seeking than all articles on average ($\left<D_\mathrm{t}\right>_\mathrm{sci} - \left<D_\mathrm{t}\right> = 0.033$), followed by articles in the opinion ($\left<D_\mathrm{t}\right>_\mathrm{op} - \left<D_\mathrm{t}\right> = 0.006$) and politics ($\left<D_\mathrm{t}\right>_\mathrm{pol} - \left<D_\mathrm{t}\right> =-0.006$) category. Articles in the opinion category show the highest similarity to the belief speaking dictionary ($\left<D_\mathrm{b}\right>_\mathrm{op} - \left<D_\mathrm{b}\right> =0.013$), followed by articles in the science ($\left<D_\mathrm{b}\right>_\mathrm{sci} - \left<D_\mathrm{b}\right> = 0.009$) and politics ($\left<D_\mathrm{b}\right>_\mathrm{pol} - \left<D_\mathrm{b}\right>=-0.007$) category. The analysis of New York Times content confirmed our expectation of articles in the science category being most similar to truth-seeking while articles in the opinion category being most similar to belief-speaking. It did not confirm our expectation of politics being more similar to truth-seeking than opinion articles and more similar to belief-speaking than science articles. 

Finally, to establish the uniqueness of our dictionaries and to differentiate the honesty conceptions from existing similar measures, we investigated the relationship between our two components to text features such as authenticity~\citep[][]{newman2003}, analytic language~\citep[][]{pennebaker2014small} and a moral component reflecting judgemental language~\citep[][]{bradyetal2020}, each measured using LIWC 2022~\citep{boyddevelopment2022} as well as positive and negative sentiment measured using VADER~\citep{Hutto2014}. We calculated scores for each of these components for every tweet in the corpus. Both belief-speaking and truth-seeking are negatively correlated with ``analytic'', although the correlation with belief-speaking ($r=-0.27$) is about twice as high as with truth seeking ($r=-0.16$). Both honesty components are positively correlated with ``authentic'', ``moral'' and negative sentiment, while the correlation with positive sentiment is positive for belief-speaking ($r=0.06$) and sightly negative for truth-seeking ($r=-0.01$). All correlations are highly significant ($p < 0.001$) but small --- the correlation with the largest magnitude ($r=-0.27$) is observed between belief-speaking similarity and ``analytic''. Details of the comparison with LIWC and VADER scores are summarised in the \osm Sections~S4. In summary, these analyses show that belief-speaking and truth-seeking do not overlap greatly with existing related measures of text features. 

\subsection*{Partisan and temporal dynamics of conceptions of honesty}\label{sec:dynamics_of_honesty_components} 
Having validated our dictionaries, we produced textual scatterplots~\citep[][]{kessler2017scattertext} (see 
%Section  ``Word and topic keyness analysis'' in the 
Methods for details) to illustrate individual terms that are characteristic of the two honesty components.  

\begin{figure}[!ht]
    \centering
    \includegraphics[width=\linewidth]{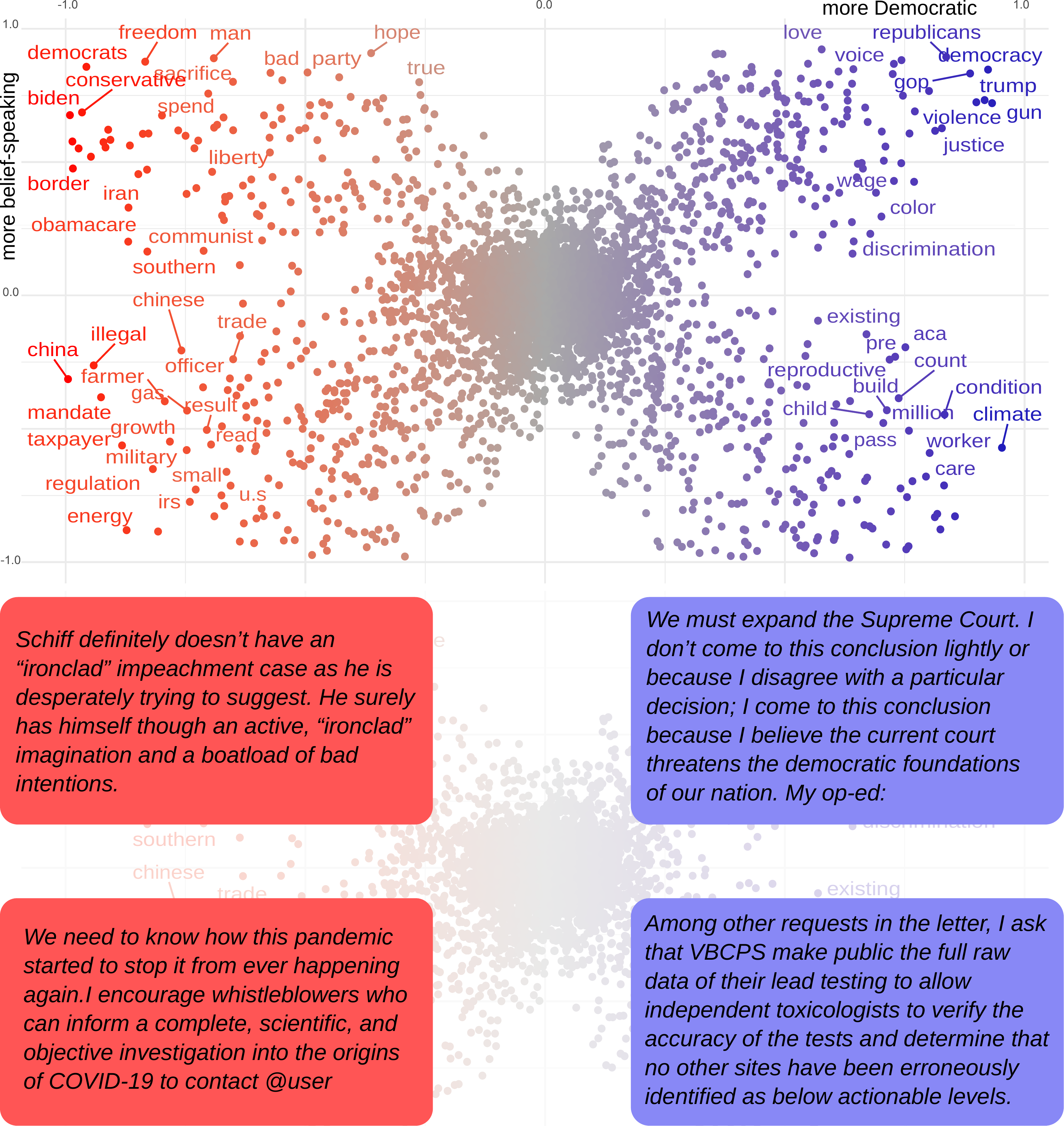}
    \caption{The figure depicts the distribution of keywords on a textual scatterplot. Every term is a dot with two coordinates associated with party ($x$-coordinate) and honesty component ($y$-coordinate) keyness. Each coordinate represents a Scaled F-Score (SFS) value ranging from -1 to 1. The word color is  associated with the party keyness. We only show word labels where $SFS > 0.65$ or $SFS < -0.65$ for readability reasons. Below the scatterplot we show four example tweets associated with the four quadrants of the scatterplot. \\ \rule{\linewidth}{1pt}
    }
    \label{fig:fig1}
\end{figure}

Figure~\ref{fig:fig1} shows diagnostic words in a two-dimensional plot, with the $x$- and $y$-axes representing party and honesty conception respectively. Each dot is a unigram from the Twitter corpus, and its colour is associated with party keyness (a word with positive party keyness occurs more often for texts from members of a given party than expected by chance). The closer to a corner a word is, the more it characterizes that particular conception of honesty and party dimension. See methods for details on how words in the figure are represented. We see that Republican belief-speaking keywords, situated in the top-left corner, often refer to political opponents or ideologies (``biden'', ``democrats'', ``conservatives'') or conservative values (``freedom'', ``liberty''). On the other hand, truth-seeking keywords by the same party are linked to economic (``energy'', ``taxpayer'', ``trade'') or foreign policy aspects (``china'', ``chinese'') and the military. On the right-hand side of the figure, we find that Democrat belief-speaking tweets also regard politicians and political ideology  (``trump'', ``democrats'', ``republicans''), and social justice (``color'', ``discrimination'', ``justice''), whereas truth-seeking texts particularly concern the climate crisis (``climate''), as well as social welfare and healthcare (``worker'', ``care'', ``pre existing condition''). 
%It is also interesting to note how ideological positions are represented across the parties. As an example, Republicans use truth-seeking texts as a medium to convey concepts such as ``regulation'', ``illegal''. By contrast, Democrats use the same honesty component to frame ideas such as ``safety'', ``resource'', ``essential'', ``care'', ``condition'', ``public'' and ``worker''.

The \osm (Section~S6) explores the topics of politicians' communications further. The analysis of some controversial topics revealed that these topics invoked more belief-speaking or truth-seeking than the average tweet, with only a few exceptions. For example, vaccine related discourse involved far less belief-speaking than other controversially discussed topics such as climate change or the opioid crisis for both parties.

\begin{figure}[!ht]
    \centering
    \includegraphics[width=\linewidth]{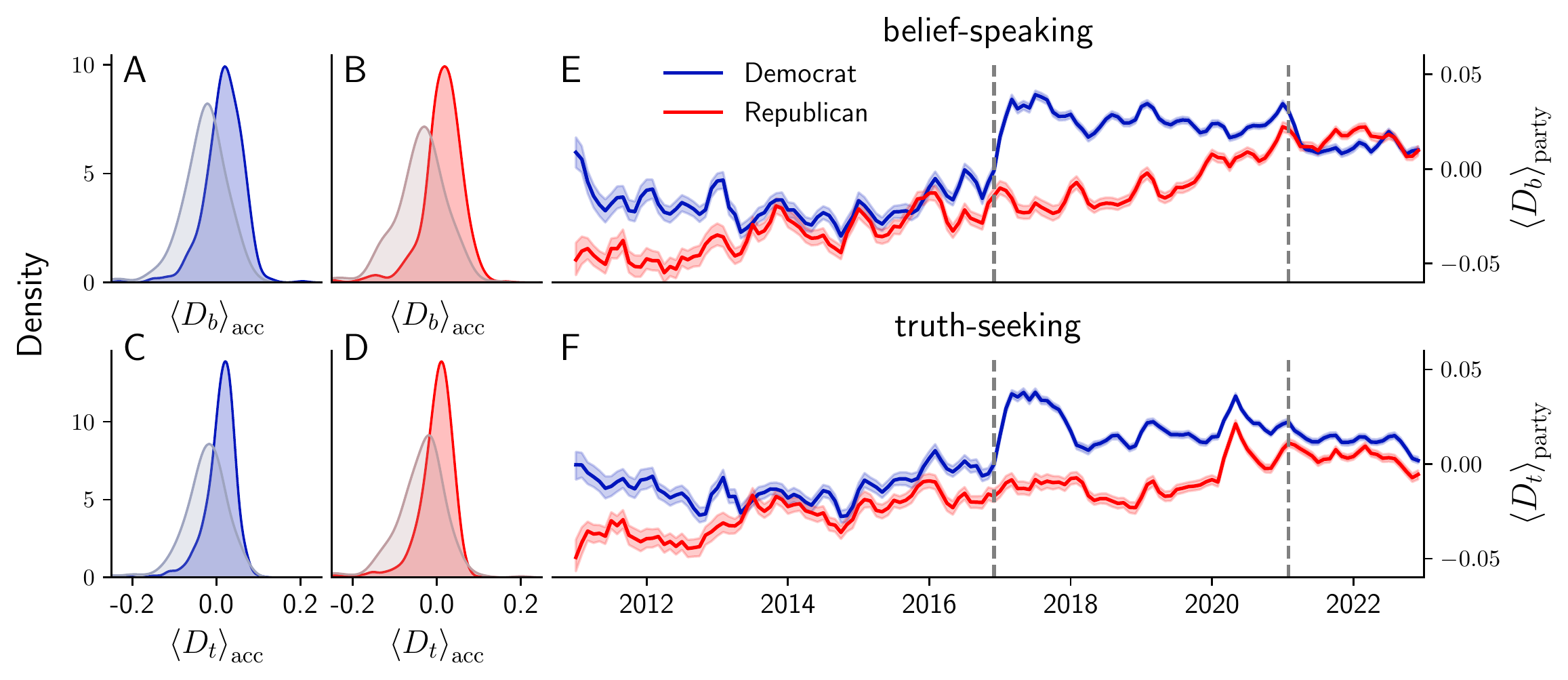}
    \caption{Belief-speaking and truth-seeking similarity in tweets by members of the U.S. Congress for the period 2011 to 2022 shown separately for members of each party. A and B distributions of the average within-politician belief-speaking similarity $\left<D'_\mathrm{b}\right>_\mathrm{acc}$ in tweets of members of the Democratic and Republican parties for the years 2011 to 2013 (grey) and 2019 to 2022, respectively. C and D distributions of the average within-politician truth-seeking similarity $\left<D'_\mathrm{t}\right>_\mathrm{acc}$. E and F show micro averages over all tweets of belief-speaking and truth-seeking similarity $\left<D'_\mathrm{b}\right>_\mathrm{party}$ and $\left<D'_\mathrm{b}\right>_\mathrm{party}$ over time. Timelines have been smoothed with a rolling average of three months. The 95\% confidence intervals were computed with bootstrap sampling over 1,000 iterations. Dashed vertical lines indicate dates of presidential elections in 2016 and 2020.}
    \label{fig:fig2}
\end{figure}

We next examined the temporal trends of the two honesty components. For the following analyses, we use the centered and length-corrected belief-speaking and truth-seeking similarity scores $D'_\mathrm{b}$ and $D'_\mathrm{t}$ (see Methods for details). To arrive at a finer-grained picture of the variability of these components between individual politicians, we calculated the average belief-speaking similarity $\left<D'_\mathrm{b}\right>_\mathrm{acc}$ and truth-seeking similarity $\left<D'_\mathrm{t}\right>_\mathrm{acc}$ of tweets for each individual politician. Note that $\left<\right>_\mathrm{acc}$ denotes an account-average. Figure~\ref{fig:fig2} A to D shows how the distribution of $\left<D'_\mathrm{b}\right>_\mathrm{acc}$ and $\left<D'_\mathrm{t}\right>_\mathrm{acc}$ shifted between the first (2011-2013, 331 Democrats, 514 Republicans) and last (2019-2022, 295 Democrats, 494 Republicans) four years of tweets contained in the corpus.

For both parties, the mean belief-speaking similarity $\left<D'_\mathrm{b}\right>_\mathrm{party}$ considerably increased from -0.031 to 0.017 for Democrats (unpaired t-test \mbox{$t=-11.317$}, \mbox{$p<0.001$}, Cohen's \mbox{$d=0.850$}) and from -0.040 to 0.012 for Republicans (\mbox{$t=-10.819$}, \mbox{$p<0.001$}, Cohen's \mbox{$d=0.854$}). Similarly, we see an increase in the similarity to truth-seeking $\left<D'_\mathrm{t}\right>_\mathrm{party}$ from -0.027 to 0.009 for Democrats (\mbox{$t=-9.753$}, \mbox{$p<0.001$}, Cohen's \mbox{$d=0.748$}) and from -0.038 to -0.003 for Republicans (\mbox{$t=-8.442$}, \mbox{$p<0.001$}, Cohen's \mbox{$d=0.671$}). This overall increase in both belief-speaking and truth-seeking similarity also becomes apparent in Figure~\ref{fig:fig2} E and F, and is especially pronounced after the presidential election in late 2016. 

This parallel increase for both belief-speaking and truth-seeking could reflect the fact that in recent years, topics concerning fake news have become increasingly central to political discourse \citep{Kozyreva20}, resulting in opposing claims and counterclaims (e.g., Donald Trump routinely accused mainstream media such as the New York Times of spreading ``fake news'', \citep{lewandowsky2020using}). Whereas those claims represented mainly belief speaking, they were accompanied by increasing attempts by the media, and other actors, to correct misinformation through truth-seeking discourse. 

\subsection*{Relation of honesty components to information trustworthiness}\label{sec:honesty_components_and_misinformation} 
To test our hypothesis that belief-speaking is preferentially associated with dissemination of misinformation, we analyzed the association between belief-speaking and truth-seeking, respectively, to the quality of the information that is being relayed. To assess information quality, we examined links to websites external to Twitter that were shared by the accounts. We followed an approach employed by similar research in this domain~\citep{grinberg2019fake, pennycook2021shifting} and used a trustworthiness assessment by professional fact checkers of the domain a link points to. We used the NewsGuard information nutrition data base~\cite{newsguard2022} as well as an independently compiled data base of domain trustworthiness labels~\cite{lasser2022misinformation} (see 
%Sections ``NewsGuard nutrition labels''  in 
Methods and \osm Sections~S7 and S8 for details). 

The NewsGuard data base as of the beginning of March 2022 indexed 6,860 English language domains. Each domain is scored on a total of 9 criteria, ranging from ``doesn't label advertising'' to ``repeatedly publishes false information''. Each category awards a varying number of points for a total of 100. Domains with less than 60 points are considered ``not trustworthy''. The majority of indexed domains (63\%) are considered trustworthy. After excluding links to other social media platforms (e.g., twitter.com, facebook.com, youtube.com and instagram.com) as well as links to search services (google.com, yahoo.com), the database covered between 20\% and 60\% of the links posted by members of the U.S. Congress, with a steadily increasing share of links covered over time and no difference in coverage between the parties --- see also Extended Data Figure~\ref{fig:ext_fig3}. 

\begin{figure}[!ht]
    \centering
    \includegraphics[width=\linewidth]{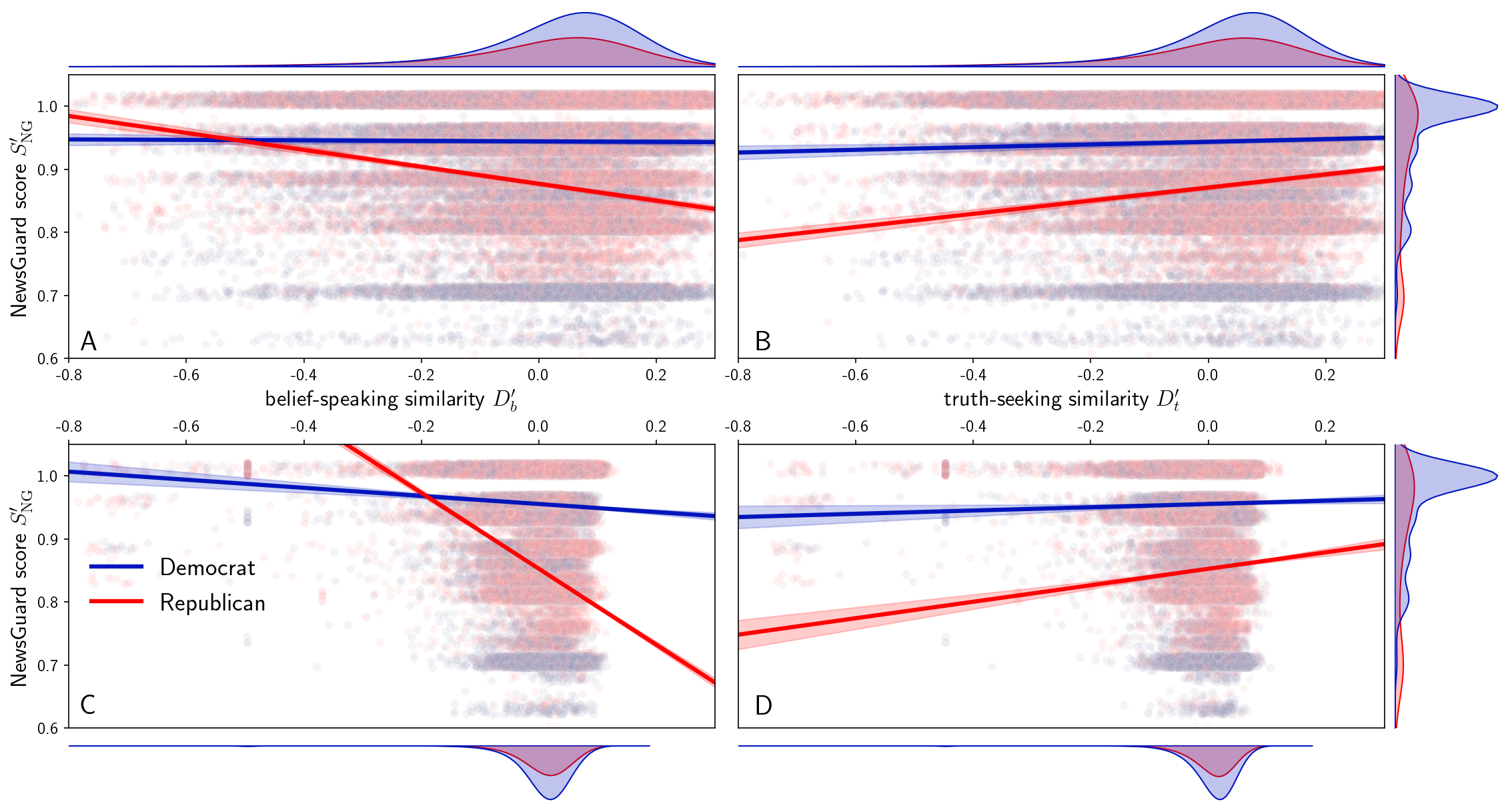}
    \caption{Relation of information quality with belief-speaking and truth-seeking. A and B show the rescaled NewsGuard score $S_\mathrm{NG}'$ of links posted by individual U.S. Congress members over belief-speaking ($D_\mathrm{b}'$) and truth-seeking ($D_\mathrm{t}'$) similarity measured in tweet texts, respectively. The lines and shaded areas indicate NewsGuard score predictions and 95\% confidence intervals from a linear mixed effects model (see Eq. (\ref{eq:regression_NGscore_tweets})). C and D show the rescaled NewsGuard score $S_\mathrm{NG}'$ over belief-speaking and truth-seeking similarity measured in article texts scraped from the tweeted links.
    The lines and shaded areas indicate NewsGuard score predictions and 95\% confidence intervals from a linear regression model (see Eq. (\ref{eq:regression_NGscore_articles})). The scatter plots show only $10^{5}$ data points per panel and vertical jitter was applied to visually separate data points. Note that we truncated the y-axis at 0.6. The full data is shown in Extended Data Figure~\ref{fig:ext_fig4}. Marginal distributions on the sides show the kernel density estimation over the full data on the respective axes, separated by party.\\ \rule{\linewidth}{1pt}}
    \label{fig:fig3}
\end{figure}

For each tweet, we calculated the belief-speaking and truth-seeking similarity $D'_\mathrm{b}$ and $D'_\mathrm{t}$. Figure~\ref{fig:fig3} A and B shows $S_\mathrm{NG}'$, the NewsGuard score rescaled to [0; 1] over the  belief-speaking and truth-seeking similarity, respectively, for each tweet posted by a member of Congress. 

To investigate the relationship between $D_\mathrm{b}'$, $D_\mathrm{t}'$ and $S_\mathrm{NG}'$, we fitted a linear mixed effects model with random slopes and intercepts for every Congress Member following Equation (\ref{eq:regression_NGscore_tweets}). The lines shown in Figure~\ref{fig:fig3} A and B show $S_\mathrm{NG}'$ predicted by the model depending on $D_\mathrm{b}'$, $D_\mathrm{t}'$, respectively, party $P$ and their interaction terms (see 
%Section ``Regression'' in the 
Methods for details). 

The analysis conducted with $P=\mathrm{Democrat}$ as baseline yielded a significant fixed effect for $D_\mathrm{t}'$ (coefficient 0.022 [0.010; 0.033], $p < 0.001$, $t=3.6$),  $P=\mathrm{Republican}$ (coefficient -0.069 [-0.074; -0.065], $p < 0.001$, $t=-29.9$), the interaction between Republican and $D_\mathrm{b}'$ (coefficient -0.128 [-0.146; -0.111], $p < 0.001$, $t=-14.4$), the interaction between Republican and $D_\mathrm{t}'$ (coefficient 0.085 [0.068; 0.103], $p < 0.001$, $t=9.6$), and the three-way interaction between $D_\mathrm{b}'$, $D_\mathrm{t}'$ and Republican (coefficient -0.085 [-0.115; -0.056], $p < 0.001$, $t=-5.6$).
See Extended Data Table~\ref{tab:ext_tab2} for the full regression statistics and Extended Data Figure~\ref{fig:ext_fig1} for a visualization of the fixed effect of the three-way interaction. 

Therefore an increase in $D_\mathrm{b}'$ of 10\% predicted a decrease in $S_\mathrm{NG}$ of 12.8, but only for members of the Republican party. An increase in $D_\mathrm{t}'$ of 10\% predicted an increase in $S_\mathrm{NG}$ of 2.1 for Democrats and of 10.6 for Republicans. For Democrats, we find no significant relationship between $S_\mathrm{NG}$ and belief-speaking similarity. Predictions of the NewsGuard score depending on belief-speaking and truth-seeking similarity based on the two-way interactions between honesty components and party are shown as lines in panels A and B of Figure~\ref{fig:fig3}, respectively.  

In the \osm Section~S9, we explore this pattern further by considering NewsGuard scores and honesty components broken down by state and party. We find that the quality of information being shared by Republicans tends to be lower in southern states (e.g., AL, TN, TX, OK, KY) than in the north (e.g., NH, AK, ME), although there are also striking exceptions (e.g., NY). For Democrats, no clearly discernible pattern across states emerges. We also find that the voting patterns during the 2020 presidential election in their home state  did not affect the quality of news being shared by members of Congress.

To exclude a dependence of these results on use of the NewsGuard data base, we validated this analysis with an independently collected list of news outlet reliability from academic and fact-checking sources. Results are reported in the \osm~(Section~S7) and are consistent with results reported in the main text. In addition, using the different outlet reliability data base, we also find a significant effect of belief-speaking similarity on the quality of shared information for Democrats that goes in the same direction as the effect for Republicans.

Finally, we wanted to know whether the content of belief-speaking and truth-seeking words in the texts found at the websites the tweets linked to was also indicative of low information quality. To this end, we attempted to scrape the text of all linked websites (see %Section ``News article collection'' in the 
Methods). We successfully collected text from about 65\% of links. We excluded texts with less than 100 words and only retained one copy of the text in the case when multiple tweets contained links to the same website. In addition, we excluded all articles collected from links that were posted by members of both parties (2462 texts, 0.91\% of articles), such that every link had a unique party designation. This resulted in a total of 271,171 unique news texts. 

We investigated the dependence of the NewsGuard score associated with the domain the text was scraped from on the belief-speaking similarity and the truth-seeking similarity of the article text (rather than in the original tweet). We fitted a linear regression model to predict the rescaled NewsGuard score $S_\mathrm{NG}'$ depending on party, the belief-speaking and truth-seeking similarities $D_\mathrm{b}'$ and $D_\mathrm{t}'$, and the two-way interaction terms (see Equation (\ref{eq:regression_NGscore_articles}) and 
%Section ``Regression'' in the 
Methods for details). 

We show both the data for individual links and the model predictions for $D_\mathrm{b}'$ and $D_\mathrm{t}'$ in Figure~\ref{fig:fig3} C and D, respectively. Again, we found a significant inverse relationship between $P=\mathrm{Republican}$ and $S_\mathrm{NG}'$ (coefficient $-0.103 \, [-0.104; -0.102]$, $p<0.001$, $t=-200.3$) as well as the interaction term between Republican and $D_\mathrm{b}'$ (coefficient $-0.538 \, [-0.560; -0.507]$, $p<0.001$, $t=-33.5$). We also confirmed the positive relationship between $S_\mathrm{NG}'$ and $D'_\mathrm{t}$ (coefficient $0.026 \, [0.004; 0.048]$, $p=0.003$, $t=2.3$), and the interaction term between Republican and $D_\mathrm{t}'$ (coefficient $0.105 \, [0.068; 0.141]$, $p<0.001$, $t=5.6$), as well as the three-way interaction term $\mathrm{party}\times D_\mathrm{b}' \times D_\mathrm{t}'$ (coefficient $-0.590 \, [-0.665; -0.508]$, $p<0.001$, $t=-14.6$). 

Different from the analysis using tweet texts, we also find a significant negative relationship for $D_\mathrm{b}'$ for Democrats (coefficient $-0.064 \, [-0.084; -0.045]$, $p<0.001$, $t=-6.5$), and a significant interaction term $D_\mathrm{b}'\times D_\mathrm{t}'$ (coefficient $0.067 \, [0.020; 0.113]$, $p=0.006$, $t=2.8$). See Extended Data Table~\ref{tab:ext_tab3} for the full regression statistics. Our analysis of article texts therefore reproduces the main results from our analysis of tweet texts.

\section{Discussion}
We curated two dictionaries that captured the distinction between an evidence-based conception of honesty (truth-seeking) and a conception based on intuition, subjective impressions, and feelings (belief-speaking). We confirmed the validity and diagnosticity of the dictionaries by soliciting ratings from human participants both for individual keywords as well as for documents, and by showing that belief-speaking prevailed in opinion pieces in the New York Times but not in their science section, whereas the reverse occurred for truth-seeking. 

Applying those dictionaries to public political discourse by members of the U.S. Congress, represented by their tweets, we find a bipartisan increase of the use of both truth-seeking and belief-speaking language over time, in particular from late 2016 onward. The use of truth-seeking and belief-speaking language is particularly intense for controversial topics, and this is also a bipartisan phenomenon. 

The parties differ considerably, however, when the quality of information being shared is considered. Overall, Republicans tend to share information of lower quality than Democrats (see also \cite{lasser2022misinformation}), and this difference is in large part driven by belief-speaking: the more Republicans engage in belief-speaking, the more likely they are to share low-quality information. This relationship is absent (or attenuated; see Section~S7 in \osmnosp) for Democrats.

Our results have several theoretical and practical implications that deserve to be explored. First, our data cast a new light on several recent analyses of the American public's information diet that have shown that conservatives are more likely to encounter and share untrustworthy information than their counterparts on the political left~\citep{grinberg2019fake,Guess20a,Guess19,lasser2022misinformation}. Several reasons have been put forward for this apparent asymmetry, for example that partisans are motivated to share derogatory content towards the political outgroup~\citep{Rathje2021-fn}. Because greater negativity towards Democrats is mostly found in lower-quality outlets, conservatives may disproportionately share untrustworthy information because it is satisfying a need for outgroup derogation~\citep{Osmundsen2021-gn}. 

Our analysis offers another explanation, namely that the public is sensitive to cues provided by the political elites which, as we have shown here, also differ considerably in the accuracy of content that they share on social media. Specifically, Republican politicians frequently, though not always, share low-quality information and are thus providing a cue to their partisan followers of the legitimacy of those outlets. Similar evidence for the sensitivity of the public to leadership cues have been observed in the climate change arena, where the growing polarization of the public along party lines mainly resulted from the Republican leadership gradually assuming a more hostile stance towards the science of climate change \cite{Brulle12}.

Our analysis furthermore identified belief-speaking as a ``gateway''  rhetorical technique for the sharing of low-quality information. The more Republican politicians appeal to beliefs and intuitions, rather than evidence, the more likely they are to share low-quality information. For Democrats, this association was absent in the main analysis using NewsGuard scores, and it was attenuated if an independent  source of domain quality was used (see \osm Section~S7). This pattern gives rise to the question why, if belief-speaking gives licence to the sharing of misinformation, is it only Republicans (or mainly Republicans) who avail themselves of that option?

A possible answer can be found in the finding that belief-speaking is associated with greater negative emotion (see \osm Section~S4). Belief-speaking may therefore result from Republican politicians' desire to derogate Democrats, as suggested by~\citep{Osmundsen2021-gn}. %In support, many Republican belief-speaking words identified in Figure~\ref{fig:fig1} pertain to their political opponents. Few if any such words appear for belief-speaking by Democrats. 
On that view, negative emotional content should be a mediator of the association between belief-speaking and low quality of shared content. Conversely, if belief-speaking were instrumental in the sharing of low-quality content for other reasons, then it should mediate the association involving negative emotionality. We report two competing mediation models in the \osm ~(Section~S10). While the models cannot definitively adjudicate between the two possibilities, the analyses suggest the former hypothesis is in a better position to explain the mediating effect on the spread of low-quality news among Republicans. Within this framework, and concordant with~\citep{Osmundsen2021-gn}, negative emotion associated with derogation of the opponent is the driving force behind the association between belief speaking and the spread of low-quality content among Republicans. Further indirect support for this possibility is provided by the fact that Republican members of Congress do not exclusively share misinformation. When they engage in truth-seeking, Republicans' accuracy of shared information rises nearly to the same level as that of Democrats. 

Finally, we return to the argument advanced at the outset, namely that belief-speaking can be a marker of ``authenticity'' which allows partisan followers to consider a politician to be honest despite them promulgating low-quality or false information. We cannot directly test this argument based on the present data because we have no way of ascertaining the perceived honesty of the politicians in our sample. We do, however, have state-level electoral data from the 2020 presidential election, which show that Republicans did not suffer an electoral penalty for their use of belief-speaking and the associated sharing of low-quality information (\osmnosp, Section~S9). There is no association between the accuracy of Republicans' shared information and the vote share for Trump, suggesting that voters were not deterred by belief-speaking based dissemination of misinformation. 

%belief-speaking is not subordinate to the need to express negative emotion but is an exogenous gateway to low-quality information. This supports the hypothesis that the current onslaught of online misinformation is in part driven by a new ontology of truth and honesty that has replaced reliance on evidence with the invocation of subjective belief.

%\section{Outlook}
Our analysis was limited to communications by the ``political class'' in the United States, and although the U.S. is the world's leading democracy, the trends uncovered here should not be considered in isolation but deserve to be contrasted to observations in other countries and cultures. A recent comparison of the overall accuracy of information shared by U.S. members of Congress found that their accuracy was lower---even among Democrats---than the information shared by parliamentarians from mainstream parties in the U.K. and Germany \cite{lasser2022misinformation}. Although there were differences between parties in those two countries as well, they were small in magnitude and European conservatives were more accurate than U.S. Republicans, underscoring that conservatism is not, per se, necessarily associated with reliance on low-quality information. Another international comparison of populist leaders (Trump in the U.S., Modi in India, Farage in the U.K. and Wilder in the Netherlands) found some commonalities among those politicians, such as the use of insults against political opponents, but also identified Trump as an outlier in the use of critical language \cite{Gonawela18}. Further examinations of belief-speaking and truth-seeking outside the U.S. context are therefore urgently needed to explore the generality of our findings and to redress the existing global imbalance in research activity \cite{LorenzSpreen22}. 

Future research
is also needed to examine the temporal stability of the patterns we observed here. Although our analysis extended to the end of 2022,
thus covering two months of Twitter activity after it was taken over by Elon Musk, there is no guarantee that the platform will remain stable in the future. Likewise, in the same way that sharing of misinformation mushroomed after 2016~\cite{lasser2022misinformation}, the long-term trend towards populism may reverse, and she sharing of misinformation may become less frequent in the future. Our analysis is therefore
best understood as a historical and contemporary picture of political discourse rather than a pointer to the future. 

Finally, future research should also address the particular role played by social media in our analysis. We de-emphasized this angle because when our analysis was extended to mainstream news articles shared by the members of Congress we found very similar results compared to the tweets. However, there may be other situations in which social media play a uniquely different role from conventional mainstream media, and those situations remain to be identified and examined. 

\section{Methods}

\subsection{U.S. Congress Member tweet corpus} 
A corpus of contemporary political communication in English was created by scraping tweets by members of both houses of the U.S. Congress on February 10, 2023. To build the corpus, lists of Twitter handles of members of congress were collected for the 114$^\mathrm{th}$ (from \url{www.socialseer.com}), 115$^\mathrm{th}$ (from \url{www.socialseer.com}), 116$^\mathrm{th}$ (from \url{https://doi.org/10.7910/DVN/MBOJNS}), and 117$^\mathrm{th}$ \& 118$^\mathrm{th}$ (from \url{https://triagecancer.org/congressional-social-media}) Congress. For the 114$^\mathrm{th}$ and 115$^\mathrm{th}$ Congress, only handles of senators were available. For the 116$^\mathrm{th}$, 117$^\mathrm{th}$ and 118$^\mathrm{th}$ Congress, Twitter handles were available for both houses of Congress. This resulted in a total of 1278 unique Twitter handles, which includes Congress member staff and Congress member campaign accounts. If a politician had multiple accounts, all were included in the dataset. No sampling was involved in collecting the data and the collected dataset is exhaustive.

For each of the Twitter handles, metadata were collected on February 10, 2023 via the Twitter API v2  using the Python package twarc~\cite{twarc}. Metadata included the account's handle, user name, creation date, location, user description, number of followers, number of accounts followed, and tweet count. Out of the 1278 accounts, 220 were not accessible because they had been deleted, suspended, or set to ``private''. 

To build the text corpus, all tweets posted by the collected Twitter accounts starting from November 6, 2010 and up to December 31, 2022 were collected, using academic access to the Twitter API. Note that following this approach we include all tweets posted by a given account in the given time span, not just tweets that were posted while a politician was in office. Earlier tweets all the way back to 2006 could be retrieved, but we chose 2010 as the earliest date due to changes in the design of retweeting in the Twitter platform at that time. The retweet button was introduced in November 2009 (previously retweeting was done by hand), and it took approximately a year for users to start using it consistently. Furthermore, the prominence of Twitter in U.S. politics emerged later, especially since 2012. 
The resulting corpus consisted of a total of 5,914,107 tweets, of which 3,463,409 were original tweets, 531,289 were quote tweets, 575,044 were replies and 1,351,346 were retweets. Note that quoting, replying and retweeting are not exclusive categories. We removed retweets from the corpus because they do not constitute original content. The number of tweets consistently increased from around 100,000 in 2011 to over 600,000 in 2020 and then declined to around 500,000 in 2022. We removed exact matches (i.e., duplicates) and included only tweets with more than 10 words. The final corpus contained 3,897,032 tweets. Next to the tweet text, the corpus contained the tweet creation date as well as a unique identifier of the account that posted the tweet. The identifier permitted linkage to the metadata collected about the user accounts, such as party affiliation.

We find a large variance in the number of tweets posted by individual accounts, ranging from only one tweet in the observed time period to 52,055 tweets, with a median number of 2876 tweets per account. To exclude a dependence of our results on highly prolific accounts, we also conducted the main analysis reported in Figure~\ref{fig:fig3} and Extended Data Table~\ref{tab:ext_tab2} using only the latest 3200 tweets per account. Results from this analysis are highly consistent with the analysis using all available tweets. See \osm Section~S11 for details. In addition, we show which accounts contribute most to the overall increase of belief-speaking and truth-seeking (see \osm Section~S12).

In addition to the perspective of individual tweets taken in the analysis presented in Section~\ref{sec:dynamics_of_honesty_components}, we also considered the perspective of individual links taken in the analysis presented in Section~\ref{sec:honesty_components_and_misinformation}. For this analysis, we only considered tweets that contained at least one link (2,700,539 tweets). Because a single tweet can contain more than one link, we expanded the dataset such that every entry referred to a single link, transferring the tweet-level honesty-component labels to the individual links. This resulted in a total of 2,844,901 links. From each link, we extracted the domain the link pointed to. If the link was shortened using a link-shortening service such as bit.ly, we followed the link to retrieve the full domain name. The domains were then matched against the NewsGuard domain trustworthiness data base as well as the independently compiled list of trustworthiness labels (described in Section~\ref{subsec:newsguard_nutrition_labels} and Section~S7 in the \osmnosp).

\subsection{Honesty component keywords and validation}\label{subsec:honesty_component_keyword_validation}
We relied on keywords  to identify the relevant subsets of tweets that involved the presumed distinct conceptions of honesty. Initially, two lists of keywords, one for each honesty component, were generated by the researchers involved in this article. The aim was to capture linguistic cues whose presence might signal that one of the components has been enacted by the speaker. To illustrate, initial keywords for truth-seeking included terms such as ``reality'', ``assess'' ``examine'', ``evidence'', ``fact'', ``truth'', ``proof'', and so on. For belief-speaking, initial keywords were terms such as ``believe'', ``opinion'', ``consider'', ``feel'', ``intuition'', or ``common sense''. 

The lists were expanded computationally using a combination of the fasttext library \citep[][]{bojanowski2016enriching} and colexification networks \citep{DiNatale21,DiNatale2023}. Using the fasttext embeddings, we expanded the seed words to include words that have a cosine similarity score above 0.75. Colexification networks connect words in a language based on their common translations to other languages, thus signalling words that can be used to express multiple concepts. For example, the words ``air'' and ``breath'' are considered to be colexifications because they both translate into the same word in multiple languages (``sukdun'' in Manchu, ``vu:jnas'' in Kildin Sami, ``jind''in Nenets; \citep{franccois2008}). Colexification networks have been used recently to study emotion structures in language \citep{jackson2019} and are predictors of word meaning ratings \citep{DiNatale21}. Including colexification networks in lexicon expansion  gives word lists with a better trade-off between precision and recall \citep{DiNatale2023} than previous approaches using wordnet or word embeddings, such as empath. We subsequently filtered the expanded lists to remove duplicates, overlapping terms appearing in more than one list, and lemma inflections (i.e., ``convey'', ``conveys'', ``conveyed''). The keywords were then used to identify texts relevant to the presumed conceptions of honesty.

To validate the keyword lists, we asked participants in an online survey to score each term on two scales reflecting the honesty components. Data were acquired on September 20, 2022, from 50 individuals (male = 15, female = 34, unlisted = 1; age M = 39.5, SD = 15.8) using the Prolific survey platform~\citep{PALAN201822}. Participants were asked to score each term on two distinct Likert scales ranging from 1 to 5, which respectively indicated low and high representativeness of the word for that honesty component. The instructions provided to participants can be found in the \osmnosp~(Section~S1). The distributions of ratings collected for each keyword are shown in the \osmnosp, Figures S1 and S2.

We next performed paired $t$-tests to see how participants sorted the terms into the two conceptions. The results of the $t$-tests are shown in the \osm (Table~S1). Out of 98 keywords, 61 were judged to belong in the category we previously assigned them to, 24 did not reach the significance threshold  ($p<0.05$) and were therefore removed, and 13 were classified by participants as belonging to the opposite category. We followed the raters' indications and moved the keywords that were classified as belonging to the opposite category from their original dictionary to the the other dictionary. The final list of keywords for both dictionaries is given in Table~\ref{tab:ext_tab1}.

\subsection{Identification of honesty components in text}
As a first preparatory step, we removed URLs and replaced user handles on Twitter with the word ``user''. We then split the tweet texts into individual tokens (words). We then created embeddings of each word contained in the honesty component dictionaries (see Table~\ref{tab:ext_tab1}) with GloVe~\cite{pennington2014glove} trained on 840B tokens from the Common Crawl corpus, following the distributed dictionary representation (DDR) approach~\cite{garten2018dictionaries}. We note that the word ``seem'' from the belief-speaking dictionary is included in the list of stopwords of GloVe. We therefore removed ``seem'' from the stopword list to include it into the dictionary embedding that was calculated using GloVe. 

We then averaged the single-word embeddings within every honesty component to create an embedded representation of the entire dictionary. Similarly, we embedded every token contained in a given tweet and calculated an average of all token embeddings to create an embedded representation of the tweet. For every tweet and both components we then calculated the cosine similarity between the embedded tweet representation and the embedded dictionary representations to arrive at a belief-speaking similarity score $D_\mathrm{b}$ and a truth-seeking similarity score $D_\mathrm{t}$ for the given tweet. Similarity scores range from -1 (not similar at all) to 1 (perfectly similar). 

We find that similarity scores correlate with the length of tweets (number of characters), with Pearson's $r=0.37$ ($p<0.001$) for belief-speaking and $r=0.42$ ($p<0.001$) for truth seeking. In addition, the length of tweets systematically increases over the years, particularly after the increase in the character limit of a tweet from 140 characters to 280 characters in 2017. To remove the trend in similarity scores due to increasing tweet length, we fit two linear models $D_\mathrm{b}\sim \mathrm{tweet\;length}$ and $D_\mathrm{t}\sim \mathrm{tweet\;length}$. We then use these linear models to predict $D_\mathrm{b}$ and $D_\mathrm{t}$ for every tweet based on its length and subtract this prediction from the measured belief-speaking and truth-seeking similarity, resulting in the centered and length-corrected similarity scores $D'_\mathrm{b}$ and $D'_\mathrm{t}$ which we report throughout the publication.

To measure belief-speaking and truth-seeking similarity in the text of the articles collected from links posted by Congress Members on Twitter (see Section~\ref{sec:newsarticles} below), we followed the same approach as described for the text of the tweets above but measure the length of an article as the number of words it contains instead of the number of characters.

To test the robustness of our results to perturbations of the dictionaries, we re-calculated belief-speaking and truth-seeking similarities using versions of the dictionaries where 7 words (20\%) were removed from the dictionary at random before embedding the words and calculating dictionary representations. We then re-ran the regression of $S_\mathrm{NG}'$ on $D_\mathrm{b}'$, $D_\mathrm{t}'$, party and the interaction terms (see Equation (\ref{eq:regression_NGscore_tweets})), where $D_\mathrm{b}'$ and $D_\mathrm{t}'$ are the belief-speaking and truth-seeking similarities, calculated using the representations of the perturbed dictionaries. The distribution of estimates for the fixed effects of the two-way interaction between party and $D_\mathrm{b}'$, and party and $D_\mathrm{t}'$ over 100 perturbations are shown in Extended Data Figure~\ref{fig:ext_fig2}. While the estimates for the effect of $D_\mathrm{b}'$ and $D_\mathrm{t}'$ on NewsGuard score vary by about 20\% between different perturbed dictionary versions, the effects never change direction and always stay significant ($p < 0.001$) for Republicans, as reported in the main text. 

In addition to GloVe~\cite{pennington2014glove} embeddings, we also calculated $D_\mathrm{b}'$ and $D_\mathrm{t}'$ using word2vec~\cite{mikolov2013efficient} and fasttext~\cite{bojanowski2016enriching} embeddings of both the dictionary keywords and the tweets, to exclude a dependence of our results on the choice of embedding. We note that similar to GloVe, the word ``seem'' is included in the stopword list of word2vec and was removed from the stopword list before computing the embeddings. Results of fitting the linear mixed effects model following Eq.~(\ref{eq:regression_NGscore_tweets}) using the alternative embeddings for the dictionaries and tweet texts are shown in the \osm Section~S13. Results are similar to the results using GloVe embeddings (see Table~\ref{tab:ext_tab2}). This shows that our results do not depend on the algorithm or the corpus (common crawl for GloVe and word2vec versus Google news for fasttext) that was used to train the embedding.

Lastly, we also investigated which individual keyword likely contributed most to the overall increases of belief-speaking and truth-seeking reported in Figure~\ref{fig:fig2}. We report the results in the \osm Section~S14.

\subsection{Honesty component document-level validation}
To validate our measures of the belief-speaking and truth-seeking honesty components on the document level, we asked human raters to rate individual tweets with respect to their similarity to the two honesty components. To this end, we sampled 20 tweets from the top belief-speaking and bottom truth-seeking quartile, as well as 20 tweets from the top truth-seeking and bottom belief-speaking quartile. In addition, we sampled 20 tweets that simultaneously belonged to the bottom belief-speaking and truth-seeking quartiles. Each sample of 20 tweets included 10 tweets from Democrats and 10 from Republicans. 

We then created a survey on Prolific \citep{PALAN201822} and asked participants ($N=51$) to rate each tweet's representativeness of the two honesty components on two separate Likert scales. We followed exactly the same setup as described in Section~\ref{subsec:honesty_component_keyword_validation} above, but presenting full tweets instead of singular keywords. The instructions provided to participants can be found in the \osm~(Section~S1). In addition, we included an attention check in the survey, with the aim of excluding all participants that failed the check. To this end, we asked all participants to select ``5'' for both categories halfway through the survey. Only one person failed the check. The responses of this person were excluded from the survey, resulting in $N=50$ total responses (male = 25, female = 24, nonbinary = 1; age M = 37.6, SD = 12.88). Data were acquired on February 10, 2023. The distributions of ratings collected for each tweet are shown in the \osm~(Section S2).

We then wanted to quantify the performance of our computed similarity scores when used as a classifier. To this end, for each honesty component we coded the 20 tweets that were selected from the top belief-speaking [truth-seeking] similarity quartile as ``belief-speaking'' [``truth-seeking''] and the 40 tweets that were selected from the bottom similarity quartile of that component as ``not belief-speaking'' [``not truth-seeking'']. We then classified every tweet for which a majority of human raters selected either a ``4'' or a ``5'' for how characteristic a tweet was for ``belief-speaking'' [``truth-seeking''] as ``belief-speaking'' [``truth-seeking''] to create a ground-truth dataset to compare our classifier against. We obtained ROC curves for belief-speaking and truth-seeking  by varying the threshold for belief-speaking [truth-seeking] similarity to categorise a tweet as ``belief-speaking'' [``truth-seeking''] (akin to varying response criteria in a behavioral study). The ROC curves are shown in the \osm (Section~S2). The area under the curve is high in both cases, with $\mathrm{AUC}=0.824$ for belief-speaking and $\mathrm{AUC}=0.772$ for truth-seeking.

\subsection{New York Times corpus}
We retrieved data from the New York Times (NYT) through their archive API (\url{https://developer.nytimes.com/docs/archive-product/1/overview}). By iterating over 
the months since the founding of the newspaper in the 19$^{th}$ century, we retrieved information on every article in the archive. The information returned by the API included the article title and an abstract that summarizes the article content, as well as additional metadata such as publication date and section of the paper. This approach is different to earlier research that used the NYT API to obtain a number of articles over time that contain certain terms, which does not yield any further text or ways to filter the data \citep{Scheffer2017}. Because we needed text to identify honesty components in articles, the archive endpoint was more suitable than the term search function of the NYT API, despite not giving us the full text of all articles but only returning a summary. 

We extracted three distinct categories of content from the NYT corpus based on the sections identified in the metadata: (i) An ``opinion'' category which comprises opinion pieces such as ``OpEds''; (ii) a ``politics'' category consisting of articles in the sections U.S., Washington, and World; and (iii) a ``science'' category which includes health, science, education, and climate articles. We chose these three clusters because we expected opinion articles to contain more belief-speaking, whereas we expected science articles to contain more truth-seeking. We expected articles in the politics cluster to fall in between. We retrieved a total of 809,271 articles consisting of 240,567 opinion articles, 518,123 politics articles, and 50,581 science articles.

\subsection{Word and topic keyness analysis}
The scatterplot in panel A of Figure~\ref{fig:fig1} was produced following the approach described in Scattertext~\cite{kessler2017scattertext}, a Python package designed to illustrate words and phrases that are more characteristic of a category such as party than others. To derive how characteristic a word is of a category, we start from raw word frequencies: for each word $w_i\in W$ and category $c_j\in C$ we define the precision of the word $w_i$ with respect to the category as
\begin{align*}
    \mathrm{prec}(i,j) = \frac{\#(w_i, c_j)}{\sum_{c\in C} \#(w_i, c)}\,.
\end{align*}
Here, the function $\#(w_i, c_j)$ represents the number of times $w_i$ occurs in a document labeled with the category $c_j$. Therefore, $\mathrm{prec}(i,j)$ represents the discriminative power of a given word across categories regardless of its frequency in the given category.

Similarly, we define the frequency a word occurs in a category $c_j$ as
\begin{align*}
    \mathrm{freq}(i,j) = \frac{\#(w_i, c_j)}{\sum_{w\in W} \#(w, c_j)}\,.
\end{align*}

To combine $\mathrm{prec}(i,j)$ and $\mathrm{freq}(i,j)$ into a single score, we scale and standardize both values using a normal cumulative density function $\Phi(z)$ and then calculate the harmonic mean between the two contributions (see~\cite{kessler2017scattertext} for details). This yields the Scaled F-Score $\mathrm{SFS}$ for every word $w_i$ and category $c_j$ that is defined as
\begin{align*}
    \mathrm{SFS}(i, j) = \mathcal{H}\left(\Phi(\mathrm{prec}(i,j)),\Phi(\mathrm{freq}(i, j))\right)\,.
\end{align*}

For our application case, we want to represent how representative a word is not only for a single category (like ``Republican'') but rather on a spectrum of representativeness that ranges from ``more Democratic'' to ``more Republican''. To this end, we need to map the two distinct scores $\mathrm{SFS}^\mathrm{D}$ for the category ``Democratic'' and $\mathrm{SFS}^\mathrm{R}$ for the category ``Republican'' to a single score that ranges from $-1$ to $+1$. For two arbitrary categories $x$ and $y$ we therefore define 
\[\mathrm{SFS} = 2 \cdot \left(-0.5 + 
\begin{cases}
\mathrm{SFS}^x \quad   \mathrm{if} \  \mathrm{SFS}^x > \mathrm{SFS}^y, \\
1 - \mathrm{SFS}^y \quad \mathrm{if} \  \mathrm{SFS}^x < \mathrm{SFS}^y, \\
0 \quad \mathrm{otherwise}\\
\end{cases} \right)\,.
\]
This maps two SFS (one for category x and one for category y) that are both defined in the range [0, 1] to a single score in the range [-1, 1]. To this end, $\mathrm{SFS}^y$ is mapped to [-1, 0], the $\mathrm{SFS}$ with the larger magnitude is selected and then rescaled to the new range. In our application case, this then yields a single Scaled F-Score $\mathrm{SFS}^\mathrm{party}$ that is -1 for more Republican tweets and +1 for more Democratic tweets.

To calculate representativeness along the ``belief-speaking -- truth-seeking'' dimension, we follow a similar approach. Before we can calculate the SFS for belief-speaking and truth-seeking, we first need to transform the continuous honesty similarity scores $D'_\mathrm{b}$ and $D'_\mathrm{t}$ into a binary honesty component label for each tweet. To this end, we divided the tweets into quantiles according to their belief-speaking [truth-seeking] similarity. We then categorized the tweets with a belief-speaking [truth-seeking] similarity in the to 20\% as ``belief-speaking'' [``truth-seeking'']. If a tweet was part of the upper quantile for both components, then the higher of the two similarity values was used to assign a category to the tweet. We then followed the approach described above to calculate a single Scaled F-Score SFS$^\mathrm{honesty}$ from SFS$^\mathrm{b}$ (for belief-speaking) and SFS$^\mathrm{t}$ (for truth-seeking). 

As a result, each word had two SFS scores: SFS$^\mathrm{party}$ and SFS$^\mathrm{honesty}$. These two scores were used as $x$- and $y$ coordinates for the scatterplot shown in panel A of Fig.~\ref{fig:fig1}. The X-shaped structure of the words in the scatterplot indicates that words that are characteristic for one dimension (e.g., party) are likely also characteristic for the other dimension (e.g., honesty component). Words that are not characteristic of any category (like stopwords) cluster in the middle. 

\subsection{NewsGuard nutrition labels}\label{subsec:newsguard_nutrition_labels}
Following precedent~\citep{grinberg2019fake, pennycook2021shifting}, we use source trustworthiness as an estimator for the trustworthiness of an individual piece of shared information. We use nutrition labels provided by NewsGuard, a company that offers professional fact checking as a service and curates a large data base of domains. The trustworthiness of a domain is assessed in nine categories, each of which awards a number of points: Does not repeatedly publish false content (up to 22 points), gathers and presents information responsibly (18), regularly corrects or clarifies errors (12.5), handles the difference between news and opinion responsibly (12.5), avoids deceptive headlines (10), website discloses ownership and financing (7.5), clearly labels advertising (7.5), reveals who is in charge, including any possible conflicts of interest (5), the site provides names of content creators, along with either contact or biographical information (5). 

NewsGuard categorizes domains with a score of 60 or higher as ``generally adheres to basic standards of credibility and transparency''~\citep{newsguard2022}. Similar to~\citep{bhadani2022political}, we use this value as a threshold below which we categorize a domain and the link pointing to it as ``not trustworthy''. 

After excluding links to other social media platforms (e.g., twitter.com, facebook.com, youtube.com, and instagram.com) as well as links to search services (google.com, yahoo.com), the NewsGuard database covers between 20\% and 60\% of the links posted by members of the U.S. Congress, with a steadily increasing share of links covered over time --- see Extended Data Figure~\ref{fig:ext_fig3} A. 

\subsection{Regression}
We performed a range of regression analyses to quantify the relationship between various manifestations of honesty components and information quality. For the predictions shown in Figure~\ref{fig:fig3} A and B we fitted the following linear mixed effects model for tweets from members of the U.S. Congress:
\begin{align}
    S_\mathrm{NG}' \sim 1 + D_\mathrm{b}' \times D_\mathrm{t}' + D_\mathrm{b}' \times D_\mathrm{t}' \times P + (1 + D_\mathrm{b}' \times D_\mathrm{t}'\;\vert\; \mathrm{user ID})\label{eq:regression_NGscore_tweets}
\end{align}
Here, $S_\mathrm{NG}'$ is the NewsGuard nutrition score of a domain a Congress member linked to in a post on Twitter, rescaled to [0; 1]. $D_\mathrm{b}'$ and $D_\mathrm{t}'$ are the centered and length-corrected belief-speaking and truth-seeking similarity of the text in the tweet with the link, respectively (see section ``Identification of Honesty components in text'' above). $P$ is the party designation of the account that posted the tweet which can be ``Republican'' or ``Democrat''. We include random slopes and intercepts for every account (userID). We fitted the model using the \texttt{lmer} function from the R library lme4~\cite{lme4}. Regression results are reported in Extended Data Table~\ref{tab:ext_tab2}. Data distribution was assumed to be normal but this was not formally tested.

For the predictions shown in Figure~\ref{fig:fig3} C and D, we fitted the following model for articles that were linked to by the U.S. Congress members:
\begin{align}
    S_\mathrm{NG}' \sim 1 + D_\mathrm{b}' \times D_\mathrm{t}' + D_\mathrm{b}' \times D_\mathrm{t}' \times P\;. \label{eq:regression_NGscore_articles}
\end{align}
Here,  $D_\mathrm{b}'$ and $D_\mathrm{t}'$ are the centered and length-corrected belief-speaking and truth-seeking similarity scores of the article text retrieved from the link. We fitted the model using an ordinary least squares fitting approach from the Python package statsmodels~\cite{seabold2010statsmodels}. Regression results are reported in Extended Data Table~\ref{tab:ext_tab3}. Data distribution was assumed to be normal but this was not formally tested. Note that we do not fit a linear mixed-effects model for the statistical analysis of the articles, since there is no clear nesting of articles within individual Twitter accounts, as a single article can be linked to from multiple accounts.

\subsection{News article collection}
\label{sec:newsarticles}
Excluding links to other social media platforms (e.g., twitter.com, facebook.com, youtube.com and instagram.com) as well as links to search services (google.com, yahoo.com), our corpus of tweets contained 1,027,050 unique links to news articles that were shared by members of Congress. Of these links, 462,853 pointed to sites that were indexed by the NewsGuard data base (see Section~\ref{subsec:newsguard_nutrition_labels} above). We scraped the text of these sites using Newspaper3k \cite{Ou-Yang_undated-ky}, a Python package for scraping and curating news articles. Some links were broken, restricted, or could not be scraped by the package. In addition, we removed all articles that contained less than 100 words or were shared only by independent politicians (i.e., not Republican or Democrat). This resulted in 65\% of total scraping coverage. When broken down by trustworthiness, the coverage for trustworthy links (N = 291,143) was 65\%, and 82\% for untrustworthy links with a NewsGuard score $<60$ (N = 7,776). We retained only one copy of each news article in case it was shared multiple times and removed from the main analysis articles that were shared by members of more than one political party (i.e., a link was shared either by Republicans or Democrats, but not both). This was done to ensure each article had only a single party designation such that our statistical analysis of articles was comparable to our statistical analysis of tweets. This resulted in the removal of 2,462 articles (0.91\% of all remaining articles), which were analyzed separately. To provide a marker for apparent bipartisan agreement, we plot the mean and standard deviation of honesty component similarity and $S'_\mathrm{NG}$ for the articles shared by both parties (gray ellipses in Extended Data Figure~\ref{fig:ext_fig4}). Removing these articles left us with a corpus of 271,171 article texts.

%A test of independence between the share of articles scraped for trustworthy and untrustworthy links did not reach statistical significance $\chi^2$(1) = 2.81, $p =0.094$. 
The distribution of NewsGuard scores as well as the  belief-speaking and truth-seeking similarity in each article is shown in Extended Data Figure~\ref{fig:ext_fig4} C and D.

\section{Data availability}
The tweet IDs of the tweet texts and URLs of the articles analysed in this study are deposited in OSF under accession code \url{https://doi.org/10.17605/OSF.IO/VNY8K}. We provide code to download tweets from tweet IDs and article texts from article URLs and process the data in the code repository that accompanies this article \url{https://doi.org/10.5281/zenodo.6826515}. 
 
Dictionaries of keywords associated with the different conceptions of honesty
are deposited in OSF under accession code \url{https://doi.org/10.17605/OSF.IO/VNY8K}.
 
The independently compiled list of domain accuracy and transparency scores is deposited on GitHub under accession code \url{https://doi.org/10.5281/zenodo.6536692}.

The NewsGuard data base used to asses domain trustworthiness is commercially available from NewsGuard
and cannot be shared publicly.

Aggregated values for information trustworthiness and honesty components for tweets and articles used to produce all figures in this article are deposited in OSF under accession code \url{https://doi.org/10.17605/OSF.IO/VNY8K}.

\section{Code availability}
Python 3.9.1 and R 4.2 were used to collect the data and perform the data analysis presented in this study. Data collection and analysis code is available under MIT license in a GitHub repository under accession code \url{https://doi.org/10.5281/zenodo.7723109}.

\section{Acknowledgments}
This report was partly funded by the Templeton Foundation through a grant awarded to Wake Forest University for the ``Honesty Project''. SL was also supported by funding from the Humboldt Foundation in Germany, and SL and DG are beneficiaries of the ERC Advanced Grant PRODEMINFO (101020961). JL was supported by the Marie Skłodowska-Curie grant No. 101026507.

We acknowledge Travis Coan for helpful feedback on the manuscript.

\section{Author contributions}
SL, DG and JL conceptualised the research. STA, FC, JL and AS developed the methodology and statistical models. FC performed the validation. JL, STA, AS and FC performed computational and statistical analyses. JL and STA collected and curated the data. JL prepared the visualizations. JL administrated the project. SL and DG acquired funding and supervised the project. SL and JL wrote the original draft of the manuscript. All authors contributed to editing the original draft of the manuscript.

\section{Inclusion and ethics}
This study is based on publicly available archival Twitter data on U.S. Members of Congress and their official staff and campaign accounts. Only public figures are analyzed and only content that was not deleted by the time of data retrieval was considered. All U.S. Members of Congress in curated Twitter account lists are included as long as their Twitter accounts were public by the retrieval data. We focused on the two major parties to have sufficient evidence for statistical analysis and our results cannot be extended to independent members of congress or members of other parties besides the Democratic and the Republican party.

\subsection{Competing interests}
The authors declare no competing interests.

\clearpage
\section{Extended data figures and tables}
\renewcommand{\figurename}{Extended Data Figure}
\renewcommand{\tablename}{Extended Data Table}
\setcounter{figure}{0}
\setcounter{table}{0}

\begin{figure}[h]
    \centering
    \includegraphics[width=\textwidth]{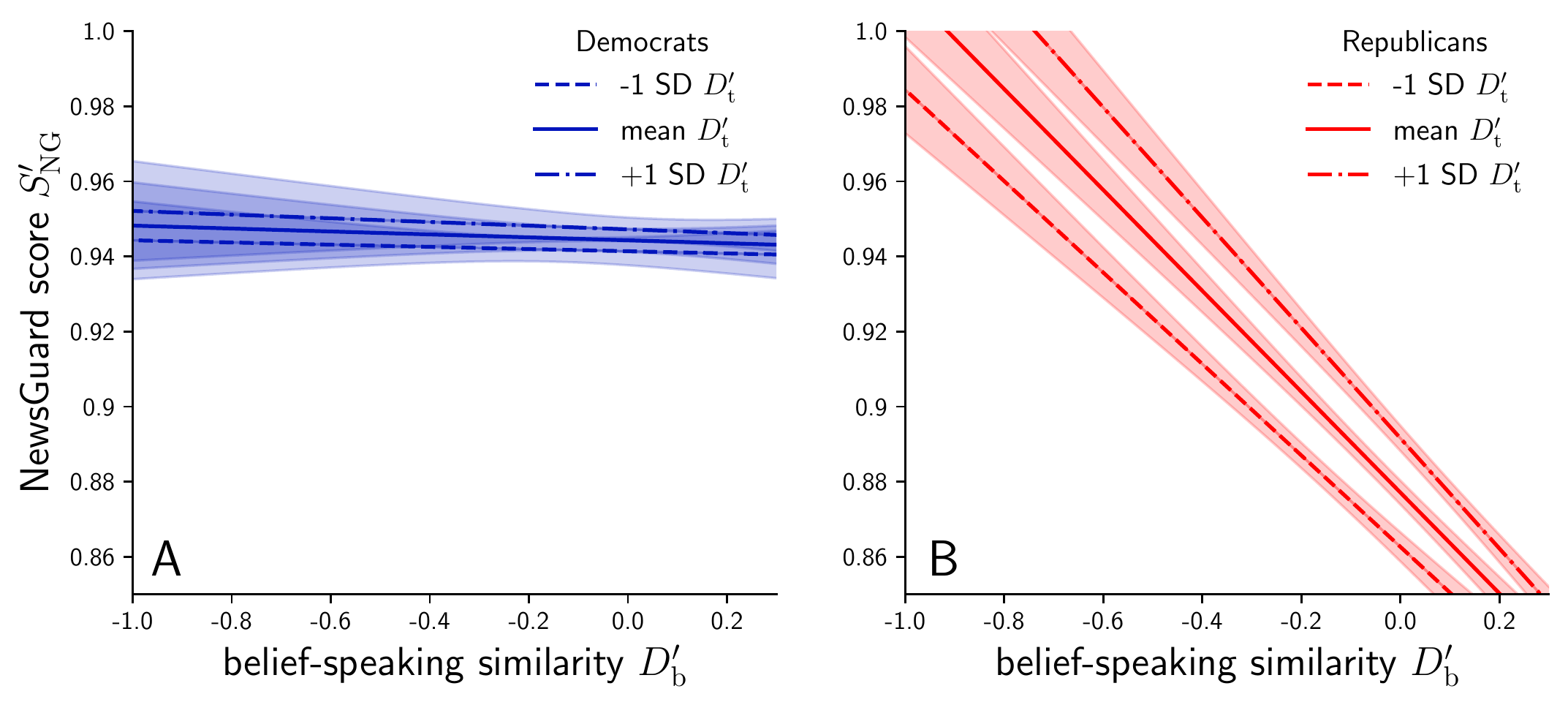}
    \caption{Prediction of rescaled NewsGuard score $S'_\mathrm{NG}$ for different values of belief-speaking similarity $D_\mathrm{b}'$ and different levels (-1\,SD, mean, +1\,SD) of truth-seeking similarity $D_\mathrm{t}'$ based on the fixed-effect estimate of the three-way interaction $P \times D_\mathrm{b}' \times D_\mathrm{t}'$ (see linear mixed effects model in Eq. (\ref{eq:regression_NGscore_tweets})) for tweets from A Democrat and B Republican members of the U.S. Congress.}
    \label{fig:ext_fig1}
\end{figure} 

\begin{figure}[h]
    \centering
    \includegraphics[width=\textwidth]{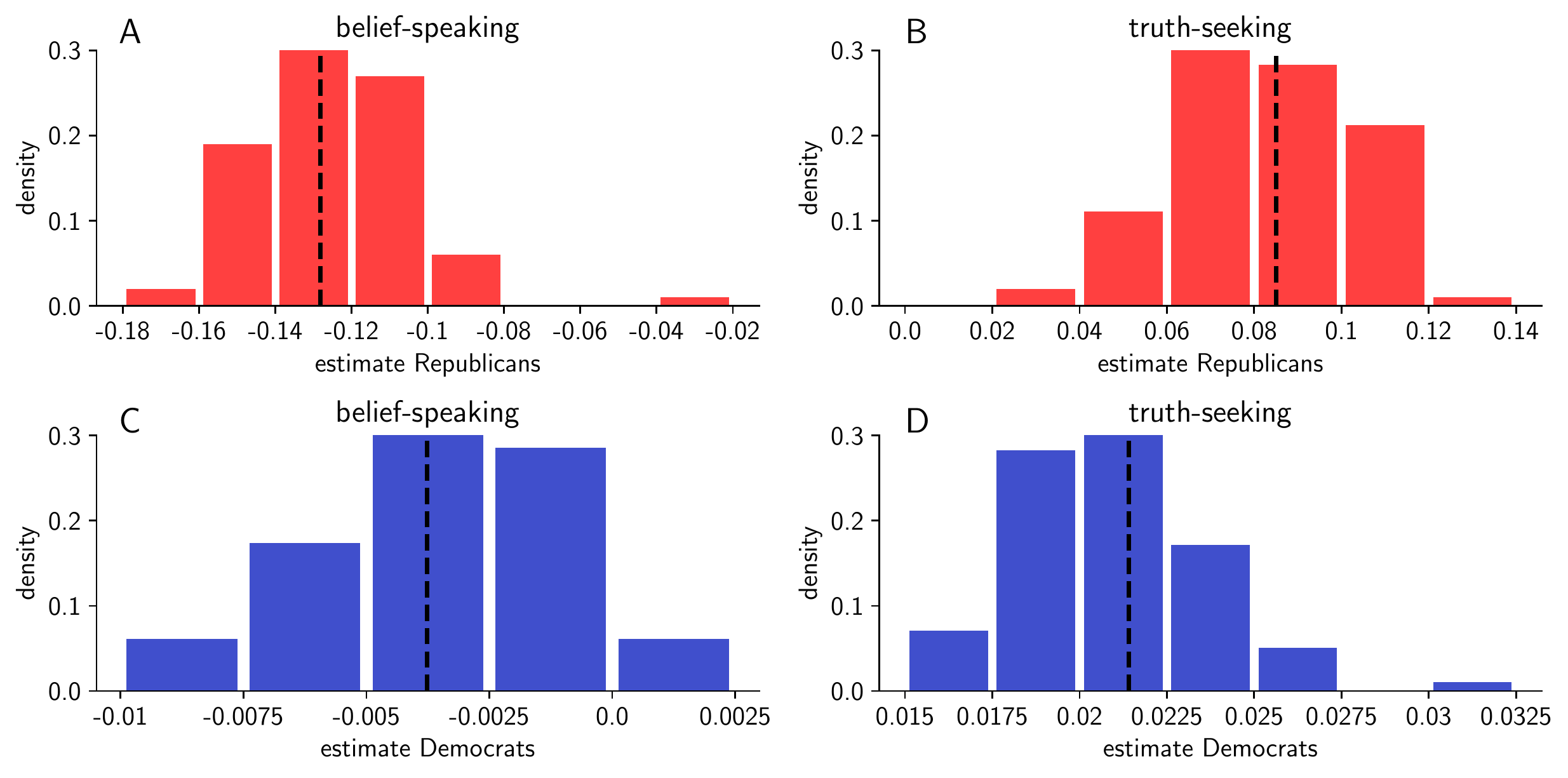}
    \caption{Dictionary robustness tests. A and B show the distribution of estimates of the effect of belief-speaking and truth-seeking similarity $D'_\mathrm{b}$ and truth-seeking similarity $D'_\mathrm{t}$ for Republicans from the linear mixed model (see Eq.(\ref{eq:regression_NGscore_tweets})), where $D'_\mathrm{b}$ and $D'_\mathrm{t}$ were calculated with a perturbed dictionary for every tweet, respectively. C and D show the distribution of estimates of the effect of $D'_\mathrm{b}$ and $D'_\mathrm{t}$ for Democrats, respectively. Distributions were calculated from 100 dictionary perturbation iterations. \\ \rule{\linewidth}{1pt}}
    \label{fig:ext_fig2}
\end{figure}

\clearpage
\begin{figure}
    \centering
    \includegraphics[width=0.9\textwidth]{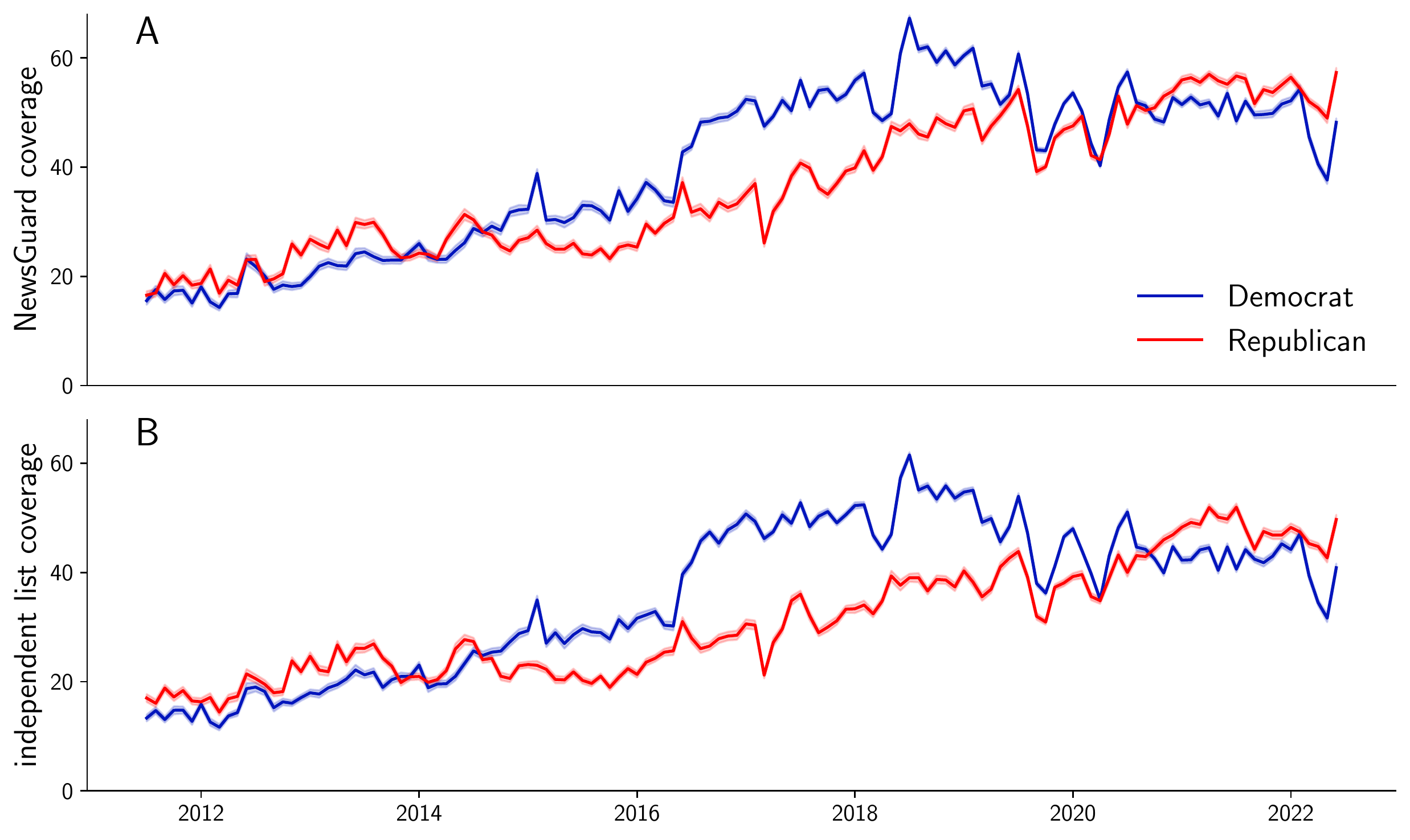}
    \caption{Share of links posted by accounts belonging to members of the U.S. Congress pointing to domains indexed in A the NewsGuard data base and B the independently compiled list.}
    \label{fig:ext_fig3}
\end{figure}  

\clearpage
\begin{figure}[t]
    \centering
    \includegraphics[width=0.9\textwidth]{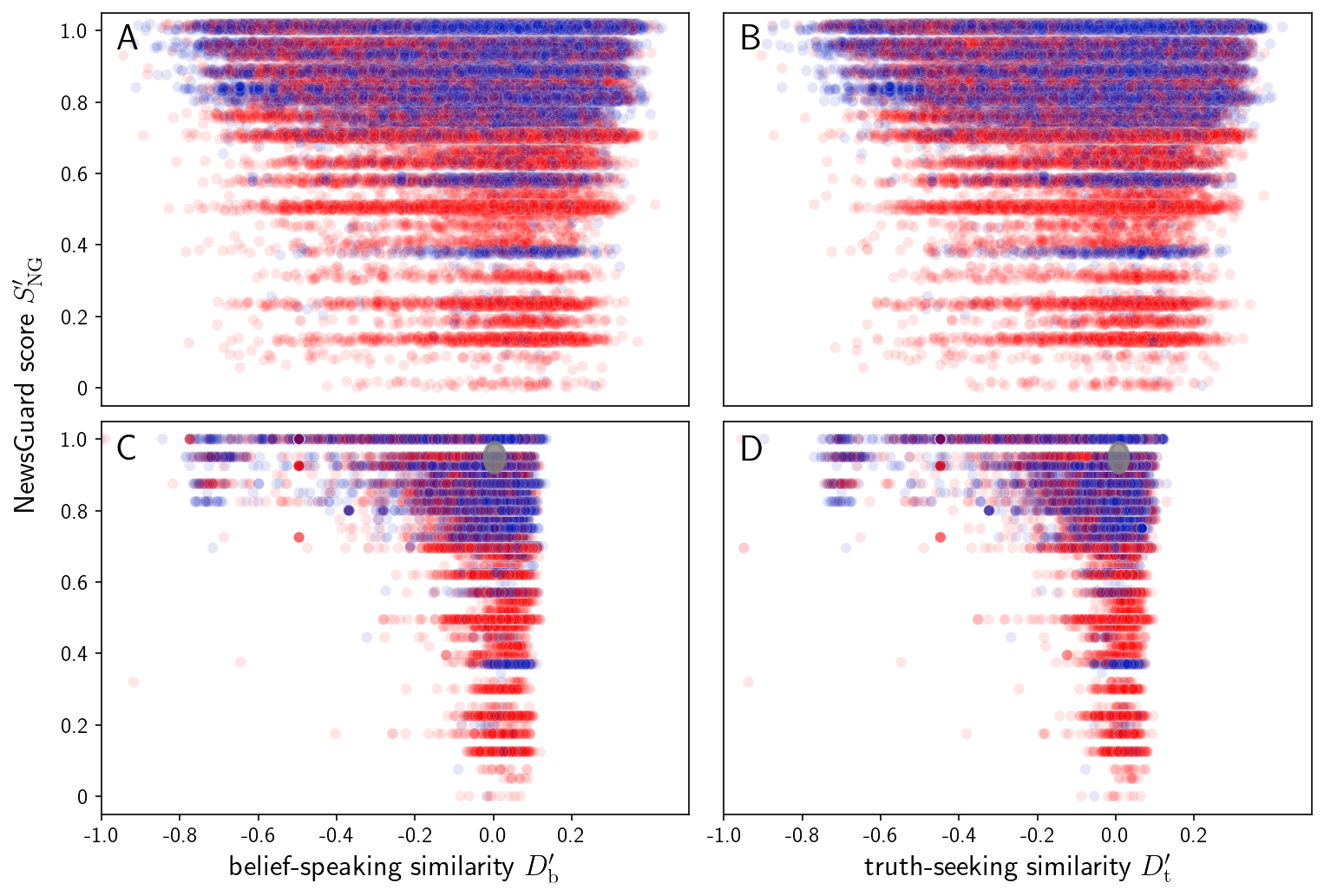}
    \caption{A and B rescaled NewsGuard score $S_\mathrm{NG}'$ of links shared in tweets by members of the U.S. congress over belief-speaking similarity $D_\mathrm{b}'$. Red and blue dots denote tweets by Democrats and Republicans, respectively. B shows $S_\mathrm{NG}'$ over truth-seeking similarity $D_\mathrm{t}'$ in tweets. C and D show the same information but with $D_\mathrm{b}'$ and $D_\mathrm{t}'$ calculated using the text of the articles that were linked instead of the tweet texts. The grey ellipses indicate the mean and standard deviation of the honesty component similarity and $S'_\mathrm{NG}$ for articles shared by members of both parties. These articles were excluded in the regression analysis.}
    \label{fig:ext_fig4}
\end{figure} 

\clearpage
\footnotesize
\begin{table}[]
\caption{Keyword lists for the two honesty components belief-speaking and truth-seeking.}
\centering
\begin{tabular}{@{}ll@{}}
\toprule
Belief-speaking & Truth-seeking \\ \midrule
basically        & actually       \\
believe          & analyze        \\
claim            & assess         \\
confide          & correct        \\
consider         & correction     \\
contemplate      & determine      \\
contention       & evaluate       \\
envisage         & evidence       \\
feel             & examine        \\
frankly          & exploration    \\
genuinely        & fact           \\
guess            & information    \\
hint             & inspect        \\
judge            & investigate    \\
look             & observe        \\
obvious          & proof          \\
obviously        & prove          \\
of course        & question       \\
opinion          & quiz           \\
plainly          & real           \\
ponder           & reality        \\
position         & rectify        \\
presume          & research       \\
probably         & revise         \\
seem             & sample         \\
sensation        & science        \\
sentiment        & scrutinize     \\
signal           & search         \\
suggest          & specify        \\
suggestion       & supervise      \\
suppose          & test           \\
surely           & trace          \\
think            & track          \\
trust            & trial          \\
try              & truth          \\
view             & validate       \\
virtually        & verify         \\ \bottomrule
\end{tabular}

\label{tab:ext_tab1}
\end{table}

\normalsize

\clearpage
\begin{table}[]
 \caption{Results of a linear mixed effects model for the dependence of the rescaled NewsGuard score of each link $S_\mathrm{NG}'$ on belief-speaking similarity $D_\mathrm{b}'$ and truth-seeking similarity $D_\mathrm{t}'$ in tweets, with party $P$ as fixed variable following Eq.(\ref{eq:regression_NGscore_tweets}). The table reports results for the fixed effects. 504,809 observations were included. Regression was performed with the function \texttt{lmer} from the R library lme4~\cite{lme4}.}
    \footnotesize
    \centering
    \begin{tabular}{l|c|c|c|c|c|c}
                                                      & coef.   & std. err. & $t$     & $P>\vert t \vert$   & [0.025  & 0.975] \\
        \toprule
		Intercept                                 & 0.9437  & 0.0016    & 582.104 & $<10^{-16}$         & 0.9406  & 0.9469 \\ 
		$D_\mathrm{b}'$                           & -0.0037 & 0.0059    & -0.626   & 0.5317              & -0.0154 & 0.0079 \\
		$D_\mathrm{t}'$                           & 0.0215  & 0.0059    & 3.656   & 0.0003              & 0.0100  & 0.0330 \\ 
		Republican                                & -0.0694 & 0.0023    & -29.894 & $<10^{-16}$         & -0.0740 & -0.0649 \\
		$D_\mathrm{b}'$ $\times$ $D_\mathrm{t}'$  & -0.0074 & 0.0099    & -0.741   & 0.4590              & -0.0268 & 0.0121 \\
		$D_\mathrm{b}'$ $\times$ Republican       & -0.1282 & 0.0089    & -14.362 & $<10^{-16}$         & -0.1457 & -0.1107 \\ 
		$D_\mathrm{t}'$ $\times$ Republican       & 0.0851  & 0.0089    & 9.598   & $<10^{-16}$         & 0.0677  & 0.1025 \\
		$D_\mathrm{b}'$ $\times$ $D_\mathrm{t}'$ $\times$ Republican & -0.0852 & 0.0151 & -5.645 & $2.6\cdot 10^{-8}$ & -0.1148 & -0.0556 \\ 
		\bottomrule 
		\multicolumn{2}{l}{Observations} & \multicolumn{1}{r}{504809} &
		\multicolumn{2}{l}{AIC} & \multicolumn{2}{r}{-800475} \\
		\multicolumn{2}{l}{Marginal R$^2$} & \multicolumn{1}{r}{0.086} &
		\multicolumn{2}{l}{log-Likelihood} & \multicolumn{2}{r}{400256} \\
	    \multicolumn{2}{l}{Conditional R$^2$} & \multicolumn{1}{r}{0.182} &  
	    \multicolumn{2}{l}{BIC} & \multicolumn{2}{r}{-800263} \\ 
        \bottomrule
    \end{tabular}
    \label{tab:ext_tab2}
\end{table}

\begin{table}[]
 \caption{Results of an ordinary least-squares regression for rescaled  NewsGuard score of each link $S_\mathrm{NG}'$ on  belief-speaking similarity $D_\mathrm{b}'$ and truth-seeking similarity $D_\mathrm{t}'$ in articles collected from links in tweets, following Eq.(\ref{eq:regression_NGscore_articles}). 296,267 observations were included. Regression was performed with the function \texttt{ols} from the Python package statsmodels~\cite{seabold2010statsmodels}, version 0.13.2.}
    \footnotesize
    \centering
    \begin{tabular}{l|c|c|c|c|c|c}
        & coef.   & std. err.  & $t$      & $P>\vert t \vert$ & [0.025  & 0.975] \\
        \toprule
		Intercept                                 & 0.9552  & 0.0003 & 3048.434 & $<10^{-16}$         & 0.95046 & 0.9559 \\ 
		$D_\mathrm{b}'$                           & -0.0641 & 0.0099 & -6.473   & $9.6\cdot 10^{-11}$ & -0.0835 & -0.0447 \\ 
		$D_\mathrm{t}'$                           & 0.0259  & 0.0113 & 2.291    & 0.0220              & 0.0037 & 0.0481 \\ 
		Republican                                & -0.1028 & 0.0005 & -200.278 & $<10^{-16}$         & -0.1038 & -0.1018 \\ 	
		$D_\mathrm{b}'$ $\times$ $D_\mathrm{t}'$  & 0.0666  & 0.0239 & 2.785    & 0.0053              & 0.0197 & 0.1134 \\ 
		  $D_\mathrm{b}'$ $\times$ Republican       & -0.5382 & 0.0161 & -33.515  & $<10^{-16}$         & -0.5596 & -0.5067 \\ 
		$D_\mathrm{t}'$ $\times$ Republican       & 0.1047  & 0.0187 & 5.638   & $1.7\cdot 10^{-8}$   & 0.0683 & 0.1411 \\ 
		$D_\mathrm{b}'$ $\times$ $D_\mathrm{t}'$ $\times$ Republican & -0.5868  & 0.0401 & -14.632 & $<10^{-16}$ & -0.6654 & -0.5082 \\  
		\bottomrule 
		\multicolumn{2}{l}{R-squared} & \multicolumn{1}{r}{0.152} & \multicolumn{2}{l}{Mean dependent var} & \multicolumn{2}{r}{0.917} \\ 
		\multicolumn{2}{l}{Adjusted R-squared} & \multicolumn{1}{r}{0.152} & \multicolumn{2}{l}{S.D. dependent var} & \multicolumn{2}{r}{0.135} \\ 
		\multicolumn{2}{l}{Model MSE} & \multicolumn{1}{r}{103.4} & \multicolumn{2}{l}{AIC} & \multicolumn{2}{r}{-348716} \\ 
		\multicolumn{2}{l}{Sum squared resid} & \multicolumn{1}{r}{4044} & \multicolumn{2}{l}{BIC} & \multicolumn{2}{r}{-348632} \\ 
		\multicolumn{2}{l}{Log-likelihood} & \multicolumn{1}{r}{174366} & \multicolumn{2}{l}{F-statistic} & \multicolumn{2}{r}{6692} \\ 
		\multicolumn{2}{l}{Durbin-Watson stat} & \multicolumn{1}{r}{1.451} & \multicolumn{2}{l}{Prob(F-statistic)} & \multicolumn{2}{r}{0.000} \\ 
            \bottomrule
    \end{tabular}
    \label{tab:ext_tab3}
\end{table}

\begin{appendices}

\renewcommand{\figurename}{Figure}
\renewcommand{\tablename}{Table}
\setcounter{figure}{0}
\setcounter{table}{0}
\renewcommand{\thefigure}{S\arabic{figure}}
\renewcommand{\thetable}{S\arabic{table}}
\renewcommand{\thesection}{S\arabic{section}}

\clearpage

\section{Instructions to participants during keyword validation}\label{sec:prolific_questionnaire_instructions}

\pagenumbering{arabic}
\setcounter{page}{2}

What follows is a verbatim 
copy of the instructions provided to the participants who rated the keywords.

\textit{People can have different ideas about what it means to be "honest".}

\textit{We are focusing on two ideas of honesty.}

\textit{One is based on intuition, "gut feeling" and authenticity. According to this idea, people speak the truth and are honest when they "say what they felt to be true in the moment". Whether or not claims are correct reflections of reality is not as important. We call this idea of honesty and truth "belief speaking".}

\textit{The other idea is based on evidence, analysis, and veracity. According to this idea, people speak the truth and are honest when their claims align with the evidence. Whether or not claims are authentic reflections of a person's feelings is not as important. We call this idea of honesty and truth "truth seeking".}

\textit{Your task is to judge, for each of the words below, which idea of honesty it is most closely related to.
If someone uses that word, does it likely reflect belief speaking? Or does the word likely reflect truth seeking?}
 
\textit{Please indicate which idea of honesty each word is closest to by selecting, for each column, a value from 1 to 5, where 1 means that the word is the least representative of that category, and 5 means that the word is highly representative of that category. 
There are no right or wrong answers, we are interested in your analysis of the meaning of those words.}

\clearpage
\section{Dictionary keyword validation results}
\begin{figure}[h]
    \centering
    \includegraphics[width=\textwidth]{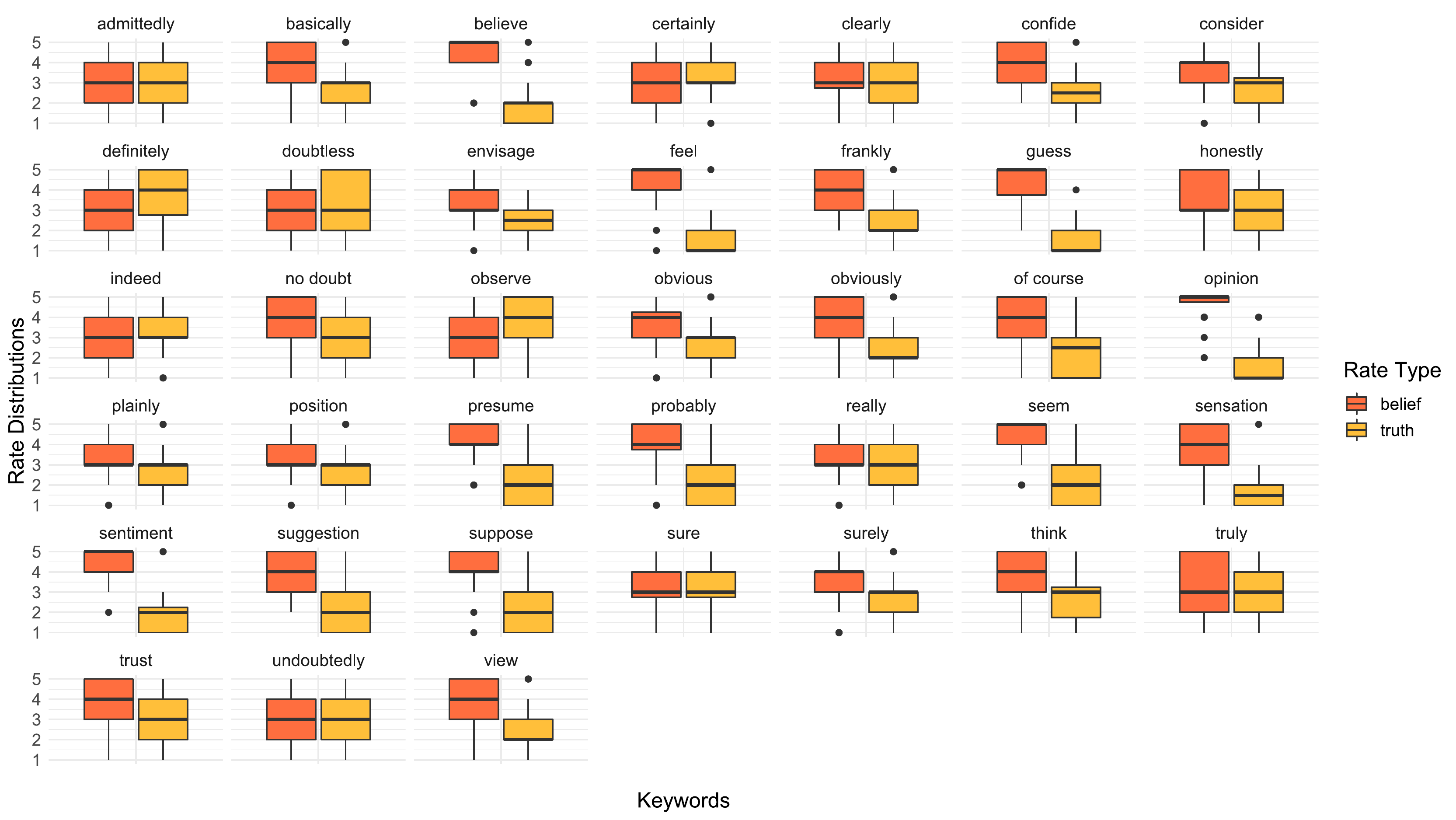}
    \caption{Boxplots of rating distributions for keywords we originally categorized as belief-speaking. Clear cases where our categorizations were confirmed are, for example, `opinion', `feel', `believe'. Examples of discarded keywords are `clearly', `undoubtedly', `sure'. The only reversed case is `observe', categorized as `truth-seeking' by the raters.\\ \rule{\linewidth}{1pt}}
    \label{fig:SI_fig1}
\end{figure}

\begin{figure}[h]
    \centering
    \includegraphics[width=\textwidth]{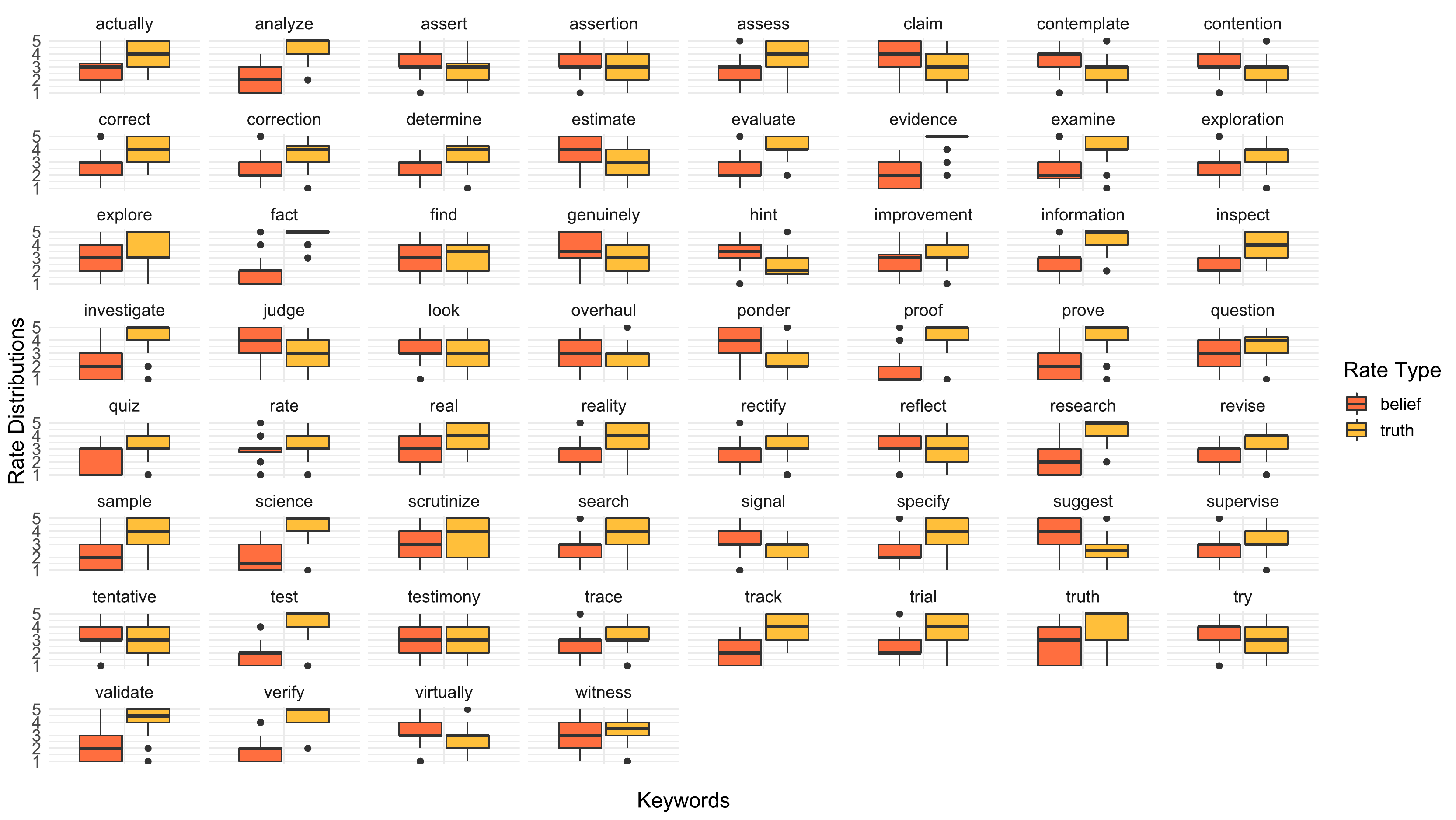}
    \caption{Boxplots of rating distributions for keywords we originally categorized as truth-seeking. Clear cases where our categorizations were confirmed are, for example, `verify', `fact', `evidence'. Examples of discarded keywords are `witness', `testimony', `overhaul'. Instances of reversed keywords are `suggest', `ponder', `judge', categorized as `belief-speaking' by the raters. \\ \rule{\linewidth}{1pt}}
    \label{fig:SI_fig2}
\end{figure}

To validate the keywords contained in the belief-speaking and truth-seeking dictionaries we asked raters on the survey platform Prolific~\citep{PALAN201822} to score each term on two scales reflecting their representativeness for belief-speaking and truth-seeking, respectively. The collected data contains responses from 50 participants and ratings from 1 to 5 for each keyword. Data were acquired September 20, 2022, the instructions provided to participants are reported in section ``Prolific Questionnaire Instructions''. The distributions of ratings collected for each keyword are shown in Figures~\ref{fig:SI_fig1} and~\ref{fig:SI_fig2}. 

To determine the validity of each keyword, we conducted t-tests between the distribution of representativeness ratings for belief-speaking and the distribution of representativeness ratings for truth-seeking for every keyword. If the difference between the distributions was significant ($\alpha=0.05$), the keyword was included in the belief-speaking dictionary if the t-value was positive, and in the truth-seeking dictionary if the t-value was negative. Results of the t-tests for each keyword are reported in Table~\ref{tab:SI_tab1}. 

\clearpage
{\footnotesize
\begin{longtable}[c]{@{}llllll@{}}
\caption{Results of the t-tests of the keyword ratings performed by 50 raters. ``component'' indicates the honesty component a given keyword was initially assigned to. The column ``valid'' is a binary variable indicating whether our initial component assignment for the keyword was confirmed by the raters, based on the t-value direction (positive for belief-speaking, negative for truth-seeking) and a significance level of $\alpha=0.05$. The column ``opposite'' indicates whether a keyword was shifted to the opposite honesty component dictionary. This happened when the t-value was significant ($\alpha=0.05$) but in the opposite direction than initially assumed. Rating distributions are shown in Figure~\ref{fig:SI_fig1} for the keywords that were initially categorised as ``belief-speaking'' and in Figure~\ref{fig:SI_fig2} for the keywords that were initially categorised as ``truth-seeking''.}\\
\toprule
keyword     & t value  & p value & component & valid & opposite \\* \midrule
\endfirsthead
\endhead
\bottomrule
\endfoot
\endlastfoot
actually    & -3.7939  & 0.0004  & truth         & yes       & no           \\
admittedly  & 1.2102   & 0.2318  & belief        & no        & no           \\
analyze     & -11.8607 & 0.0000  & truth         & yes       & no           \\
assert      & 1.6488   & 0.1053  & truth         & no        & no           \\
assertion   & 1.9003   & 0.0630  & truth         & no        & no           \\
assess      & -6.6167  & 0.0000  & truth         & yes       & no           \\
basically   & 5.6661   & 0.0000  & belief        & yes       & no           \\
believe     & 12.9276  & 0.0000  & belief        & yes       & no           \\
certainly   & -1.7321  & 0.0893  & belief        & no        & no           \\
claim       & 3.7398   & 0.0005  & truth         & no        & yes          \\
clearly     & 0.2989   & 0.7663  & belief        & no        & no           \\
confide     & 5.5550   & 0.0000  & belief        & yes       & no           \\
consider    & 2.6606   & 0.0104  & belief        & yes       & no           \\
contemplate & 3.3981   & 0.0013  & truth         & no        & yes          \\
contention  & 2.0449   & 0.0460  & truth         & no        & yes          \\
correct     & -4.5756  & 0.0000  & truth         & yes       & no           \\
correction  & -4.7842  & 0.0000  & truth         & yes       & no           \\
definitely  & -1.8134  & 0.0757  & belief        & no        & no           \\
determine   & -5.4070  & 0.0000  & truth         & yes       & no           \\
doubtless   & -0.5534  & 0.5824  & belief        & no        & no           \\
envisage    & 5.5751   & 0.0000  & belief        & yes       & no           \\
estimate    & 1.6797   & 0.0991  & truth         & no        & no           \\
evaluate    & -9.2428  & 0.0000  & truth         & yes       & no           \\
evidence    & -13.5218 & 0.0000  & truth         & yes       & no           \\
examine     & -7.2276  & 0.0000  & truth         & yes       & no           \\
exploration & -3.7341  & 0.0005  & truth         & yes       & no           \\
explore     & -1.4402  & 0.1559  & truth         & no        & no           \\
fact        & -14.9015 & 0.0000  & truth         & yes       & no           \\
feel        & 13.3212  & 0.0000  & belief        & yes       & no           \\
find        & -1.9767  & 0.0535  & truth         & no        & no           \\
frankly     & 5.3732   & 0.0000  & belief        & yes       & no           \\
genuinely   & 2.1898   & 0.0331  & truth         & no        & yes          \\
guess       & 11.8937  & 0.0000  & belief        & yes       & no           \\
hint        & 3.9430   & 0.0002  & truth         & no        & yes          \\
honestly    & 1.0163   & 0.3143  & belief        & no        & no           \\
improvement & -1.0674  & 0.2908  & truth         & no        & no           \\
indeed      & -0.5380  & 0.5929  & belief        & no        & no           \\
information & -7.8184  & 0.0000  & truth         & yes       & no           \\
inspect     & -8.3901  & 0.0000  & truth         & yes       & no           \\
investigate & -10.0865 & 0.0000  & truth         & yes       & no           \\
judge       & 4.4555   & 0.0000  & truth         & no        & yes          \\
look        & 2.1598   & 0.0355  & truth         & no        & yes          \\
no doubt    & 0.8743   & 0.3860  & belief        & no        & no           \\
observe     & -4.3294  & 0.0001  & belief        & no        & yes          \\
obvious     & 2.7386   & 0.0085  & belief        & yes       & no           \\
obviously   & 4.6009   & 0.0000  & belief        & yes       & no           \\
of course   & 3.9459   & 0.0002  & belief        & yes       & no           \\
opinion     & 15.1750  & 0.0000  & belief        & yes       & no           \\
overhaul    & 1.1481   & 0.2563  & truth         & no        & no           \\
plainly     & 2.5317   & 0.0145  & belief        & yes       & no           \\
ponder      & 4.9805   & 0.0000  & truth         & no        & yes          \\
position    & 2.0462   & 0.0459  & belief        & yes       & no           \\
presume     & 8.7004   & 0.0000  & belief        & yes       & no           \\
probably    & 5.7093   & 0.0000  & belief        & yes       & no           \\
proof       & -12.3100 & 0.0000  & truth         & yes       & no           \\
prove       & -8.3425  & 0.0000  & truth         & yes       & no           \\
question    & -3.2428  & 0.0021  & truth         & yes       & no           \\
quiz        & -4.4351  & 0.0000  & truth         & yes       & no           \\
rate        & -1.1864  & 0.2409  & truth         & no        & no           \\
real        & -4.3970  & 0.0001  & truth         & yes       & no           \\
reality     & -5.6908  & 0.0000  & truth         & yes       & no           \\
really      & 1.0758   & 0.2871  & belief        & no        & no           \\
rectify     & -2.5995  & 0.0122  & truth         & yes       & no           \\
reflect     & 1.9660   & 0.0548  & truth         & no        & no           \\
research    & -10.5963 & 0.0000  & truth         & yes       & no           \\
revise      & -4.6863  & 0.0000  & truth         & yes       & no           \\
sample      & -7.0234  & 0.0000  & truth         & yes       & no           \\
science     & -12.6170 & 0.0000  & truth         & yes       & no           \\
scrutinize  & -2.1898  & 0.0331  & truth         & yes       & no           \\
search      & -3.7338  & 0.0005  & truth         & yes       & no           \\
seem        & 8.6065   & 0.0000  & belief        & yes       & no           \\
sensation   & 11.1959  & 0.0000  & belief        & yes       & no           \\
sentiment   & 11.0784  & 0.0000  & belief        & yes       & no           \\
signal      & 2.5428   & 0.0141  & truth         & no        & yes          \\
specify     & -5.4430  & 0.0000  & truth         & yes       & no           \\
suggest     & 5.1004   & 0.0000  & truth         & no        & yes          \\
suggestion  & 6.6150   & 0.0000  & belief        & yes       & no           \\
supervise   & -2.5412  & 0.0141  & truth         & yes       & no           \\
suppose     & 8.6284   & 0.0000  & belief        & yes       & no           \\
sure        & 0.3841   & 0.7025  & belief        & no        & no           \\
surely      & 3.0461   & 0.0037  & belief        & yes       & no           \\
tentative   & 1.6450   & 0.1061  & truth         & no        & no           \\
test        & -9.2804  & 0.0000  & truth         & yes       & no           \\
testimony   & 0.1301   & 0.8970  & truth         & no        & no           \\
think       & 4.3846   & 0.0001  & belief        & yes       & no           \\
trace       & -2.7584  & 0.0080  & truth         & yes       & no           \\
track       & -8.4954  & 0.0000  & truth         & yes       & no           \\
trial       & -6.3374  & 0.0000  & truth         & yes       & no           \\
truly       & 1.1579   & 0.2523  & belief        & no        & no           \\
trust       & 2.4280   & 0.0187  & belief        & yes       & no           \\
truth       & -4.9316  & 0.0000  & truth         & yes       & no           \\
try         & 3.0329   & 0.0038  & truth         & no        & yes          \\
undoubtedly & -0.3104  & 0.7575  & belief        & no        & no           \\
validate    & -8.0957  & 0.0000  & truth         & yes       & no           \\
verify      & -15.4471 & 0.0000  & truth         & yes       & no           \\
view        & 5.0381   & 0.0000  & belief        & yes       & no           \\
virtually   & 2.6190   & 0.0116  & truth         & no        & yes          \\
witness     & -0.8494  & 0.3996  & truth         & no        & no           \\
\bottomrule

\label{tab:SI_tab1}
\end{longtable}}

\clearpage

\clearpage
\section{Document-level validation results}
To validate the belief-speaking and truth-seeking measures, we asked raters on the survey platform Prolific~\citep{PALAN201822} to score tweets on two scales reflecting their representativeness for belief-speaking and truth-seeking, respectively. 

Tweets shown to the participants were sampled from the full corpus of tweets with the aim of sampling tweets with high and low honesty component similarity $D'_\mathrm{b}$ and $D'_\mathrm{t}$. We thus sampled 20 tweets from the top belief-speaking and bottom truth-seeking quartile, as well as 20 tweets from the top truth-seeking and bottom belief-speaking quartile. In addition, we sampled 20 tweets that simultaneously belonged to the bottom belief-speaking and truth-seeking quartiles. Each sample of 20 tweets included 10 tweets from Democrats and 10 from Republicans. 

The collected data contains responses from 50 participants (one participant from the initial 51 participants was excluded due to failing the attention check) and ratings from 1 to 5 for each tweet for belief-speaking and truth-seeking, respectively. Data were acquired February 10, 2023. The instructions provided to participants are the same as those reported in Section~\ref{sec:prolific_questionnaire_instructions} with the only adaptation that the term ``word'' was replaced with the term ``tweet''.

We then classified every tweet for which a majority of human raters selected either a ``4'' or a ``5'' for how characteristic a tweet was for ``belief-speaking'' [``truth-seeking''] as ``belief-speaking'' [``truth-seeking''] to create a ground-truth dataset to compare our classifier against. This resulted in 27 tweets that were classified as ``belief-speaking'', 21 tweets that were classified as ``truth-seeking'' and 12 tweets that were classified as neither by human raters.

\begin{figure}[h]
    \centering
    \includegraphics[width=\textwidth]{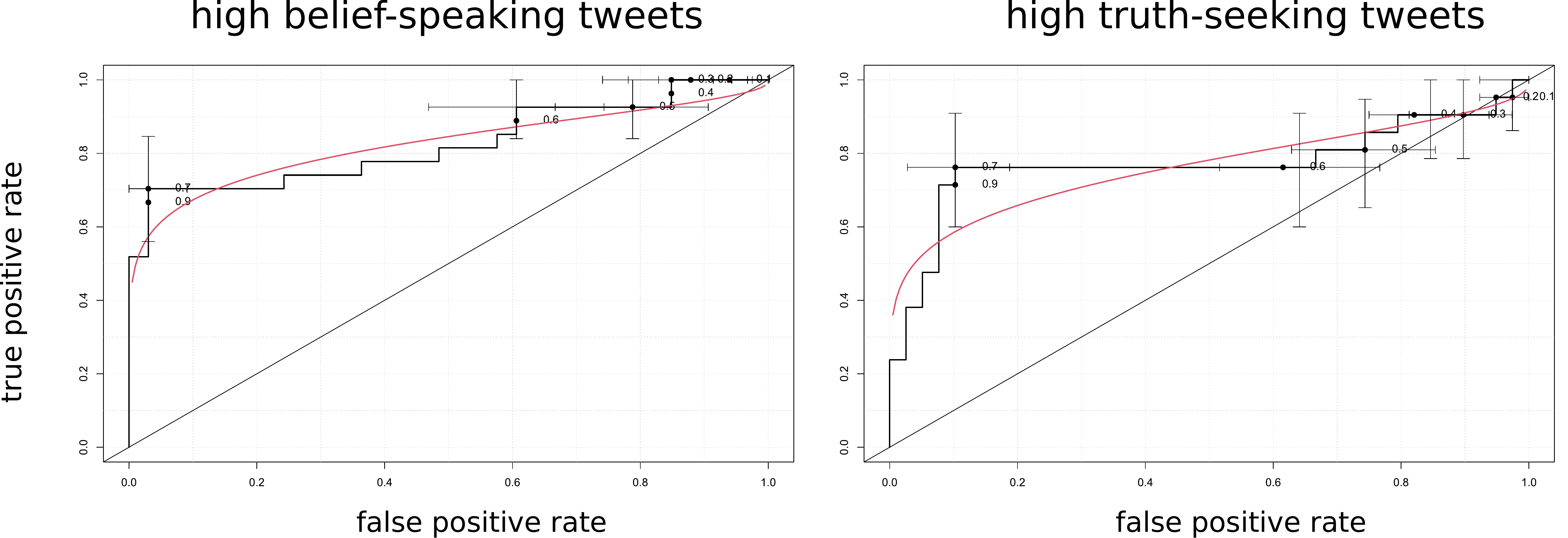}
    \caption{ROC curves for the classification of individual tweets into belief-speaking (left) and truth-seeking (right).\\ \rule{\linewidth}{1pt}}
    \label{fig:SI_fig3}
\end{figure}

\begin{figure}[h]
    \centering
    \includegraphics[width=\textwidth]{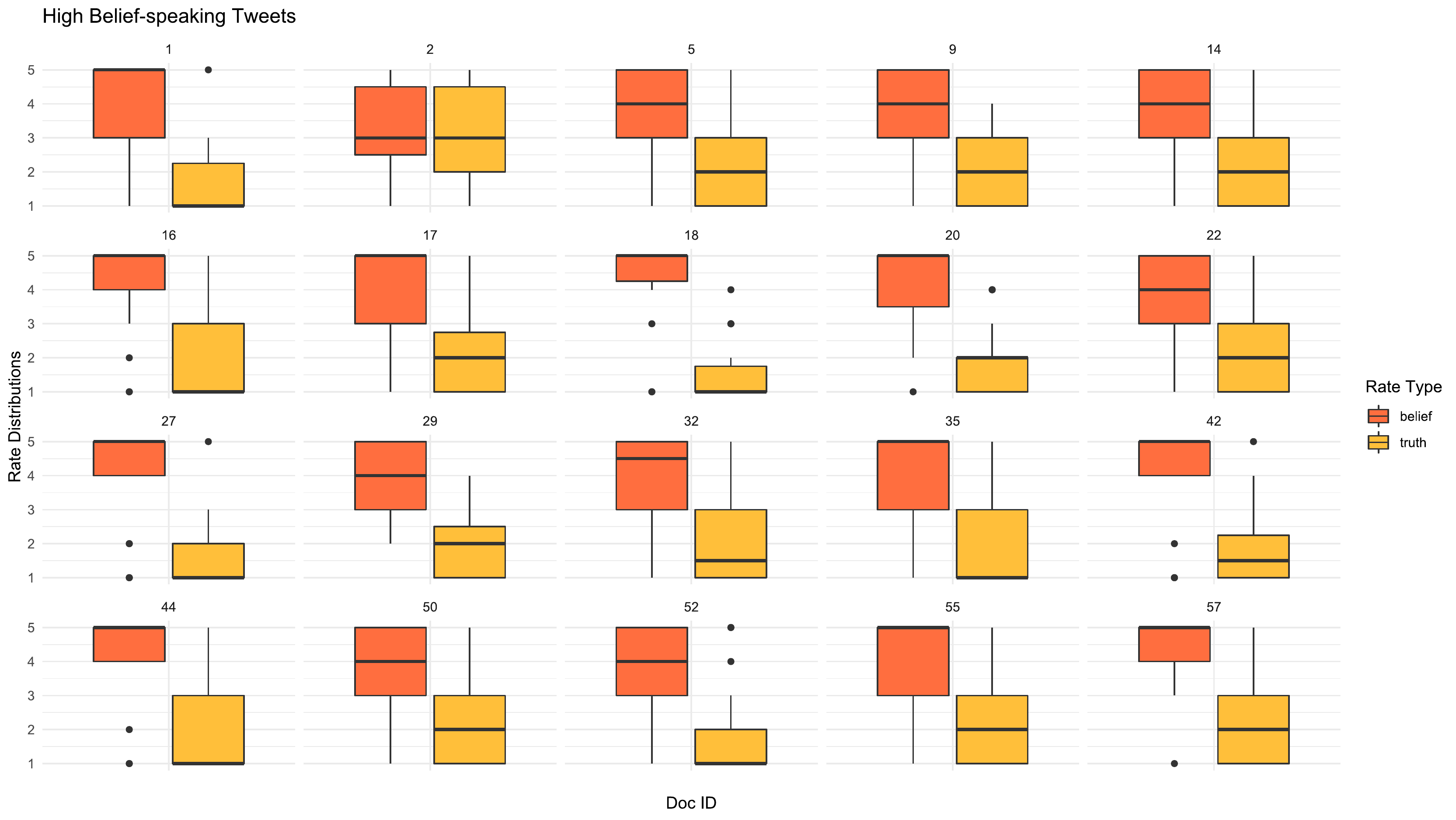}
    \caption{Boxplots of rating distributions for tweets sampled from the top belief-speaking and bottom truth-seeking quantiles.\\ \rule{\linewidth}{1pt}}
    \label{fig:SI_fig4}
\end{figure}

\begin{figure}[h]
    \centering
    \includegraphics[width=\textwidth]{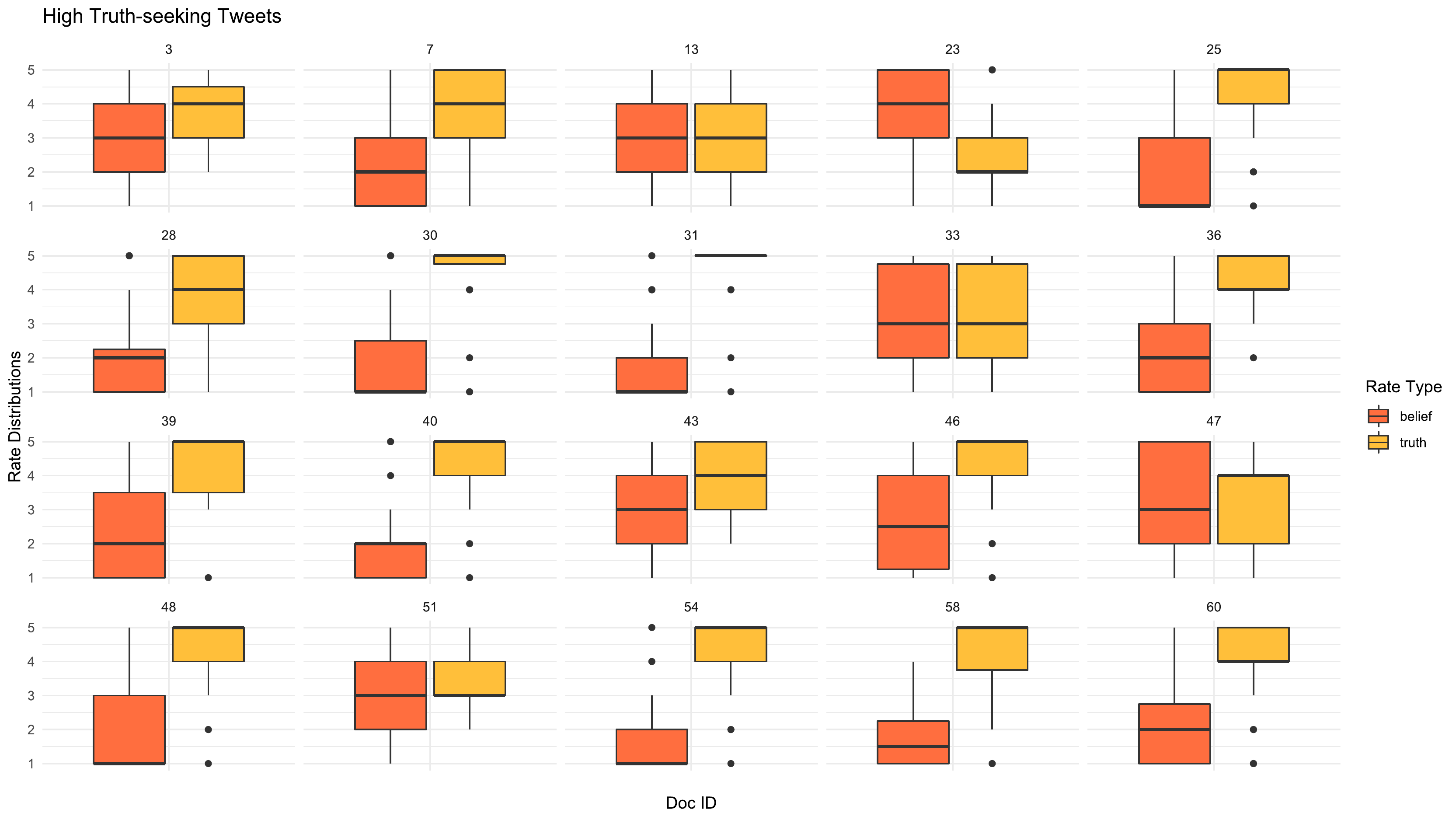}
    \caption{Boxplots of rating distributions for tweets sampled from the top truth-seeking and bottom belief-speaking quantiles.\\ \rule{\linewidth}{1pt}}
    \label{fig:SI_fig5}
\end{figure}

\begin{figure}[h]
    \centering
    \includegraphics[width=\textwidth]{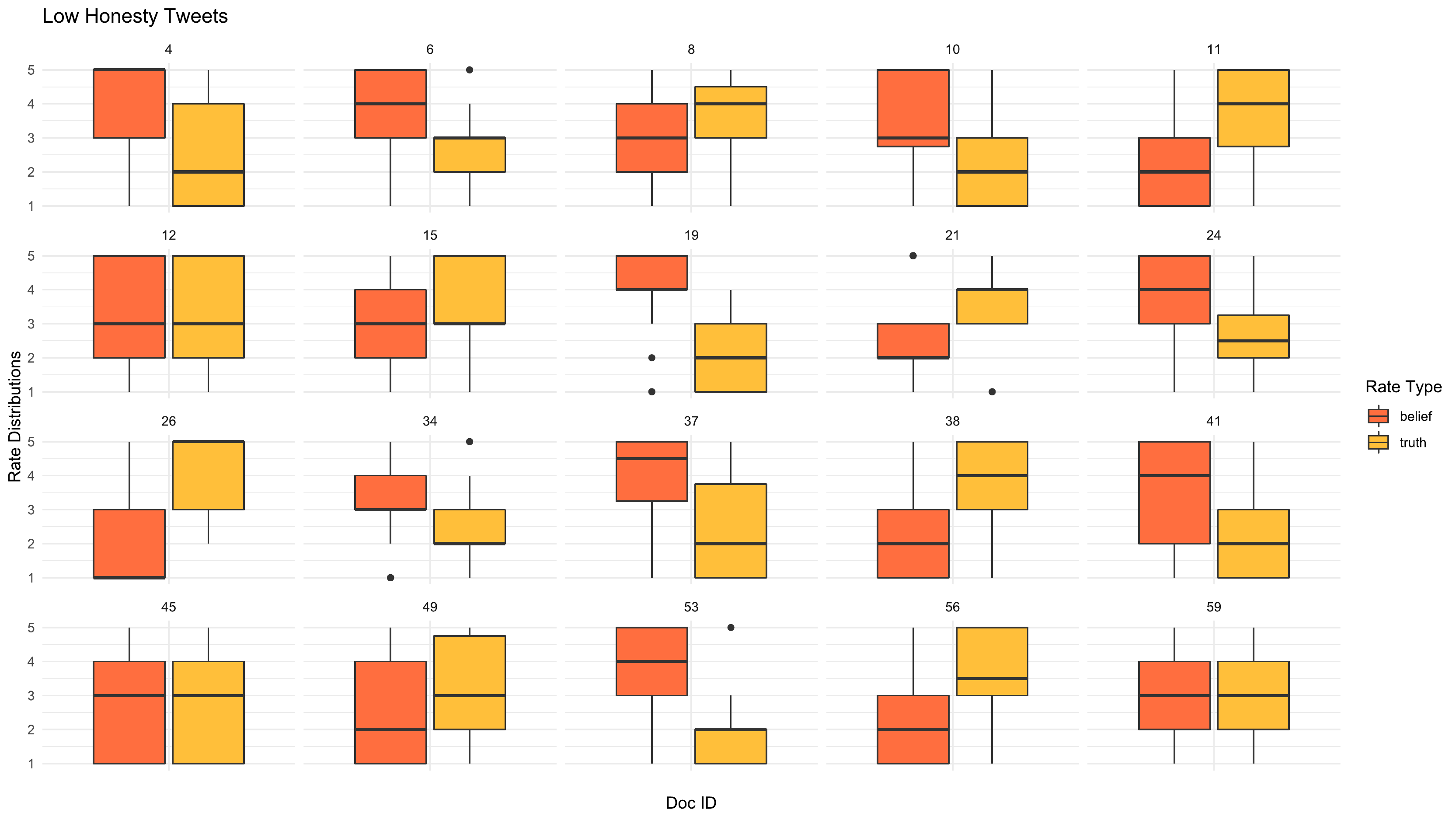}
    \caption{Boxplots of rating distributions for tweets sampled from the bottom belief-speaking and truth-seeking quantiles.\\ \rule{\linewidth}{1pt}}
    \label{fig:SI_fig6}
\end{figure}

To assess the performance of our similarity-based classifier, we calculate the ROC curves for belief-speaking as the threshold for the belief-speaking similarity $D'_\mathrm{b}$ to classify a tweet as ``belief-speaking'' is varied (see Figure~\ref{fig:SI_fig3}, left panel). The ROC curve for the truth-seeking similarity $D'_\mathrm{t}$  is shown in the right panel of Figure~\ref{fig:SI_fig3}. The area under the curve is high in both cases, with $\mathrm{AUC}=0.824$ for belief-speaking and $\mathrm{AUC}=0.772$ for truth-seeking. The distributions of ratings collected for each keyword are shown in Figures~\ref{fig:SI_fig4}, \ref{fig:SI_fig5} and~\ref{fig:SI_fig6}. 

\clearpage
\section{VADER text analysis}
We explored the content of the tweet texts within the two honesty components using Valence  Aware  Dictionary  for sEntiment Reasoning (VADER) \citep{Hutto2014}. VADER is a lexicon and rule-based sentiment analysis tool that is specifically attuned to sentiments expressed in social media. VADER computes sentiment polarity of a text and provides a ``positive'' and ``negative'' sentiment score, as well as a ``neutral'' and ``compound'' score. 

\begin{figure}  
    \includegraphics[width=\textwidth]{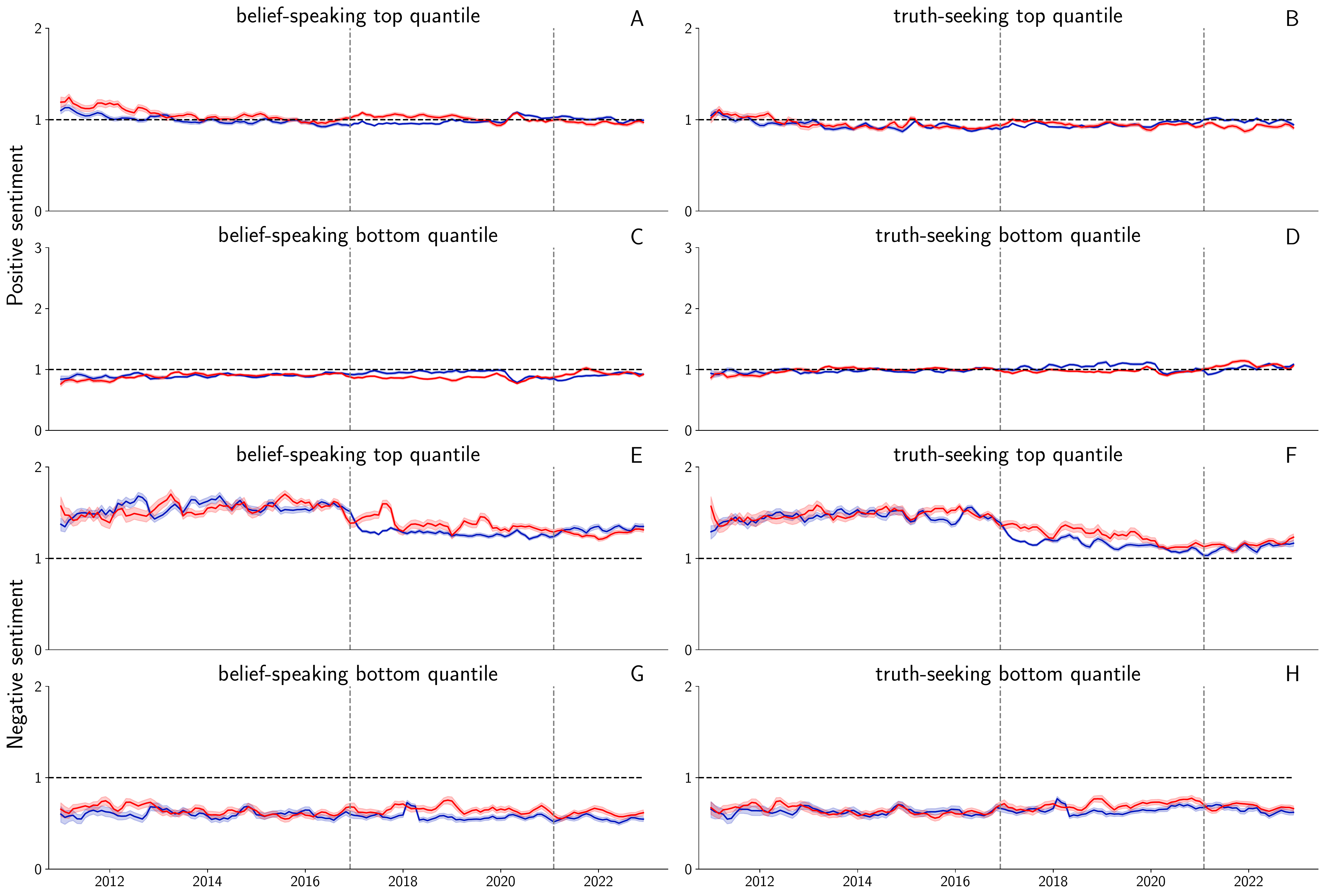}
    \caption{ \footnotesize Time-development of VADER scores of positive and negative sentiment in tweets of members of the U.S. Congress. Panels \textbf{A} and \textbf{B} show the score for positive sentiment for tweets that belong to the top belief-speaking and truth-seeking similarity quantile, while panels \textbf{C} and \textbf{D} show the positive sentiment score for the bottom similarity quantiles. Timelines are normalized by the overall sentiment score (baseline) for positive sentiment measured in the full corpus. The dashed horizontal line at 1.0 corresponds to prevalence equal to baseline. Red and blue lines correspond to tweets by Republicans and Democrats, respectively. Panels \textbf{E}, \textbf{F}, \textbf{G} and \textbf{H} show the same information as panels \textbf{A}, \textbf{B}, \textbf{C} and \textbf{D}, but for the negative sentiment score instead of the positive sentiment score. The 95\% confidence intervals (indicated by shading) were computed with bootstrap sampling over 1,000 iterations. Dashed vertical lines indicate dates of presidential elections in 2016 and 2020. Timelines are smoothed, using a rolling average over three months. \\ \rule{\linewidth}{1pt}}
    \label{fig:SI_fig7}
\end{figure}

Correlations between VADER scores and belief-speaking and truth-seeking similarity are given in Table~\ref{tab:SI_tab2}. In addition, we show the time-development of the positive and negative scores broken down for the top and bottom quantiles of belief-speaking and truth-seeking similarity in Figure~\ref{fig:SI_fig7}.

\begin{table}[!ht]
    \caption{Pearson correlation between belief-speaking and truth-seeking similarity and LIWC scores measuring the prevalence of ``analytic'', ``authentic'' and ``moral'' language, as well as positive and negative sentiment measured with VADER.}
    \centering
    \begin{tabular}{c|c|c|c|c|c}
        Honesty component & Analytic & Authentic & Moral & Pos. sentiment & Neg. sentiment \\
        \toprule
         Belief-speaking & -0.27 & 0.10 & 0.07 & 0.06 & 0.19 \\
         Truth-seeking   & -0.16 & 0.06 & 0.02 & -0.01 & 0.15 \\
         \bottomrule
    \end{tabular}
    \label{tab:SI_tab2}
\end{table}

\section{LIWC text analysis}
We also explored the content of the tweet texts within the two honesty components using the Linguistic Inquiry and Word Count (LIWC) program \citep{Boyd22}. LIWC is a text processing software that has been continuously developed for more than two decades and computes several indicator variables from text based on word lists generated by psychologists and validated in various experiments --- similar to our approach in generating the word lists for the belief-speaking and truth-seeking word lists.

With the Beta version of LIWC-2022 software (\url{https://www.liwc.app/}), we computed the scores for each tweet text for the following LIWC categories: authenticity, analytic, and moral. Authenticity indicates to what extent the language used is perceived as honest and genuine \citep[][]{newman2003}. Analytic is linked to logical and formal thinking \citep[][]{pennebaker2014small}. Finally, moral reflects the judgmental language expressed by positive or negative evaluation of someone's behavior or character \citep[][]{bradyetal2020}. The scores provide an efficient summary of those attributes in each text. 

\begin{figure}  
\includegraphics[width=\textwidth]{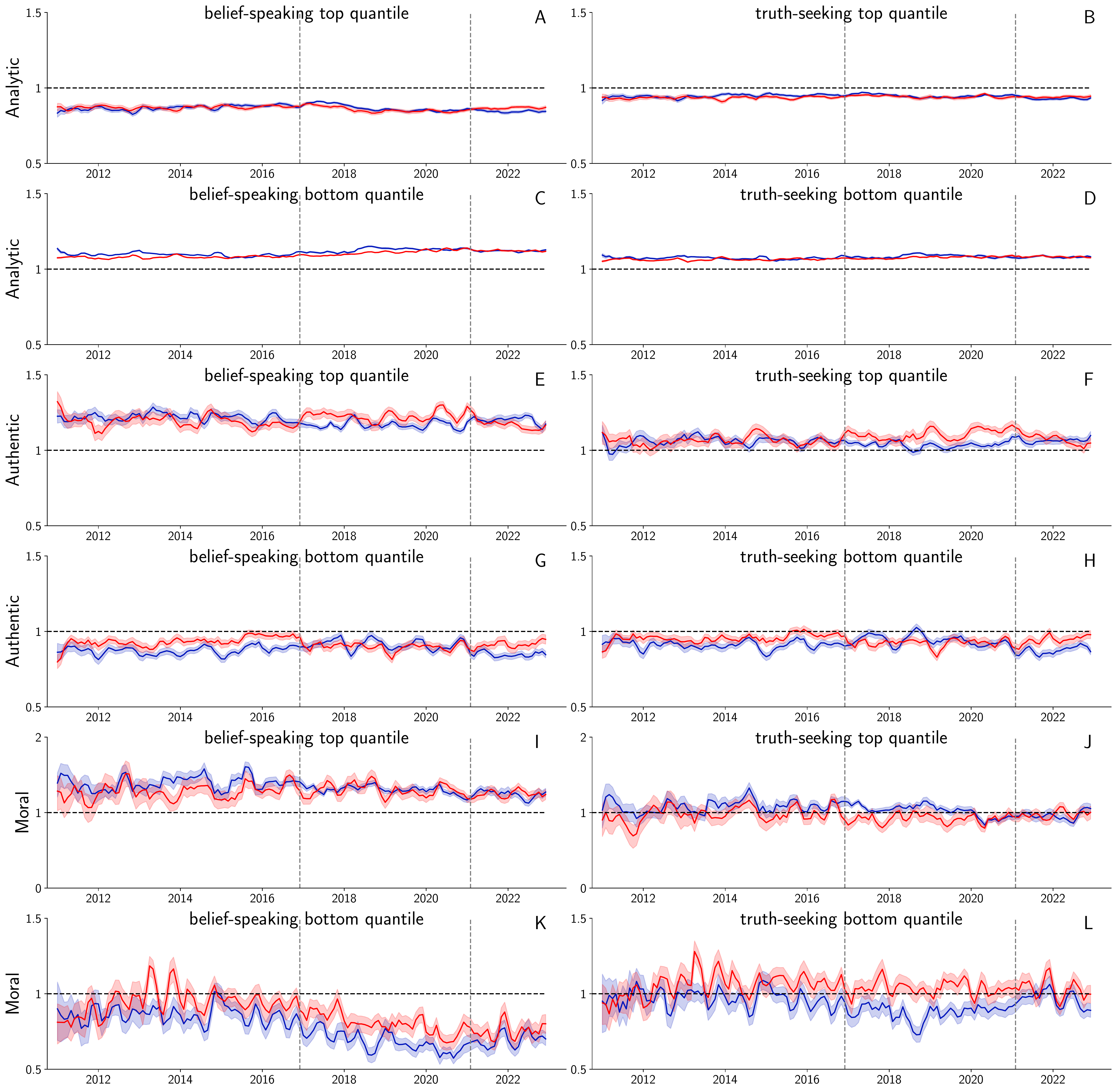}
\caption{ \footnotesize Time-development of LIWC scores of ``analytic'', ``authentic'' and ``moral'' language in tweets of members of the U.S. Congress. Panels \textbf{A} and \textbf{B} show the ``analytic'' score for tweets that belong to the top belief-speaking and truth-seeking similarity quantile, while panels \textbf{C} and \textbf{D} show the ``analytic'' score for the bottom similarity quantiles. Timelines are normalized by the overall ``analytic'' score (baseline) measured in the full corpus. Red and blue lines correspond to tweets by Republicans and Democrats, respectively. Panels \textbf{E} to \textbf{H} show the same information as panels \textbf{A} to \textbf{D}, but for ``authentic'' language, while panels \textbf{I} to \textbf{L} show the same information for ``moral'' language. The 95\% confidence intervals (indicated by shading) were computed with bootstrap sampling over 1,000 iterations. Dashed vertical lines indicate dates of presidential elections in 2016 and 2020. Timelines are smoothed, using a rolling average over three months. \label{fig:SI_fig8} \\ \rule{\linewidth}{1pt}}
\end{figure}

Correlations between LIWC scores and belief-speaking and truth-seeking similarity are given in Table~\ref{tab:SI_tab2}. In addition, we show the time-development of the scores broken down for the top and bottom quantiles of belief-speaking and truth-seeking similarity for the ``analytic'', ``authentic'' and ``moral'' components in  Figure~\ref{fig:SI_fig8}.

Figure~\ref{fig:SI_fig7} shows the timelines of LIWC scores for positive and negative emotions for the top and bottom quantile for belief-speaking and truth-seeking similarity. We performed the same analysis for ``authentic'', ``analytic'' and ``moral'' language, using LIWC dictionaries as described in the Methods Section ``LIWC text analysis''. The time development of ``analytic'' language broken down by honesty component is shown in Figure~\ref{fig:SI_fig8}, panels \textbf{A} to \textbf{D}, the time development of ``authentic'' language is shown in Figure~\ref{fig:SI_fig8} panels \textbf{E} to \textbf{H} and the time development of ``moral'' language is shown in Figure~\ref{fig:SI_fig8} panels \textbf{I} to \textbf{L}.

\clearpage
\section{Topic analysis}
To investigate the prevalence of belief-speaking and truth-seeking, we performed topic modelling using the Python package BERTopic \citep[][]{grootendorst2022bertopic}. Following a three-step approach, the package uses the Sentence-BERT (SBERT) framework to create the embeddings for each document, then uses the Uniform Manifold Approximation and Projection (UMAP) technique~\cite{mcinnes2018umap} to decrease the  dimensionality of embeddings and identify clusters through HDBSCAN~\cite{campello2013density}. Finally, it creates topic representations using class-based term-frequency inverse-document-frequency (TF-IDF). We opted for BERTopic rather than other techniques such as Latent Dirichlet Allocation (LDA) because the former performs better when modelling short and unstructured texts as in the case of Twitter data when compared to the latter \citep[][]{egger2022topic, alhaj2022improving}. Since BERTopic relies on an embedding approach, data was only minimally preprocessed to keep the original sentence structure. This means we lemmatized the entire dataset to produce cleaner topic representations, and only removed URLs from the texts. 

Since the number of documents was too large to fit a topic model of all documents, we restricted the corpus to the last 3200 tweets from each account. We also applied thresholds to the topic modelling: The document minimum frequency was set to 200 in order to reduce the number of small topics. The number of neighboring sample points used when making the manifold approximation was set to 100 to produce a more global view of the embedding structure. Finally, the minimum document frequency for the c-TF-IDF was set to 50 to reduce the topic-term matrix size and decrease memory-related issues during the computation. With these settings, the model was able to identify 363 topics.

To check whether this was an optimal number of topics, we used ldatuning \citep[][]{nikita2016package}, an R package that trains multiple models and calculates validation metrics. Despite the fact that \textit{ldatuning} does not employ embeddings but Latent Dirichlet allocation and that the data it modelled was preprocessed by removing stopwords and irrelevant text (numbers, unknown characters, URLs, Twitter handles), it indicated 300 as an optimal number of topics for the dataset, thus converging towards the BERTopic results. 

\begin{sidewaysfigure}[!ht]
    \centering
    \includegraphics[width=\linewidth]{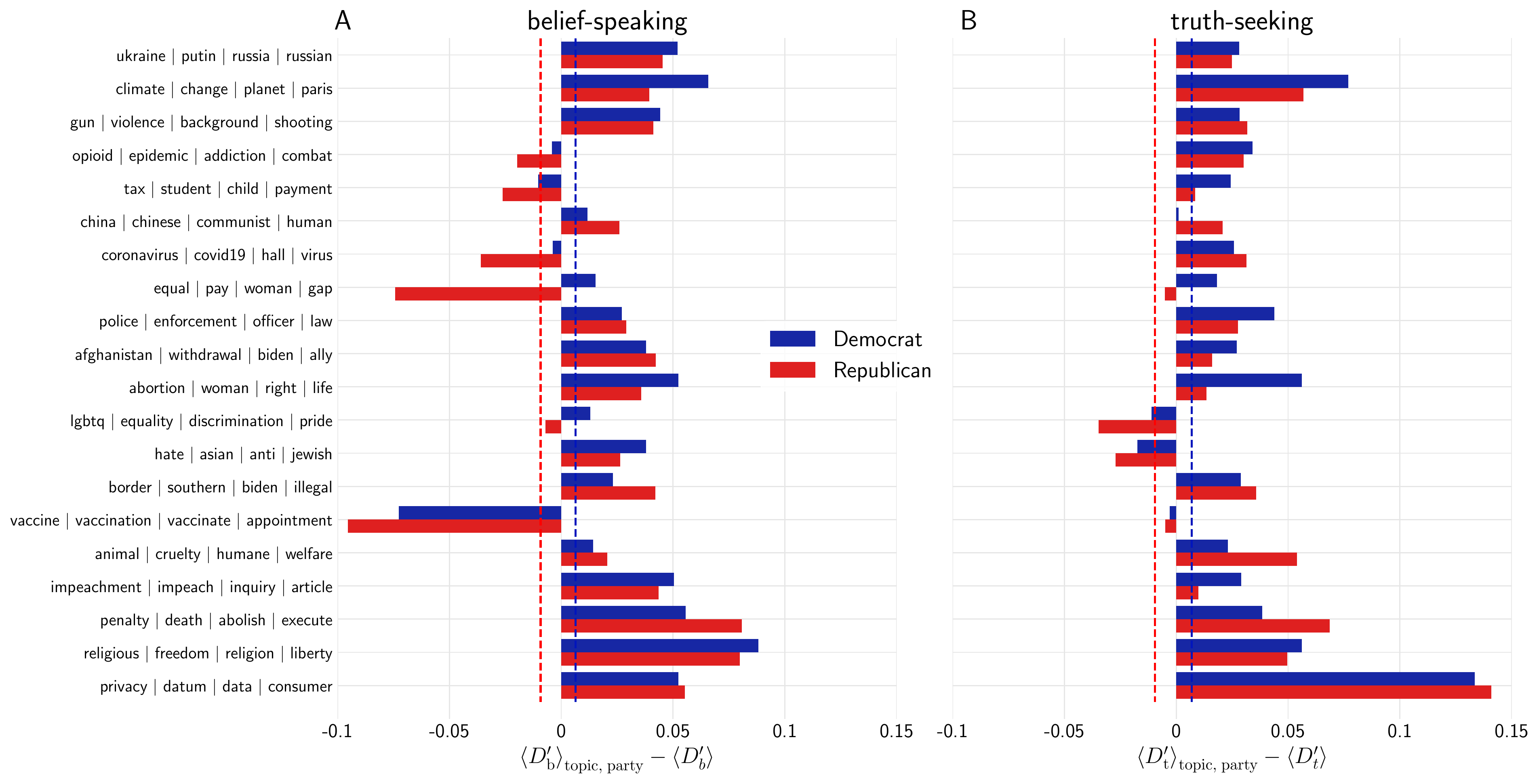}
    \caption{ 
    Panels \textbf{A} and \textbf{B} show the difference between within-topic within-party belief-speaking and truth-seeking similarity $\left<D'_\mathrm{b}\right>_\mathrm{topic,\;party}$ and $\left<D'_\mathrm{t}\right>_\mathrm{topic,\;party}$ and mean corpus belief-speaking and truth-seeking similarity $\left<D'_\mathrm{b}\right>$ and $\left<D'_\mathrm{t}\right>$, respectively for a range of hand-picked controversial topics. Values for Democrats and Republicans correspond to blue and red bars, respectively. Dashed lines indicate the mean belief-speaking and truth-seeking similarity for each party in the full corpus $\left<D'_\mathrm{b}\right>_\mathrm{party} - \left<D'_\mathrm{b}\right>$ and $\left<D'_\mathrm{t}\right>_\mathrm{party} - \left<D'_\mathrm{t}\right>$, respectively. 
    \\ \rule{\linewidth}{1pt}}
    \label{fig:SI_fig9}
\end{sidewaysfigure}

Building on the topic modelling, we investigated the difference between belief-speaking and truth-seeking in communication about controversial topics in U.S. politics, such as foreign policy, climate change, or the death penalty, and how this differs by party. The selection of controversial topics presented here is inspired by other research in the same area, e.g.~\cite{cinelli2021echo} and current research topics of non-partisan think-tanks, e.g.~\cite{pewresearch}. By default, BERTopic assigns each document to a single topic. Therefore, we used this information to calculate how particular controversial topics were distributed across parties and components, as shown in Figure~\ref{fig:SI_fig9}. 
To do this, we grouped the tweets by the topic they were assigned to as well as by the party the politician that created them was affiliated with. We then averaged their belief-speaking and truth-seeking similarity scores to calculate $\left<D'_\mathrm{b}\right>_\mathrm{topic,\;party}$ and $\left<D'_\mathrm{t}\right>_\mathrm{topic,\;party}$, respectively. We repeated this procedure for all 20 topics of interest. We also calculated the average belief-speaking similarity score $\left<D'_\mathrm{b}\right>$ and truth-seeking similarity score $\left<D'_\mathrm{t}\right>$ for all 363 topics found by BERTopic. Finally, we subtracted the specific component averages of a topic \textit{t} from the full corpus component averages to highlight how parties differ in honesty-speech when talking about controversial matters. 

In Figure~\ref{fig:SI_fig9} \textbf{A} and \textbf{B} we show the average belief-speaking and truth-seeking similarity within a given topic $\left<D'_\mathrm{b}\right>_\mathrm{topic,\;party}$ and $\left<D'_\mathrm{t}\right>_\mathrm{topic,\;party}$, minus the average belief-speaking and truth-seeking similarity calculated over the full corpus $\left<D'_\mathrm{b}\right>$ and $\left<D'_\mathrm{t}\right>$ for members of the Democratic and Republican parties, respectively. 
Each horizontal bar in the figure thus represents the deviation from the average score across the entire corpus. A value greater than zero implies that a topic involved more belief-speaking or truth-seeking than expected on average, and a value less than zero implies below-average invocation of belief-speaking or truth-seeking. It is immediately apparent that most of these controversial topics invoked more belief-speaking or truth-seeking than the average tweet, with only a few exceptions. For example, vaccine related discourse involved far less belief-speaking than any other topic for both parties. 

There is, however, also considerable heterogeneity in the amount of belief-speaking and truth-seeking used between the topics: Topics such as impeachment, religious freedom and Putin\,/\,Ukraine show a large amount of belief-speaking in both parties, whereas topics such as vaccines show little. Similarly, for truth-seeking the topics climate change, impeachment and religious freedom show a large share of this honesty component for both parties whereas the LGBTQ topic shows little. 

There are also marked differences in the balance of belief-speaking and truth-seeking within a topic and between the parties. The topics of climate change, gun violence, COVID-19 and the gender pay gap have the largest difference in belief-speaking, with tweets by Democrats containing more belief-speaking than those by Republicans. The topics of climate change, police, Afghanistan and abortion have the largest difference in truth-seeking with tweets by Democrats containing more truth-seeking while for the topic of animal cruelty, tweets by Republicans contain more truth-seeking.

\clearpage
\section{Validation using an independently compiled list of unreliable news sources}
To exclude a dependence of the main results reported in Section ``Relation of honesty components to information trustworthiness'' on use of the NewsGuard data base, we validated this analysis with an independently collected list of news outlet reliability from academic and fact-checking sources. Details on how this list was compiled are reported in Section ``Independent list of untrustworthy sources'' below. Using this list, we can assign an accuracy score $S_\mathrm{a}$ ranging from 1 to 5 as well as a transparency score $S_\mathrm{t}$, ranging from 1 to 3 to each domain. In addition, a domain with an accuracy score of $\leq 2$ and/or a transparency score of 1 will be labelled as ``unreliable''. Similar to the analysis above, we analyse the dependency of the accuracy score $S_\mathrm{a}'$ rescaled to [0; 1] and the transparency score $S_\mathrm{t}'$ rescaled to [0; 1] on the centered and length-corrected belief-speaking and truth-seeking similarity measured in tweet texts $D_\mathrm{b}'$ and $D_\mathrm{t}'$, respectively. We fit a linear mixed effects model with party as fixed variable and random slopes and intercepts for every Congress Member for each of the two scores:
\begin{align}
    S_\mathrm{a}' &\sim 1 + D_\mathrm{b}' \times D_\mathrm{t}' + D_\mathrm{b}' \times D_\mathrm{t}' \times P + (1 + D_\mathrm{b}' \times D_\mathrm{t}'\;\vert\; \mathrm{user ID})\label{eq:regression_accuracyScore_tweets} \\
    S_\mathrm{t}' &\sim 1 + D_\mathrm{b}' \times D_\mathrm{t}' + D_\mathrm{b}' \times D_\mathrm{t}' \times P + (1 + D_\mathrm{b}' \times D_\mathrm{t}'\;\vert\; \mathrm{user ID})\label{eq:regression_transparencyScore_tweets}
\end{align}

Again, we found a significant positive fixed effect of $D_\mathrm{t}'$ (coefficient 0.097 [0.079; 0.115], $p<0.001$, $t=10.7$) and accuracy $S_\mathrm{a}'$ as well as for party $P=\mathrm{Republican}$ (coefficient -0.071 [-0.079; -0.064], $p<0.001$, $t=-18.8$). We reproduce the negative effect of the interaction term between $D_\mathrm{b}'$ and Republican (coefficient -0.059 [-0.085; -0.033], $p<0.001$, $t=-4.4$), the interaction term between $D_\mathrm{t}'$ and Republican (coefficient 0.040 [0.013; 0.067], $p<0.001$, $t=2.9$), and the three-way interaction between $D'_\mathrm{b}$, $D_\mathrm{t}'$ and Republican (coefficient -0.196 [-0.249; -0.144], $p<0.001$, $t=-7.3$). 

Different from the main analysis, we also find a significant negative effect for $D_\mathrm{b}'$ (coefficient -0.120 [-0.137; -0.103], $p<0.001$, $t=-13.6$). 

We see the same pattern for the transparency score $S_\mathrm{t}'$, where we see a significant negative relation with $D_\mathrm{b}'$ and a significant positive relation with $D_\mathrm{t}'$ for both parties, as well as a significant effect of party, the interaction terms party $\times D_\mathrm{b}'$ and party $\times D_\mathrm{t}'$, and three-way interaction $D_\mathrm{b}' \times D_\mathrm{t}' \times \mathrm{party}$.

Full regression statistics are reported in Tables~\ref{tab:SI_tab3} and~\ref{tab:SI_tab4}. We note that there is extensive agreement between the trustworthiness labels in the NewsGuard data base and the alternative data base: An account that is labelled ``untrustworthy'' in the NewsGuard data base has a high chance of being labelled ``unreliable'' in the alternative database as well (Krippendorff's $\alpha$ of 0.84). This is also shown in a recent preprint~\cite{Lin2022} that compares both data bases.

\begin{table}[]
    \caption{Results of a linear mixed effects model for the dependence of the rescaled accuracy score of each link $S_\mathrm{a}'$ on belief-speaking similarity $D_\mathrm{b}'$ and truth-seeking similarity $D_\mathrm{t}'$ in tweet texts, with party $P$ as fixed variable following Eq.(\ref{eq:regression_accuracyScore_tweets}). The table reports results for the fixed effects. 442,500 observations were included. Regression was performed with the function \texttt{lmer} from the R library lme4~\cite{lme4}.}
    \footnotesize
    \centering
    \begin{tabular}{l|c|c|c|c|c|c}
                                                     & coef.   & std. err. & $t$     & $P>\vert t \vert$ & [0.025  & 0.975] \\
        \toprule
		Intercept                                & 0.8148  & 0.0026    & 308.592 & $<10^{-16}$       & 0.8097  & 0.8200 \\ 
		$D_\mathrm{b}'$                          & -0.1198 & 0.0088    & -13.577 & $<10^{-16}$       & -0.1371 & -0.1025 \\
		$D_\mathrm{t}'$                          & 0.0967  & 0.0090    & 10.692  & $<10^{-16}$       & 0.0790  & 0.1145 \\ 
		Republican                               & -0.0711 & 0.0038    & -18.784 & $<10^{-16}$       & -0.0785 & -0.0637 \\
		$D_\mathrm{b}'$ $\times$ $D_\mathrm{t}'$ & -0.0293 & 0.0176    & -1.662  & 0.0972            & -0.0638 & 0.0052 \\
		$D_\mathrm{b}'$ $\times$ Republican      & -0.0590 & 0.0134    & -4.411  & $1.2\cdot 10^{-5}$& -0.0852 & -0.0328 \\ 
		$D_\mathrm{t}'$ $\times$ Republican      & 0.0397  & 0.0138    & 2.885   & 0.0041             & 0.0127  & 0.0667 \\
		$D_\mathrm{b}'$ $\times$ $D_\mathrm{t}'$ $\times$ Republican & -0.1964 & 0.0268 & -7.340 & $6.9\cdot 10^{-13}$ & -0.2489 & -0.1440 \\ 
		\bottomrule 
		\multicolumn{2}{l}{Observations} & \multicolumn{1}{r}{442500} &
		\multicolumn{2}{l}{AIC} & \multicolumn{2}{r}{-399392} \\
		\multicolumn{2}{l}{Marginal R$^2$} & \multicolumn{1}{r}{0.049} &
		\multicolumn{2}{l}{log-Likelihood} & \multicolumn{2}{r}{199715} \\
	    \multicolumn{2}{l}{Conditional R$^2$} & \multicolumn{1}{r}{0.179} &  
	    \multicolumn{2}{l}{BIC} & \multicolumn{2}{r}{-399184} \\ 
        \bottomrule
    \end{tabular}

    \label{tab:SI_tab3}
\end{table}

\begin{table}[]
    \caption{Results of a linear mixed effects model for the dependence of the rescaled transparency score of each link $S_\mathrm{t}'$ on the belief-speaking similarity $D_\mathrm{b}'$ and truth-seeking similarity $D_\mathrm{t}'$ in tweet texts, with party $P$ as fixed variable following Eq.(\ref{eq:regression_transparencyScore_tweets}). The table reports results for the fixed effects. 442,500 observations were included. Regression was performed with the function \texttt{lmer} from the R library lme4~\cite{lme4}.}
    \footnotesize
    \centering
    
    \begin{tabular}{l|c|c|c|c|c|c}
                                                     & coef.   & std. err. & $t$     & $P>\vert t \vert$ & [0.025  & 0.975] \\
        \toprule
        Intercept                                & 0.9585  & 0.0025    & 380.958 & $<10^{-16}$          & 0.9536  & 0.9634 \\ 
        $D_\mathrm{b}'$                          & -0.0631 & 0.0081    & -7.804  & $2.9\cdot 10^{-14}$  & -0.0789 & -0.0473 \\
        $D_\mathrm{t}'$                          & 0.0646  & 0.0084    & 7.638   & $9.5 \cdot 10^{-14}$ & 0.0481  & 0.0811 \\ 
        Republican                               & -0.0944 & 0.0036    & -26.178 & $<10^{-16}$          & -0.1015 & -0.0874 \\
        $D_\mathrm{b}'$ $\times$ $D_\mathrm{t}'$ & -0.0382 & 0.0165    & -2.339  & 0.0207               & -0.0705 & -0.0059 \\
        $D_\mathrm{b}'$ $\times$ Republican      & -0.0859 & 0.0123    & -6.982  & $7.5\cdot 10^{-12}$  & -0.1100 & -0.0618 \\ 
        $D_\mathrm{t}'$ $\times$ Republican      & 0.0458  & 0.0129    & 3.565   & 0.0004               & 0.0206  & 0.0711 \\
        $D_\mathrm{b}'$ $\times$ $D_\mathrm{t}'$ $\times$ Republican & -0.1901   & 0.0252  & -7.560 & $1.4\cdot 10^{-13}$ & -0.2394 & -0.1408 \\ 
        \bottomrule 
        \multicolumn{2}{l}{Observations} & \multicolumn{1}{r}{442500} &
        \multicolumn{2}{l}{AIC} & \multicolumn{2}{r}{-400759} \\
        \multicolumn{2}{l}{Marginal R$^2$} & \multicolumn{1}{r}{0.080} &
        \multicolumn{2}{l}{log-Likelihood} & \multicolumn{2}{r}{200399} \\
        \multicolumn{2}{l}{Conditional R$^2$} & \multicolumn{1}{r}{0.196} &  
        \multicolumn{2}{l}{BIC} & \multicolumn{2}{r}{-400551} \\ 
        \bottomrule
    \end{tabular}
    \label{tab:SI_tab4}
\end{table}

\clearpage
\section{Independent list of untrustworthy sources}\label{subsec:independent_list}
We compiled a list of trustworthiness ratings from a range of academic sources and fact-checking sites. Most of these sources were also used by~\cite{gallotti2020assessing}. The list includes Bufale~\cite{bufale}, Bufalopedia~\cite{bufalopedia}, Butac~\cite{butac}, Buzzfeed News~\cite{buzzfeednews}, Columbia Journalism Review~\cite{columbiajournalism}, Fake News Watch~\cite{fakenewswatch},
Media Bias Fact Check~\cite{mediabiasfactcheck}, Politifact~\cite{politifact}, and Melissa Zimdars~\cite{zimdars}. After removing duplicates, our list contained 4,767 domains,  1,677 of which were also contained in the NewsGuard data base, as of March 1, 2022. 

The main challenge in combining lists from different fact checkers lies in unifying the labels the fact checkers assign to the domains. To address this, we devised a scheme where we rated each domain on two dimensions that we consider to be important to assess reliability and trustworthiness of information: ``accuracy'' and ``transparency''. 
We devise an accuracy score $S_\mathrm{a}$ that varies from 1 (false information) to 5 (scientific) and a transparency score $S_\mathrm{t}$ that varies from 1 (no transparency) to 3 (transparent). We provide a more detailed description of the five accuracy and three transparency levels in 
Tables~\ref{tab:SI_tab5} and \ref{tab:SI_tab6}. Mappings of the labels of individual fact checking sites to accuracy and transparency scores as well as the full list of domains are provided at~\cite{lasser2022misinformation}.

\begin{table}[!ht]
    \centering
       \caption{Description of accuracy scores.}
    \begin{tabular}{p{1cm}|p{2.75cm}|p{6cm}}
         Score & Label & Description \\
         \toprule
         1 & False Information & No or very little accuracy (e.g. fake news, conspiracy, satire) \\ 
         2 & Clickbait & Might contain smatterings of facts but is mostly misleading or clickbait \\ 
         3 & Biased & Mixed accuracy, half-truths, left/right bias \\ 
         4 & Mainstream & Low bias, mainstream media \\ 
         5 & Scientific & No reporting bias, scientific information \\ 
         \bottomrule
    \end{tabular}
 
    \label{tab:SI_tab5}
\end{table}

\begin{table}[!ht]
    \centering
        \caption{Description of transparency scores.}
    \begin{tabular}{p{1cm}|p{2.75cm}|p{6cm}}
         Score & Label & Description \\
         \toprule
         1 & No Transparency & Intentionally misleading or no information about editorial process (e.g. fake news, conspiracy) \\ 
         2 & Mixed Transparency & Sites with (partially) transparent intention, but can still be misunderstood because of the way articles are written (e.g. bias, clickbait, satire) \\ 
         3 & Transparent & Sites with a transparent editorial process and legal notice (e.g. mainstream, scientific news) \\ 
         \bottomrule
    \end{tabular}

    \label{tab:SI_tab6}
\end{table}

\clearpage
After mapping all individual lists to the accuracy and transparency dimensions, we label every domain that has an accuracy score of 1 (False Information) or 2 (Clickbait) and/or a transparency score of 1 (No Transparency) as ``unreliable''. This results in a total of 2,170 domains being labelled as ``unreliable'' and 2,597 as ``reliable''. For the 1,677 domains that are contained in both data bases, the Krippendorff's $\alpha$ between ``untrustworthy'' (score $< 60$ in NewsGuard) and ``unreliable'' in the independently compiled data base is 0.84, which shows a very high agreement between the two databases. The independently compiled domain list including the unified labels is openly accessible at \url{https://doi.org/10.5281/zenodo.6536692}.

After excluding links to other social media platforms (e.g.,  twitter.com, facebook.com, youtube.com, and instagram.com) as well as links to search services (google.com, yahoo.com), the database covers a very similar share of links as the NewsGuard data base (between 20\% and 60\%) --- see also Extended Data Figure~3 \textbf{B} in the main article. 

\clearpage
\section{Honesty components by state}
\begin{figure}[h]
    \centering
    \includegraphics[width=0.91\textwidth]{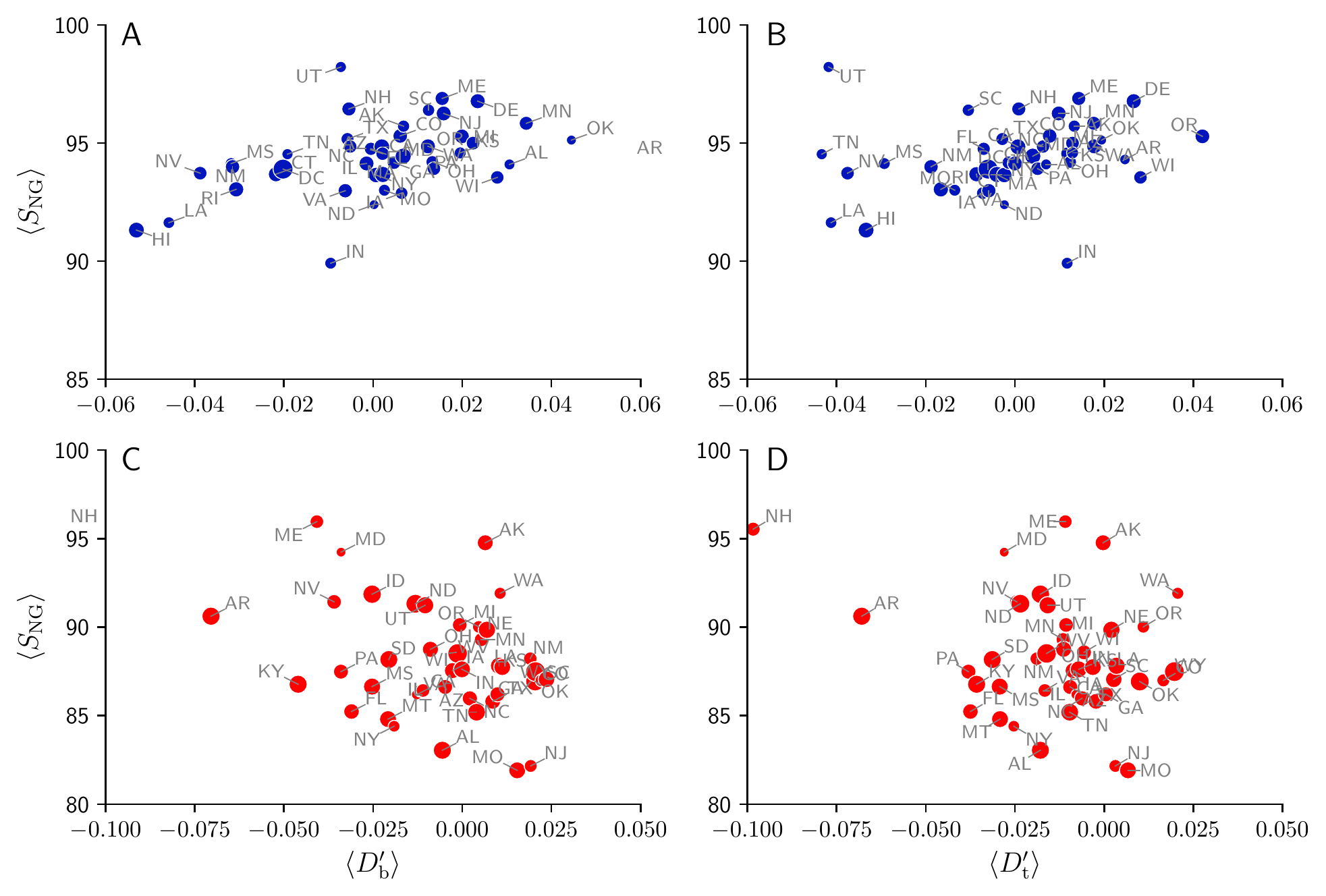}
    \caption{Honesty components by state. Panels \textbf{A}, \textbf{C} show the NewsGuard score $S_\mathrm{NG}$ over belief-speaking similarity $D'_\mathrm{b}$ averaged by state for Democratic and Republican members of Congress, respectively. Panels \textbf{B}, \textbf{D} show  $S_\mathrm{NG}$ over $D'_\mathrm{t}$ averaged by state for Democratic and Republican members of Congress, respectively. Marker sizes are scaled with the percentage of votes for Biden in the 2020 presidential election for the panels showing Democratic Congress members, and with the percentage of votes for Trump in the panels showing Republican Congress members. Note that the axes are scaled separately for each panel to reduce visual density of the point cloud.}
    \label{fig:SI_fig10}
\end{figure}

To examine geographical heterogeneity, we averaged NewsGuard scores across representatives and senators within each state, broken down by party. The results are shown in Figure~\ref{fig:SI_fig10}, plotting each state's NewsGuard score against average belief-speaking similarity (left panels) and truth-seeking similarity (right panels), respectively. The size of plotting symbols additionally represents the vote share for Trump (in the bottom panels) and for Biden (top panels) during the 2020 presidential election. It can be seen that quality of information being shared by Republicans tends to be lower in southern states (e.g., AL, TN, TX, OK, KY) than in the north (e.g., NH, AK, ME). For democrats, no clearly discernible pattern emerges. 

We also considered the outcome of the 2020 presidential election and compared the states that were called for Trump and Biden, respectively.
In states that were called for Biden, Democrat members of Congress on average have a NewsGuard score of 94.5 whereas Republicans have 88.6. In states that were called for Trump, the NewsGuard scores were 94.2 (Democrats) and 87.7 (Republicans), respectively. These differences were small, suggesting that the electoral pattern in their home states did not affect the quality of information shared by members of Congress.

\clearpage
\section{Mediation analysis}
Why is it the case that belief speaking is the preferred means to spread low-quality information? One possibility is that belief-speaking is the result of Republican politicians' desire to disparage Democrats, as suggested by ~\cite{Osmundsen2021-gn}, given that belief speaking was found to be associated with greater negative sentiment (see Figure~\ref{fig:SI_fig7}), and given that lower-quality information tends to be biased towards negativity~\cite{Soroka19}. According to this theory, the relationship between belief-speaking and low-quality shared information should be mediated by negative sentiment. On the other hand, if belief speaking were involved in the dissemination of poor quality content for other reasons, it should mediate the association involving negative sentiment.

To test these opposing predictions, we examined separately for Democrats and Republicans whether (1) negative sentiment mediated the effects of belief speaking on sharing low-quality information, or (2) belief speaking mediated the effects of negative sentiment on sharing low-quality information. For each user, we computed mean scores of negative sentiment (measured via VADER, see Section ``VADER text analysis''), belief speaking similarity, and prevalence of sharing low-quality news (average NewsGuard score of the shared articles).  We conducted a causal mediation analysis using the `mediation' R package~\cite{tingley2014mediation} and a bootstrap method with 10,000 iterations.  

Among Republicans, when considering negative sentiment as a mediator, 
the effect of the direct path was not statistically significant 
(mean direct effect = $-9.56$, 95\% CI of bootstrapped samples = [$-21.35, 1.10$], $p = .077$). The mediation, however,  was significant (average causal mediation effect = $-19.48$, 95\% CI = [$-26.92, -13.43$], $p < .001$), accounting for 67\% of the total effect. When considering belief speaking as a mediator, the opposite pattern emerged: the direct effect was statistically significant (mean direct effect = $-125.87$, 95\% CI = [$-159.83, -93.33$], $p = .001$), 
but the average causal mediation effect did not reach statistical significance, 
(average causal mediation effect = $-11.15$, 95\% CI = [$-24.08, 0.83$], $p = .067$). 
See Tables~\ref{tab:SI_tab10} and~\ref{tab:SI_tab11} for the full details. These results align with the findings of ~\cite{Osmundsen2021-gn}, suggesting that the relationship between belief-speaking and low-quality shared information is indeed driven by negative sentiment.

\begin{table}[]
\centering
 \caption{Mediation analysis with belief speaking similarity as mediator. ACME = average causal mediation effect; ADE = average direct effect. 519 observations were included for Republicans and 525 for Democrats. Mediation analysis was performed using the function \texttt{mediate} from the R package mediation, version 4.5.0.}
\begin{tabular}{@{}llllll@{}}
\toprule
                             & Parameter      & Estimate & \textit{P} & [0.025   & 0.975]   \\ \midrule
\multirow{4}{*}{Republicans} & ACME           & -11.146  & 0.067      & -24.084  & 0.830    \\
                             & ADE            & -125.870 & \textless{}.001          & -159.827 & -93.327  \\
                             & Total Effect   & -137.016 & \textless{}.001          & -168.150 & -108.366 \\
                             & Prop. Mediated & 0.0813   & 0.067      & -0.006   & 0.185    \\
                             &                &          &            &          &          \\
\multirow{4}{*}{Democrats}   & ACME           & 3.985    & 0.074      & -0.387   & 9.366    \\
                             & ADE            & 6.045    & 0.287      & -5.020   & 17.609   \\
                             & Total Effect   & 10.030   & 0.047      & 0.123    & 20.331   \\
                             & Prop. Mediated & 0.397    & 0.119      & -0.236   & 2.619    \\ \cmidrule(l){2-6} 
\end{tabular}
\label{tab:SI_tab10}
\end{table}

\begin{table}[]
\centering
            \caption{Mediation analysis with negative sentiment as mediator. ACME = average causal mediation effect; ADE = average direct effect. 519 observations were included for Republicans and 525 for Democrats. Mediation was performed using the function \texttt{mediate} from the R package mediation, version 4.5.0.}
\begin{tabular}{@{}llllll@{}}
\toprule
                             & Parameter      & Estimate & \textit{P} & [0.025  & 0.975]  \\ \midrule
\multirow{4}{*}{Republicans} & ACME           & -19.480  & \textless{}.001      & -26.916 & -13.433 \\
                             & ADE            & -9.558   & 0.077      & -21.352 & 1.097   \\
                             & Total Effect   & -29.038  & \textless{}.001      & -42.276 & -18.018 \\
                             & Prop. Mediated & 0.671    & \textless{}.001      & 0.442   & 1.055   \\
                             &                &          &            &         &         \\
\multirow{4}{*}{Democrats}   & ACME           & 0.698    & 0.290      & -0.655  & 2.462   \\
                             & ADE            & 3.518    & 0.0732     & -0.297  & 8.682   \\
                             & Total Effect   & 4.216    & 0.007      & 0.924   & 8.990   \\
                             & Prop. Mediated & 0.166    & 0.296      & -0.162  & 1.158   \\ \cmidrule(l){2-6} 
\end{tabular}
        \label{tab:SI_tab11}
\end{table}

\clearpage
\section{Robustness analysis using only a restricted number of tweets per account}
The number of tweets posted by an individual account varies widely: while the median number of tweets posted by an account is 2876, the mean is 4278, with the most prolific account posting 52,055 tweets and 10\% of the accounts posting 9800 tweets or more in the observed time span (November 6, 2010 to December 31, 2022). 

To assess whether our results are driven by accounts that post a large number of tweets, we repeat our main analysis analysis reported in Figure 3, including only the latest 3200 tweets from every account. The results of fitting the linear mixed effects model following Eq.~(1) in Table~\ref{tab:SI_tab7} show only minute deviations from the results presented in the main text where we used all tweets to fit the model (see Extended Data Table 2).

\begin{table}[!ht]
 \caption{Results of a linear mixed effects model for the dependence of the rescaled NewsGuard score of each link $S_\mathrm{NG}'$ on belief-speaking similarity $D_\mathrm{b}'$ and truth-seeking similarity $D_\mathrm{t}'$ in tweets, with party $P$ as fixed variable following Eq.(1). The table reports results for the fixed effects. Observations were restricted to the latest 3200 tweets for every accounts. A total of 247,947 observations were included. Regression was performed with the function \texttt{lmer} from the R library lme4~\cite{lme4}.}
    \footnotesize
    \centering
    \begin{tabular}{l|c|c|c|c|c|c}
                                                & coef.   & std. err. & $t$     & $P>\vert t \vert$   & [0.025  & 0.975] \\
        \toprule
		Intercept                               & 0.9443  & 0.0016    & 584.076 & $<10^{-16}$         & 0.9411  & 0.9475 \\ 
		$D_\mathrm{b}'$                          & -0.0002 & 0.0064    & 0.030   & 0.9763              & -0.0124 & 0.0128 \\
		$D_\mathrm{t}'$                          & 0.0165  & 0.0064    & 2.582   & 0.0101              & 0.0040  & 0.0290 \\ 
		Republican                              & -0.0647 & 0.0024    & -27.293 & $<10^{-16}$         & -0.0694 & -0.0601 \\
		$D_\mathrm{b}'$ $\times$ $D_\mathrm{t}'$  & 0.0062  & 0.0139    & 0.447   & 0.6548              & -0.0211 & 0.0335 \\
		$D_\mathrm{b}'$ $\times$ Republican      & -0.1372 & 0.0098    & -13.966 & $<10^{-16}$         & -0.1564 & -0.1179 \\ 
		$D_\mathrm{t}'$ $\times$ Republican      & 0.0794  & 0.0098    & 8.114   & $2.2\cdot 10^{-15}$ & 0.0602  & 0.0986 \\
		$D_\mathrm{b}'$ $\times$ $D_\mathrm{t}'$ $\times$ Republican & -0.1732 & 0.0200 & -8.664 & $<10^{-16}$ & -0.2124 & -0.1340 \\ 
		\bottomrule 
		\multicolumn{2}{l}{Observations} & \multicolumn{1}{r}{247947} &
		\multicolumn{2}{l}{AIC} & \multicolumn{2}{r}{-388518} \\
		\multicolumn{2}{l}{Marginal R$^2$} & \multicolumn{1}{r}{0.081} &
		\multicolumn{2}{l}{log-Likelihood} & \multicolumn{2}{r}{194278} \\
	    \multicolumn{2}{l}{Conditional R$^2$} & \multicolumn{1}{r}{0.174} &  
	    \multicolumn{2}{l}{BIC} & \multicolumn{2}{r}{-388320} \\ 
        \bottomrule
    \end{tabular}
   
    \label{tab:SI_tab7}
\end{table}

\clearpage
\section{Increase of belief-speaking and truth-seeking similarity by account}
To investigate the overall increase of both belief-speaking and truth-seeking reported in Fig.~2 in the main text, we investigated which politicians contributed most to the overall increase in both honesty components. We show the top 10 accounts with the largest change in belief-speaking and truth-seeking similarity between the 2010–2013 and the 2019–2022 period for both Democrats and Republicans in Tables~\ref{tab:SI_tab12} and~\ref{tab:SI_tab13}. 

\begin{table}[!ht]
    \caption{Twitter accounts of Democratic and Republican representatives with the highest increase in average belief-speaking similarity $\left<D'_\mathrm{b}\right>_\mathrm{acc}$ between the period 2011–2013 and 2019–2022.}
    \centering
    \begin{tabular}{l|c|c|c}
         account handle & $\left<D'_\mathrm{b}\right>_\mathrm{acc}$ 2010–2013 & $\left<D'_\mathrm{b}\right>_\mathrm{acc}$ 2019–2022 & difference\\
         \toprule
         \multicolumn{4}{c}{Democrats}\\
         \midrule
         SenatorLujan    & -0.43 & 0.01 & 0.44\\
         SenStabenow     & -0.33 & 0.04 & 0.36\\
         SenBooker       & 0.24  & 0.04 & 0.28\\
         aguilarpete     & -0.01 & 0.21 & 0.22\\
         WilliamKeating  & -0.17 & 0.03 & 0.21\\
         USRepKeating    & -0.16 & 0.04 & 0.20\\
         pallonefornj    & -0.14 & 0.05 & 0.19\\
         BobbyScott4VA3  & -0.28 & -0.10 & 0.18\\
         TulsiPress      & -0.15 & 0.03 & 0.17\\
         Matsui4Congress & -0.16 & 0.01 & 0.17\\
         \midrule
         \multicolumn{4}{c}{Republicans}\\
         \midrule
         SenBobCorker    & -0.14 & 0.16 & 0.31\\
         GrassleyPress   & -0.28 & 0.00 & 0.29\\
         McCaulforTexas  & -0.25 & 0.02 & 0.26\\
         krhern          & -0.15 & 0.09 & 0.24\\
         votetimscott    & -0.12 & 0.09 & 0.21\\
         MaElviraSalazar & -0.51 & -0.31 & 0.20\\
         congbillposey   & -0.31 & -0.12 & 0.19\\
         MacTXPress      & -0.12 & 0.07 & 0.19\\
         MikeKellyforPA  & -0.13 & 0.06 & 0.19\\
         JohnKennedyLA   & -0.12 & 0.07 & 0.19\\
         \bottomrule         
    \end{tabular}
    \label{tab:SI_tab12}
\end{table}

\begin{table}[!ht]
    \caption{Twitter accounts of Democratic and Republican representatives with the highest increase in average truth-seeking similarity $\left<D'_\mathrm{t}\right>_\mathrm{acc}$ between the period 2011–2013 and 2019–2022.}
    \centering
    \begin{tabular}{l|c|c|c}
         account handle & $\left<D'_\mathrm{t}\right>_\mathrm{acc}$ 2010–2013 & $\left<D'_\mathrm{t}\right>_\mathrm{acc}$ 2019–2022 & difference\\
         \toprule
         \multicolumn{4}{c}{Democrats}\\
         \midrule
         SenatorLujan    & -0.33 & 0.02 & 0.35\\
         SenStabenow     & -0.25 & 0.04 & 0.29\\
         WilliamKeating  & -0.19 & 0.01 & 0.20\\
         TulsiPress      & -0.14 & 0.05 & 0.19\\
         DeGette5280     & -0.15 & 0.03 & 0.18\\
         USRepKeating    & -0.14 & 0.04 & 0.18\\
         Matsui4Congress & -0.18 & -0.02 & 0.16\\
         SenBooker       & -0.12 & 0.04 & 0.16\\
         RepJoseSerrano  & -0.21 & -0.06 & 0.15\\
         BobbyScott4VA3  & -0.25 & -0.11 & 0.15\\
         \midrule
         \multicolumn{4}{c}{Republicans}\\
         \midrule
         McCaulforTexas  & -0.25 & 0.00 & 0.26\\
         SenBobCorker    & -0.11 & 0.12 & 0.23\\
         GrassleyPress   & -0.17 & 0.04 & 0.21\\
         TeamCMR         & -0.14 & 0.05 & 0.19\\
         congbillposey   & -0.25 & -0.06 & 0.19\\
         cindyhydesmith  & -0.14 & 0.03 & 0.16\\
         stephaniebice   & -0.15 & 0.01 & 0.16\\
         votetimscott    & -0.13 & 0.03 & 0.16\\
         CurtisUT        & -0.13 & 0.03 & 0.16\\
         MikeKellyforPA  & -0.15 & 0.01 & 0.15\\
         \bottomrule         
    \end{tabular}
    \label{tab:SI_tab13}
\end{table}

\clearpage
\section{Robustness analysis using different embeddings}
In addition to GloVe~\cite{pennington2014glove} embeddings used for the results presented in the main text, we also calculated $D_\mathrm{b}'$ and $D_\mathrm{t}'$ using word2vec~\cite{mikolov2013efficient} and fasttext~\cite{bojanowski2016enriching} embeddings to exclude a dependence of our results on the choice of embedding. We note that both GloVe and fasttext were trained on the ``common crawl'' corpus, whereas word2vec was trained on Google news, a corpus with a more restricted scope. Results for the linear mixed effects modeling following Eq.~(1) using word2vec and fasttext embeddings are shown in Tables~\ref{tab:SI_tab8} and~\ref{tab:SI_tab9}, respectively. Results for both word2vec and fasttext are similar to the results using GloVe reported in Extended Data Table 2. 

\begin{table}[!ht]
 \caption{Results of a linear mixed effects model for the dependence of the rescaled NewsGuard score of each link $S_\mathrm{NG}'$ on belief-speaking similarity $D_\mathrm{b}'$ and truth-seeking similarity $D_\mathrm{t}'$ in tweets, with party $P$ as fixed variable following Eq.(1). In contrast to Tables~\ref{tab:SI_tab7} and Extended Data Tables 2 and 3 in the main text, the belief-speaking and truth-seeking similarities have been calculated using word2vec~\cite{mikolov2013efficient} embeddings. The table reports results for the fixed effects. A total of 504,809 observations were included. Regression was performed with the function \texttt{lmer} from the R library lme4~\cite{lme4}.}
    \footnotesize
    \centering
    \begin{tabular}{l|c|c|c|c|c|c}
                                                     & coef.   & std. err. & $t$     & $P>\vert t \vert$   & [0.025  & 0.975] \\
        \toprule
		Intercept                                & 0.9435  & 0.0016    & 592.953 & $<10^{-16}$         & 0.9403  & 0.9466 \\ 
		$D_\mathrm{b}'$                          & 0.0038  & 0.0088    & 0.430   & 0.6672              & -0.0135 & -0.0211 \\
		$D_\mathrm{t}'$                          & 0.0243  & 0.0087    & 2.804   & 0.0052              & 0.0073  & 0.0413 \\ 
		Republican                               & -0.0671 & 0.0023    & -29.463 & $<10^{-16}$         & -0.0716 & -0.0627 \\
		$D_\mathrm{b}'$ $\times$ $D_\mathrm{t}'$ & 0.0061  & 0.0158    & 0.383   & 0.7018              & -0.0250 & 0.0371 \\
		$D_\mathrm{b}'$ $\times$ Republican      & -0.2043 & 0.0131    & -15.590 & $<10^{-16}$         & -0.2300 & -0.1787 \\ 
		$D_\mathrm{t}'$ $\times$ Republican      & 0.1031 & 0.0130    & 7.911  & $1.2\cdot 10^{-14}$   & 0.0775  & 0.1286 \\
		$D_\mathrm{b}'$ $\times$ $D_\mathrm{t}'$ $\times$ Republican & -0.2765 & 0.0239 & -11.580 & $<10^{-16}$ & -0.3233 & -0.2297 \\ 
		\bottomrule 
		\multicolumn{2}{l}{Observations} & \multicolumn{1}{r}{504809} &
		\multicolumn{2}{l}{AIC} & \multicolumn{2}{r}{-801648} \\
		\multicolumn{2}{l}{Marginal R$^2$} & \multicolumn{1}{r}{0.087} &
		\multicolumn{2}{l}{log-Likelihood} & \multicolumn{2}{r}{400843} \\
	    \multicolumn{2}{l}{Conditional R$^2$} & \multicolumn{1}{r}{0.184} &  
	    \multicolumn{2}{l}{BIC} & \multicolumn{2}{r}{-801436} \\ 
        \bottomrule
    \end{tabular}
   
    \label{tab:SI_tab8}
\end{table}

\begin{table}[!ht]
 \caption{Results of a linear mixed effects model for the dependence of the rescaled NewsGuard score of each link $S_\mathrm{NG}'$ on belief-speaking similarity $D_\mathrm{b}'$ and truth-seeking similarity $D_\mathrm{t}'$ in tweets, with party $P$ as fixed variable following Eq.(1). In contrast to Tables~\ref{tab:SI_tab7} and Extended Data Tables 2 and 3 in the main text, the belief-speaking and truth-seeking similarities have been calculated using fasttext~\cite{bojanowski2016enriching} embeddings. The table reports results for the fixed effects. A total of 504,809 observations were included. Regression was performed with the function \texttt{lmer} from the R library lme4~\cite{lme4}.}
    \footnotesize
    \centering
    \begin{tabular}{l|c|c|c|c|c|c}
                                                     & coef.   & std. err. & $t$     & $P>\vert t \vert$   & [0.025  & 0.975] \\
        \toprule
		Intercept                                & 0.9438  & 0.0016    & 577.940 & $<10^{-16}$         & 0.9406  & 0.9470 \\ 
		$D_\mathrm{b}'$                          & 0.0093  & 0.0081    & 1.151   & 0.251               & -0.0065 & 0.0252 \\
		$D_\mathrm{t}'$                          & -0.0061 & 0.0076    & -0.793  & 0.427               & -0.0210  & 0.0089 \\ 
		Republican                               & -0.0716 & 0.0023    & -30.611 & $<10^{-16}$         & -0.0762 & -0.0670 \\
		$D_\mathrm{b}'$ $\times$ $D_\mathrm{t}'$ & 0.0452  & 0.0486    & 0.929   & 0.353               & -0.0501 & 0.1405 \\
		$D_\mathrm{b}'$ $\times$ Republican      & -0.1919 & 0.0122    & -15.736 & $<10^{-16}$         & -0.2158 & -0.1680 \\ 
		$D_\mathrm{t}'$ $\times$ Republican      & 0.0796  & 0.0116    & 6.836   & $2.0\cdot 10^{-11}$ & 0.0568  & 0.1024 \\
		$D_\mathrm{b}'$ $\times$ $D_\mathrm{t}'$ $\times$ Republican & -0.0730 & 0.0751 & -0.972 & 0.331 & -0.2201 & 0.0742 \\ 
		\bottomrule 
		\multicolumn{2}{l}{Observations} & \multicolumn{1}{r}{504809} &
		\multicolumn{2}{l}{AIC} & \multicolumn{2}{r}{-799568} \\
		\multicolumn{2}{l}{Marginal R$^2$} & \multicolumn{1}{r}{0.086} &
		\multicolumn{2}{l}{log-Likelihood} & \multicolumn{2}{r}{399803} \\
	    \multicolumn{2}{l}{Conditional R$^2$} & \multicolumn{1}{r}{0.182} &  
	    \multicolumn{2}{l}{BIC} & \multicolumn{2}{r}{-799358} \\ 
        \bottomrule
    \end{tabular}
    \label{tab:SI_tab9}
\end{table}

\clearpage
\section{Increase of belief-speaking and truth-seeking similarity by keyword}
\begin{figure}[!ht]
    \centering
    \includegraphics[width=\textwidth]{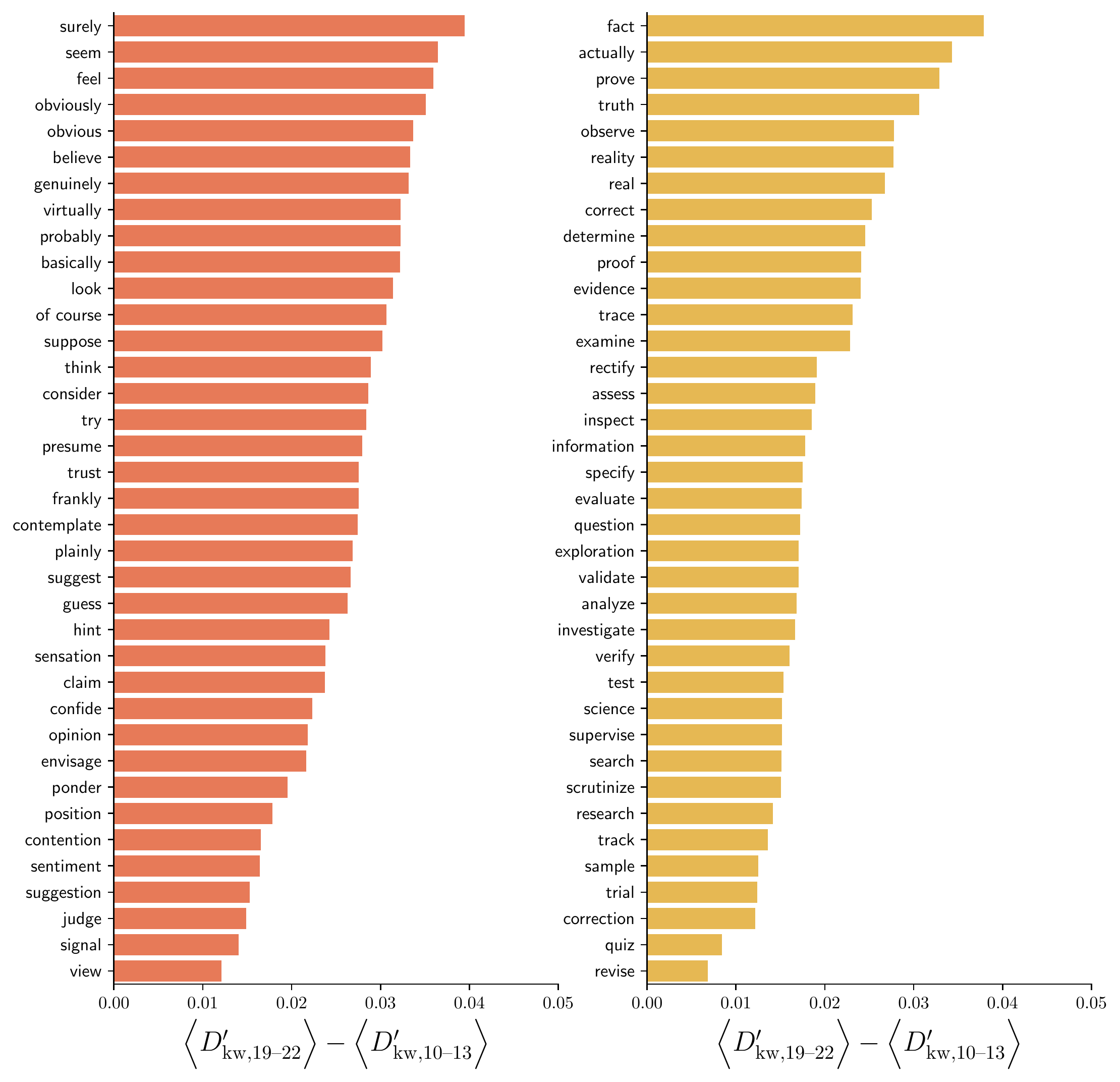}
    \caption{Increase of similarity score of tweets of individual keywords from the belief-speaking (left) and truth-seeking (right) dictionaries between the time-periods 2010--2013 and 2019--2022.}
    \label{fig:SI_fig11}
\end{figure}

To asses which keywords in the belief-speaking and truth-seeking dictionaries contributed most to the increase of overall belief-speaking and truth-seeking similarity, we created embeddings of single keywords to calculate the centered and length-corrected similarity $D'_\mathrm{kw}$ of tweets to a given keyword. For every keyword, we then calculated the mean similarity for tweets from the years 2010 to 2013 and for tweets from the years 2019 to 2022. We show the increase in similarity for every keyword in Figure~\ref{fig:SI_fig11}.

\clearpage
\bibliography{temp.bib}

\begin{thebibliography}{10}
\providecommand{\url}[1]{{#1}}
\providecommand{\urlprefix}{URL }
\providecommand{\doi}[1]{\url{https://doi.org/#1}}
\bibcommenthead

\bibitem{FreedomHouse20}
{Freedom~House}, Freedom in the world 2020. {A} leaderless struggle for
  democracy.
\newblock Tech. rep., {Freedom House} (2020)

\bibitem{FreedomHouse21}
{Freedom~House}, Freedom in the world 2021. {D}emocracy under siege.
\newblock Tech. rep., {Freedom House} (2021)

\bibitem{lewandowsky2020wilful}
S.~Lewandowsky, in \emph{Deliberate {I}gnorance: {C}hoosing {N}ot to {K}now},
  vol.~29 ({MIT} Press, 2020).
\newblock \doi{10.7551/mitpress/13757.003.0011}

\bibitem{Loomba21}
S.~Loomba, A.~de~Figueiredo, S.J. Piatek, K.~de~Graaf, H.J. Larson, Measuring
  the impact of {COVID}-19 vaccine misinformation on vaccination intent in the
  {UK} and {USA}.
\newblock Nature Human Behaviour \textbf{5} (2021).
\newblock \doi{10.1038/s41562-021-01056-1}

\bibitem{Cantarella20}
M.~Cantarella, N.~Fraccaroli, R.G. Volpe, Does fake news affect voting
  behaviour?
\newblock {SSRN} Electronic Journal  (2020).
\newblock \doi{10.2139/ssrn.3629666}

\bibitem{muller2021fanning}
K.~M{\"u}ller, C.~Schwarz, Fanning the flames of hate: Social media and hate
  crime.
\newblock {Journal of the European Economic Association} \textbf{19},
  2131--2167 (2021).
\newblock \doi{10.1093/jeea/jvaa045}

\bibitem{LorenzSpreen22}
P.~Lorenz-Spreen, L.~Oswald, S.~Lewandowsky, R.~Hertwig, A systematic review of
  worldwide causal and correlational evidence on digital media and democracy.
\newblock Nature Human Behaviour pp. 1--28 (2022).
\newblock \doi{10.1038/s41562-022-01460-1}

\bibitem{Lewandowsky17b}
S.~Lewandowsky, U.K.H. Ecker, J.~Cook, Beyond misinformation: {U}nderstanding
  and coping with the post-truth era.
\newblock Journal of Applied Research in Memory and Cognition \textbf{6},
  353--369 (2017).
\newblock \doi{10.1016/j.jarmac.2017.07.008}

\bibitem{Wood18}
T.~Wood, E.~Porter, {The Elusive Backfire Effect: Mass Attitudes' Steadfast
  Factual Adherence}.
\newblock Political Behavior \textbf{41}, 135--163 (2018).
\newblock \doi{10.1007/s11109-018-9443-y}

\bibitem{Swire17}
B.~Swire, A.J. Berinsky, S.~Lewandowsky, U.K.H. Ecker, Processing political
  misinformation: comprehending the {T}rump phenomenon.
\newblock Royal Society Open Science \textbf{4}, 160,802 (2017).
\newblock \doi{10.1098/rsos.160802}

\bibitem{Swire19}
B.~Swire-Thompson, U.K.H. Ecker, S.~Lewandowsky, A.J. Berinsky, {They Might Be
  a Liar But They're My Liar: Source Evaluation and the Prevalence of
  Misinformation}.
\newblock Political Psychology \textbf{41}, 21--34 (2020).
\newblock \doi{10.1111/pops.12586}

\bibitem{Hahl18}
O.~Hahl, M.~Kim, E.W.Z. Sivan, {The Authentic Appeal of the Lying Demagogue:
  Proclaiming the Deeper Truth about Political Illegitimacy}.
\newblock American Sociological Review \textbf{83}, 1--33 (2018).
\newblock \doi{10.1177/0003122417749632}

\bibitem{Lewandowsky17}
S.~Lewandowsky, E.A. Lloyd, S.~Brophy, When {THUNC}ing {T}rumps thinking: What
  distant alternative worlds can tell us about the real world.
\newblock Argumenta \textbf{3}, 217--231 (2018).
\newblock \doi{10.23811/52.arg2017.lew.llo.bro}

\bibitem{McCright17}
A.M. McCright, R.E. Dunlap, {Combatting Misinformation Requires Recognizing Its
  Types and the Factors That Facilitate Its Spread and Resonance}.
\newblock Journal of Applied Research in Memory and Cognition \textbf{6},
  389--396 (2017).
\newblock \doi{10.1016/j.jarmac.2017.09.005}

\bibitem{Cooper21}
B.~Cooper, T.R. Cohen, E.~Huppert, E.~Levine, W.~Fleeson, {Honesty in
  Organizations: A Systematic Review and New Conceptual Framework}.
\newblock Academy of Management Annals  (2021).
\newblock \doi{10.17605/OSF.IO/PCG7M}

\bibitem{Varshizky12}
A.~Varshizky, Alfred {R}osenberg: The {N}azi \emph{Weltanschauung} as modern
  gnosis.
\newblock Politics, Religion {\&} Ideology \textbf{13}, 311--331 (2012).
\newblock \doi{10.1080/21567689.2012.698977}

\bibitem{Voegelin00}
E.~Voegelin, in \emph{{The collected works of Eric Voegelin (Volume 5)}}
  (University of Missouri Press, 2000), pp. 21--73

\bibitem{vanZoonen12}
L.~{van~Zoonen}, I-{Pistemology}: {Changing} truth claims in popular and
  political culture.
\newblock European Journal of Communication \textbf{27}, 56--67 (2012).
\newblock \doi{10.1177/0267323112438808}

\bibitem{Edis20}
T.~Edis, A revolt against expertise: {P}seudoscience, right-wing populism, and
  post-truth politics.
\newblock Disputatio \textbf{9} (2020).
\newblock \doi{10.5281/zenodo.3567166}

\bibitem{Waisbord18}
S.~Waisbord, The elective affinity between post-truth communication and
  populist politics.
\newblock Communication Research and Practice \textbf{4}, 17--34 (2018).
\newblock \doi{10.1080/22041451.2018.1428928}

\bibitem{Farrell18a}
H.~Farrell, B.~Schneier, Common-knowledge attacks on democracy.
\newblock Tech. rep., Berkman Klein Center for Internet \& Society (2018)

\bibitem{Uscinski13}
J.E. Uscinski, R.W. Butler, The epistemology of fact checking.
\newblock Critical Review \textbf{25}, 162--180 (2013).
\newblock \doi{10.1080/08913811.2013.843872}

\bibitem{Amazeen15}
M.A. Amazeen, Revisiting the epistemology of fact-checking.
\newblock Critical Review \textbf{27}, 1--22 (2015).
\newblock \doi{10.1080/08913811.2014.993890}

\bibitem{Jacobson21}
G.C. Jacobson, Driven to {Extremes}: {Donald} {Trump}'s {Extraordinary}
  {Impact} on the 2020 {Elections}.
\newblock Presidential Studies Quarterly \textbf{51}, 492--521 (2021).
\newblock \doi{10.1111/psq.12724}

\bibitem{Williams+2002}
B.~Williams, \emph{Truth and Truthfulness: An Essay in Genealogy} (Princeton
  University Press, Princeton, 2002).
\newblock \doi{doi:10.1515/9781400825141}

\bibitem{Graham20}
M.H. Graham, M.W. Svolik, Democracy in {America}? {Partisanship},
  {Polarization}, and the {Robustness} of {Support} for {Democracy} in the
  {United} {States}.
\newblock American Political Science Review \textbf{114}, 392--409 (2020).
\newblock \doi{10.1017/S0003055420000052}

\bibitem{barbera2019leads}
P.~Barber{\'a}, A.~Casas, J.~Nagler, P.J. Egan, R.~Bonneau, J.T. Jost, J.A.
  Tucker, Who leads? who follows? measuring issue attention and agenda setting
  by legislators and the mass public using social media data.
\newblock {American Political Science Review} \textbf{113}, 883--901 (2019).
\newblock \doi{10.1017/S0003055419000352}

\bibitem{lewandowsky2020using}
S.~Lewandowsky, M.~Jetter, U.K. Ecker, Using the president’s tweets to
  understand political diversion in the age of social media.
\newblock Nature Communications \textbf{11}, 1--12 (2020).
\newblock \doi{10.1038/s41467-020-19644-6}

\bibitem{nelson2020computational}
L.K. Nelson, Computational grounded theory: {A} methodological framework.
\newblock Sociological Methods \& Research \textbf{49}, 3--42 (2020).
\newblock \doi{10.1177/004912411772970}

\bibitem{garten2018dictionaries}
J.~Garten, J.~Hoover, K.M. Johnson, R.~Boghrati, C.~Iskiwitch, M.~Dehghani,
  Dictionaries and distributions: Combining expert knowledge and large scale
  textual data content analysis.
\newblock Behavior Research Methods \textbf{50}, 344--361 (2018).
\newblock \doi{10.3758/s13428-017-0875-9}

\bibitem{newman2003}
M.L. Newman, J.W. Pennebaker, D.S. Berry, J.M. Richards, {Lying Words:
  Predicting Deception from Linguistic Styles}.
\newblock Personality and Social Psychology Bulletin \textbf{29}, 665--675
  (2003).
\newblock \doi{10.1177/0146167203029005010}

\bibitem{pennebaker2014small}
J.W. Pennebaker, C.K. Chung, J.~Frazee, G.M. Lavergne, D.I. Beaver, {When Small
  Words Foretell Academic Success: The Case of College Admissions Essays}.
\newblock PloS one \textbf{9}, e115,844 (2014).
\newblock \doi{10.1371/journal.pone.0115844}

\bibitem{bradyetal2020}
W.J. Brady, M.J. Crockett, J.J.V. Bavel, The {MAD} {M}odel of {M}oral
  {C}ontagion: {T}he {R}ole of {M}otivation, {A}ttention, and {D}esign in the
  {S}pread of {M}oralized {C}ontent {O}nline.
\newblock Perspectives on Psychological Science \textbf{15}, 978--1010 (2020).
\newblock \doi{10.1177/1745691620917336}

\bibitem{boyddevelopment2022}
R.L. Boyd, A.~Ashokkumar, S.~Seraj, J.W. Pennebaker, The {D}evelopment and
  {P}sychometric {P}roperties of {LIWC-22}.
\newblock Tech. rep., University of Texas at Austin Austin TX (2022).
\newblock \urlprefix\url{https://www.liwc.app}

\bibitem{Hutto2014}
C.~Hutto, E.~Gilbert, {VADER}: A parsimonious rule-based model for sentiment
  analysis of social media text.
\newblock Proceedings of the International {AAAI} Conference on Web and Social
  Media \textbf{8}, 216--225 (2014).
\newblock \doi{10.1609/icwsm.v8i1.14550}

\bibitem{kessler2017scattertext}
J.S. Kessler, in \emph{Proceedings of {ACL}-2017 System Demonstrations}
  (Association for Computational Linguistics, Vancouver, Canada, 2017).
\newblock \doi{10.48550/arXiv.1703.00565}

\bibitem{Kozyreva20}
A.~Kozyreva, S.~Lewandowsky, R.~Hertwig, Citizens {Versus} the {Internet}:
  {Confronting} {Digital} {Challenges} {With} {Cognitive} {Tools}.
\newblock Psychological Science in the Public Interest \textbf{21}, 103--156
  (2020).
\newblock \doi{10.1177/1529100620946707}

\bibitem{grinberg2019fake}
N.~Grinberg, K.~Joseph, L.~Friedland, B.~Swire-Thompson, D.~Lazer, Fake news on
  {T}witter during the 2016 {US} presidential election.
\newblock Science \textbf{363}, 374--378 (2019).
\newblock \doi{10.1126/science.aau2706}

\bibitem{pennycook2021shifting}
G.~Pennycook, Z.~Epstein, M.~Mosleh, A.A. Arechar, D.~Eckles, D.G. Rand,
  Shifting attention to accuracy can reduce misinformation online.
\newblock Nature \textbf{592}, 590--595 (2021).
\newblock \doi{10.1038/s41586-021-03344-2}

\bibitem{newsguard2022}
I.~{NewsGuard}.
\newblock Rating process and criteria.
\newblock Internet Archive,
  \url{https://web.archive.org/web/20200630151704/https://www.newsguardtech.com/ratings/rating-process-criteria/}
  (2020).
\newblock Accessed: 2022-04-20

\bibitem{lasser2022misinformation}
J.~Lasser.
\newblock List of domain accuracy and transparency scores v1.1 (2022).
\newblock \doi{10.5281/ZENODO.6536692}

\bibitem{Guess20a}
A.M. Guess, B.~Nyhan, J.~Reifler, Exposure to untrustworthy websites in the
  2016 {U.S.} election.
\newblock Nature Human Behavior \textbf{4}, 472--480 (2020).
\newblock \doi{10.1038/s41562-020-0833-x}

\bibitem{Guess19}
A.M. Guess, J.~Nagler, J.~Tucker, Less than you think: Prevalence and
  predictors of fake news dissemination on {F}acebook.
\newblock Science Advances \textbf{5}, eaau4586 (2019).
\newblock \doi{10.1126/sciadv.aau4586}

\bibitem{Rathje2021-fn}
S.~Rathje, J.J. Van~Bavel, S.~van~der Linden, Out-group animosity drives
  engagement on social media.
\newblock Proceedings of the National Academy of Sciences \textbf{118} (2021).
\newblock \doi{10.1073/pnas.202429211}

\bibitem{Osmundsen2021-gn}
M.~Osmundsen, A.~Bor, P.B. Vahlstrup, A.~Bechmann, M.B. Petersen, {Partisan
  Polarization Is the Primary Psychological Motivation behind Political Fake
  News Sharing on Twitter}.
\newblock American Political Science Review \textbf{115}, 999--1015 (2021).
\newblock \doi{10.1017/S0003055421000290}

\bibitem{Brulle12}
R.J. Brulle, J.~Carmichael, J.C. Jenkins, Shifting public opinion on climate
  change: an empirical assessment of factors influencing concern over climate
  change in the {U.S.}, 2002--2010.
\newblock Climatic Change \textbf{114}, 169--188 (2012).
\newblock \doi{10.1007/s10584-012-0403-y}

\bibitem{Gonawela18}
A.~Gonawela, J.~Pal, U.~Thawani, E.~van~der Vlugt, W.~Out, P.~Chandra, Speaking
  their {Mind}: {Populist} {Style} and {Antagonistic} {Messaging} in the
  {Tweets} of {Donald} {Trump}, {Narendra} {Modi}, {Nigel} {Farage}, and
  {Geert} {Wilders}.
\newblock Computer Supported Cooperative Work (CSCW) \textbf{27}, 293--326
  (2018).
\newblock \doi{10.1007/s10606-018-9316-2}

\bibitem{twarc}
E.~Summers, I.~Brigadir, S.~Hames, H.~Van~Kemenade, P.~Binkley, {Tinafigueroa},
  N.~Ruest, {Walmir}, D.~Chudnov, {Recrm}, {, Celeste}, {Hause Lin}, A.~Chosak,
  {R. Miles McCain}, I.~Milligan, A.~Segerberg, {Daniyal Shahrokhian},
  M.~Walsh, L.~Lausen, N.~Woodward, F.V. M\"{u}nch, {Eggplants}, {Ashwin
  Ramaswami}, D.~Hereñú, D.~Milajevs, F.~Elwert, K.~Westerling, {Rongpenl},
  S.~Costa, {, Shawn}.
\newblock {DocNow}/twarc: v2.10.4 (2022).
\newblock \doi{10.5281/ZENODO.6503180}

\bibitem{bojanowski2016enriching}
P.~Bojanowski, E.~Grave, A.~Joulin, T.~Mikolov, Enriching {Word} {Vectors} with
  {Subword} {Information}.
\newblock Transactions of the Association for Computational Linguistics
  \textbf{5}, 135--146 (2017).
\newblock \doi{10.1162/tacl_a_00051}

\bibitem{DiNatale21}
A.~{Di Natale}, M.~Pellert, D.~Garcia, Colexification {Networks} {Encode}
  {Affective} {Meaning}.
\newblock Affective Science \textbf{2}, 99--111 (2021).
\newblock \doi{10.1007/s42761-021-00033-1}

\bibitem{DiNatale2023}
A.~Di~Natale, D.~Garcia, {LEXpander: Applying colexification networks to
  automated lexicon expansion}.
\newblock {Behaviour Research Methods}  (2023).
\newblock \doi{10.3758/s13428-023-02063-y}

\bibitem{franccois2008}
A.~Fran{\c{c}}ois, in \emph{From polysemy to semantic change: {T}owards a
  typology of lexical semantic associations} (John Benjamins Publishing
  Company, Philadelphia, {PA}, 2008), p. 163.
\newblock \doi{10.1075/slcs.106.09fra}

\bibitem{jackson2019}
J.C. Jackson, J.~Watts, T.R. Henry, J.M. List, R.~Forkel, P.J. Mucha, S.J.
  Greenhill, R.D. Gray, K.A. Lindquist, Emotion semantics show both cultural
  variation and universal structure.
\newblock Science \textbf{366}, 1517--1522 (2019).
\newblock \doi{10.1126/science.aaw8160}

\bibitem{PALAN201822}
S.~Palan, C.~Schitter, Prolific.ac -- a subject pool for online experiments.
\newblock Journal of Behavioral and Experimental Finance \textbf{17}, 22--27
  (2018).
\newblock \doi{10.1016/j.jbef.2017.12.004}

\bibitem{pennington2014glove}
J.~Pennington, R.~Socher, C.D. Manning, in \emph{Proceedings of the 2014
  conference on empirical methods in natural language processing ({EMNLP})}
  (2014), pp. 1532--1543.
\newblock \doi{10.3115/v1/D14-1162}

\bibitem{mikolov2013efficient}
T.~Mikolov, K.~Chen, G.~Corrado, J.~Dean, Efficient estimation of word
  representations in vector space.
\newblock arXiv  (2013).
\newblock \doi{10.48550/arXiv.1301.3781 Focus to learn more}

\bibitem{Scheffer2017}
M.~Scheffer, I.~van~de Leemput, E.~Weinans, J.~Bollen, The rise and fall of
  rationality in language.
\newblock Proceedings of the National Academy of Sciences \textbf{118} (2021).
\newblock \doi{10.1073/pnas.2107848118}

\bibitem{bhadani2022political}
S.~Bhadani, S.~Yamaya, A.~Flammini, F.~Menczer, G.L. Ciampaglia, B.~Nyhan,
  Political audience diversity and news reliability in algorithmic ranking.
\newblock Nature Human Behaviour \textbf{6}, 495--505 (2022).
\newblock \doi{10.1038/s41562-021-01276-5}

\bibitem{lme4}
D.~Bates, M.~M{\"a}chler, B.~Bolker, S.~Walker, Fitting linear mixed-effects
  models using {lme4}.
\newblock Journal of Statistical Software \textbf{67}, 1--48 (2015).
\newblock \doi{10.18637/jss.v067.i01}

\bibitem{seabold2010statsmodels}
S.~Seabold, J.~Perktold, in \emph{9th Python in Science Conference} (2010)

\bibitem{Ou-Yang_undated-ky}
L.~Ou-Yang.
\newblock Newspaper3k: Article scraping \& curation.
\newblock https://github.com/codelucas/newspaper

\bibitem{Boyd22}
R.L. Boyd, A.~Ashokkumar, S.~Seraj, J.W. Pennebaker, The development and
  psychometric properties of {LIWC-22}.
\newblock Tech. rep., University of Texas at Austin, Austin, TX (2022)

\bibitem{grootendorst2022bertopic}
M.~Grootendorst, {BERTopic}: Neural topic modeling with a class-based {TF-IDF}
  procedure.
\newblock {arXiv}  (2022).
\newblock \doi{10.48550/arXiv.2203.05794}

\bibitem{mcinnes2018umap}
L.~McInnes, J.~Healy, J.~Melville, {UMAP}: Uniform manifold approximation and
  projection for dimension reduction.
\newblock {arXiv}  (2018).
\newblock \doi{10.48550/arXiv.1802.03426}

\bibitem{campello2013density}
R.J. Campello, D.~Moulavi, J.~Sander, in \emph{Pacific-Asia conference on
  knowledge discovery and data mining} (Springer, 2013), pp. 160--172.
\newblock \doi{10.1007/978-3-642-37456-2_14}

\bibitem{egger2022topic}
R.~Egger, J.~Yu, A {T}opic {M}odeling {C}omparison {B}etween {LDA}, {NMF},
  {Top2Vec}, and {BERTopic} to {D}emystify {T}witter {P}osts.
\newblock Frontiers in Sociology \textbf{7} (2022).
\newblock \doi{10.3389/fsoc.2022.886498}

\bibitem{alhaj2022improving}
F.~Alhaj, A.~Al-Haj, A.~Sharieh, R.~Jabri, Improving {A}rabic cognitive
  distortion classification in {T}witter using {BERTopic}.
\newblock International Journal of Advanced Computer Science and Applications
  \textbf{13}, 854--860 (2022).
\newblock \doi{10.14569/IJACSA.2022.0130199}

\bibitem{nikita2016package}
M.~Nikita, M.M. Nikita.
\newblock Package ‘ldatuning’ (2016).
\newblock \urlprefix\url{https://CRAN.R-project.org/package=ldatuning}

\bibitem{cinelli2021echo}
M.~Cinelli, G.~De~Francisci~Morales, A.~Galeazzi, W.~Quattrociocchi,
  M.~Starnini, The echo chamber effect on social media.
\newblock Proceedings of the National Academy of Sciences \textbf{118},
  e2023301,118 (2021).
\newblock \doi{10.1073/pnas.2023301118}

\bibitem{pewresearch}
{Pew Research Center}.
\newblock https://www.pewresearch.org/ (2022).
\newblock Accessed 2022-06-15

\bibitem{Lin2022}
H.~Lin, J.~Lasser, S.~Lewandowsky, R.~Cole, A.~Gully, D.G. Rand, G.~Pennycook,
  High level of agreement across different news domain quality ratings.
\newblock PsyArXiv  (2022).
\newblock \doi{10.31234/osf.io/qy94s}

\bibitem{gallotti2020assessing}
R.~Gallotti, F.~Valle, N.~Castaldo, P.~Sacco, M.~De~Domenico, Assessing the
  risks of ‘infodemics’ in response to {COVID-19} epidemics.
\newblock Nature Human Behaviour \textbf{4}, 1285--1293 (2020).
\newblock \doi{10.1038/s41562-020-00994-6}

\bibitem{bufale}
Bufale.
\newblock https://www.bufale.net/.
\newblock Accessed: 2022-05-01

\bibitem{bufalopedia}
Bufalopedia.
\newblock Un catalogo di indagini e risorse antibufala.
\newblock https://bufalopedia. blogspot.com/p/siti-creatori-di-bufale.html.
\newblock Accessed: 2022-05-01

\bibitem{butac}
Butac.
\newblock The black list.
\newblock https://www.butac.it/the-black-list/.
\newblock Accessed: 2022-05-01

\bibitem{buzzfeednews}
{Buzzfeed News}.
\newblock Inside the partisan fight for your news feed.
\newblock https://www.buzzfeednews.com/article/craigsilverman/
  inside-the-partisan-fight-for-your-news-feed.
\newblock Accessed: 2022-05-01

\bibitem{columbiajournalism}
{Columbia Journalism Review}.
\newblock {CJR} index of fake-news, clickbait, and hate.
\newblock http://web.archive.org/web/20210720140548.
\newblock Accessed: 2022-05-01

\bibitem{fakenewswatch}
{Fake News Watch}.
\newblock https://web.archive.org/web/20180213181029.
\newblock Accessed: 2022-05-01

\bibitem{mediabiasfactcheck}
{Media Bias Fact Check}.
\newblock Media bias fact check.
\newblock https://mediabiasfactcheck.com.
\newblock Accessed: 2022-05-01

\bibitem{politifact}
Politifact.
\newblock Politifact’s guide to fake news websites and what they peddle.
\newblock
  https://www.politifact.com/article/2017/apr/20/politifacts-guide-fake-news-websites-and-what-they.
\newblock Accessed: 2022-05-01

\bibitem{zimdars}
M.~Zimdars.
\newblock My "fake news list" went viral. but made-up stories are only part of
  the problem.
\newblock
  https://www.washingtonpost.com/posteverything/wp/2016/11/18/my-fake-news-list-went-viral-but-made-up-stories-are-only-part-of-the-problem.
\newblock Accessed: 2022-05-01

\bibitem{Soroka19}
S.~Soroka, P.~Fournier, L.~Nir, Cross-national evidence of a negativity bias in
  psychophysiological reactions to news.
\newblock Proceedings of the National Academy of Sciences \textbf{116},
  18,888--18,892 (2019).
\newblock \doi{10.1073/pnas.1908369116}

\bibitem{tingley2014mediation}
D.~Tingley, T.~Yamamoto, K.~Hirose, L.~Keele, K.~Imai, Mediation: {R} package
  for causal mediation analysis.
\newblock Tech. rep., {UCLA} Statistics/American Statistical Association (2014)

\end{thebibliography}

\end{appendices}

\end{document}